\documentclass[10pt]{article}
\usepackage{amsmath,amssymb,graphicx}
\usepackage{amsthm}
\usepackage{euscript}
\usepackage{eufrak}

\batchmode

\newtheorem{lem}{Lemma}
\newtheorem{thm}{Theorem}

\newcommand{\pr}{\noindent{\bf Proof}. }
\newcommand{\rem}{\noindent{\bf Remark}. }
\newcommand{\rems}{\noindent{\bf Remarks}. }

\newcommand{\da}{\dagger}
\newcommand{\pa}{\partial}
\newcommand{\one}{\cO(1)}
\newcommand{\bpsi}{\bar \psi}
\newcommand{\bPsi}{\bar \Psi}
\newcommand{\const}{\textrm{const}}

\newcommand{\supp}{ \mathrm{ supp  }}
\newcommand{\hs}{ \hspace{1cm}}
\newcommand{\tr}{\textrm{ tr }}
\newcommand{\Tr}{\textrm{ Tr }}

\newcommand{\Vol}{\textrm{Vol}}

\newcommand{\tk}{\bbT^{-k}_{N -k}}
\newcommand{\tz}{\bbT^0_{N -k}}

\newcommand{\B}{\Big}
\newcommand{\blan}{\Big  \langle} 
\newcommand{\bran}{\Big  \rangle} 
 
\newcommand{\uA}{\underline{\cA}} 

\newcommand{\crit}{\textrm{crit}}
\newcommand{\sq}{\square}
\newcommand{\st}{\star}

\newcommand{\be}{\begin{equation}}
\newcommand{\ee}{\end{equation}}
\newcommand{\bs}{\begin{split}}
\newcommand{\es}{\end{split}}

\newcommand{\bom}{\mathbf{\Omega}}

\newcommand{\ba}{\mathbf{a}}

\newcommand{\bb}{\mathbf{b}}

\newcommand{\sZ}{\mathsf{Z}}

\newcommand{\sN}{\mathsf{N}}

\newcommand{\sB}{\mathsf{B}}
\newcommand{\sx}{\mathsf{x}}

\newcommand{\al}{\alpha}
\newcommand{\De}{\Delta}
\newcommand{\de}{\delta}
\newcommand{\ga}{\gamma}
\newcommand{\Ga}{\Gamma}
\newcommand{\ka}{\kappa}

\newcommand{\la}{\lambda}
\newcommand{\Om}{\Omega}
\newcommand{\om}{\omega}

\newcommand{\ep}{\epsilon}
\newcommand{\si}{\sigma}
\newcommand{\Si}{\Sigma}

\newcommand{\vep}{\varepsilon}

\newcommand{\cA}{ \EuScript{A} }
\newcommand{\fD}{\EuFrak{D}} 
 
\newcommand{\fS}{\EuFrak{S}} 
 
\newcommand{\fA}{\EuFrak{A}}
\newcommand{\fG}{\EuFrak{G}}

\newcommand{\cB}{{\cal B}}

\newcommand{\cD}{{\cal D}}

\newcommand{\cO}{{\cal O}}

\newcommand{\cH}{{\cal H}}
\newcommand{\cS}{{\cal S}}

\newcommand{\cR}{{\cal R}}

\newcommand{\cF}{{\cal F}}

\newcommand{\cG}{{\cal G}}
\newcommand{\cM}{{\cal M}}
\newcommand{\cN}{{\cal N}}
\newcommand{\cW}{{\cal W}}
\newcommand{\cQ}{{\cal Q}}

\newcommand{\cZ}{{\cal Z}}

\newcommand{\bbR}{{\mathbb{R}}}
\newcommand{\bbZ}{{\mathbb{Z}}}

\newcommand{\bbT}{{\mathbb{T}}}

\begin{document}

\title{Multiscale block averaging  for QED in d=3}
\author{ 
J. Dimock
\thanks{dimock@buffalo.edu} \thanks{A version of this paper was originally posted with the title "Ultraviolet Stability for QED in d=3 - part I"}
 \\
Dept. of Mathematics \\
SUNY at Buffalo \\
Buffalo, NY, USA,  14260 }
\maketitle

\begin{abstract}  
We   continue  the   study of  the ultraviolet problem  for  QED in d=3.    The model is defined on a fine  toroidal  lattice and  we  seek  control as the lattice spacing goes to zero.
The problem is analyzed   using Balaban's formulation of the renormalization group.  This involves a sequence of transformations
consisting of a split into large and small field regions, then  block averaging, and then scaling.   The the effective actions generated by this method depend 
strongly on certain multi-scale propagators and minimizers.  The study of these objects both for fermions and for gauge fields  is content of this paper.   Earlier work on the
subject is reviewed.  In addition for fermions  a polymer expansion is obtained for the determinants of the fermion propagators.  
For the gauge field a detailed local regularity result is obtained for the minimizers. 
\end{abstract} 

 
 \section{Introduction}

 We study quantum electrodynamics (QED)  on   a Euclidean space-time of dimension  $d=3$, continuing the analysis of  \cite{Dim15b}.  
 We  work on the toroidal lattices     $  ( L^{-N} \bbZ / \bbZ )^3$ where   
where  $L$  is  a    (large)  positive odd   integer.  These have   lattice spacing  
  $\ep  = L^{-N}$   and unit volume.     
 We  consider the partition function   
 \be  
\sZ(N,e)    =     \int    \exp \B(  -  \frac12   \| d \cA  \|^2 -  <  \bar \psi,   (\fD _{\cA}    +  \bar   m) \psi >    +   m^N  <  \bar \psi,   \psi > +   \vep^N  \B)
    \   D  \bpsi \      D  \psi      \ D  \cA      
 \ee
 Here  $\bpsi_{\al}(x), \psi_{\al}(x)$  are the fermion fields, elements of a Grassmann algebra.   The abelian gauge    field     $\cA$ is the electromagnetic potential.
 It   is a function on  bonds in  the lattice,  and has a   field strength $d \cA$ which is a function on  plaquettes.   The      $\bar m$ is  the  bare fermion   mass
 and   
\be      \fD_{ \cA} =   \ga \cdot \nabla_{\cA}  -   \frac12   \ep  \De_{\cA}
\ee
is the Dirac operator  on the lattice  with    Wilson  correction.   Here $\ga = (\ga_0, \ga_1, \ga_2)$ with the anti commutator $\{ \ga_{\mu}, \ga_{\nu} \} =2 \de_{\mu \nu} $.   The  operator  $\nabla_{\cA}$  is  the   covariant symmetric lattice derivative  defined with   coupling constant  (charge)   $e$,   and  $\De_{\cA}$ is the associated  covariant Laplacian.   The     $m^N, \vep^N$  are counter terms. ($N$ is a superscript, not a power.)
For precise definitions see \cite{Dim15b}.

In this series of  papers   the goal is to     prove that   one can choose the counter terms   so that 
 an     ultraviolet stability  bound holds.    The bound says   that if  $e$  is sufficiently small    there are constant  $K_{\pm}$ independent of  $N$ such that  the  relative partition function satisfies
\be      K_-   \leq  \frac{\sZ(N,e)  }{\sZ(N,0)}      \leq  K_+  
\ee
Here    $\sZ(N,0)$ is the free field partition function defined with $\fD_0$.

This result would be  a key first step  in  controlling the ultraviolet properties of the model.   With  modifications  one can expect to get
 uniform    bounds on  the correlation functions,   and to show that the  correlation functions have
limits as $N \to \infty$.

Scaling will play  an important role in the  analysis,  so  we  consider general toroidal lattices
    \be
   \bbT^{-N}_{N'}   = ( L^{-N} \bbZ /  L^{N'} \bbZ )^3
\ee
 with lattice spacing   $L^{-N}$ and  linear dimension  $L^{N'}$.   Our starting point is then   $\bbT^{-N}_0$,  but   
 we  immediately scale up   to     $\bbT^0_N$, which has unit lattice spacing and linear dimension  $L^N$.
 Up to a  multiplicative  constant   we have  with fields    $  \bPsi_0 ,  \Psi_0,  A_0$  defined on  $\bbT^0_N$ 
  \be    \label{plum}
\sZ(N,e)    =     \int    \exp \B( - \frac12   \|  d A_0  \|^2   +  \blan  \bar \Psi_0,    (\fD _{A_0}    +  \bar   m^N_0)      \Psi_0 \bran    +   m^N_0  \blan  \bar \Psi_0,   \Psi_0   \bran 
+   \vep^N_0   \Vol(\bbT^0_N) \B)    \      D  \Psi_0      \ D  A_0     
 \ee
 The covariant derivatives are now defined with  coupling constant   $e^N_0$
 and   all coupling constants and masses  have scaled to the tiny quantities  
 \be    
   e^{N}_0  =   L^{-\frac12  N} e  \hs     \bar m^{N}_0  =  L^{-N}   \bar m   \hs    m^{N}_0  =  L^{-N}  m^N  
\hs           \vep^{N}_0 =L^{-3N}  \vep^N  
 \ee 
  In the following we omit the superscript $N $  writing  $e_0,  \bar   m_0, m_0, \vep_0 $.  
 \bigskip

The   strategy is to use a renormalization group method pioneered by Balaban   \cite{Bal82a} - \cite{Bal98b}.  
In this formulation   renormalization group transformations are defined by   block 
averaging  over blocks of size $L$ followed by  scaling down by $L$ to return to a unit lattice.   The  issue is to control the flow of  these transformations.        In  a previous  paper  \cite{Dim15b} this was accomplished with the approximation that  the  field  strength  $dA$   is bounded.   
For the full model the strategy is that at each renormalization group step one  breaks down the effective density  as a sum over  small field blocked  regions where the field strength
 is bounded as in \cite{Dim15b}  and complementary large field regions   where the bound is violated at at least one point in each block.    The bound in the
small field region is generous    and  after $k$ steps roughly   has the form $|dA| \leq   (\log e_k^{-1} ) ^p$   where $e_k = L^{\frac12k}e_0  = L^{ - \frac12 (N-k)} e$ is a tiny running coupling constant
and $p$ is a positive integer. 
The small field region is analyzed  as in \cite{Dim15b}.  The effective action after $k$ steps  has leading  terms of   the form of (\ref{plum}) now with new parameters $e_k, \bar m_k, m_k, \vep_k$.  Then there are  also 
extensive local corrections.    In the large field region one  identifies a factor  $e^{-\frac12 \| dA \|^2} $ in each block.   This is smaller  than any power of 
$e_k$ so the large field region gives a tiny contribution and  this  controls   the sum over regions.

In this paper we establish some key technical results needed to carry out this plan.   The division into large and small field regions at each step together with
the block averaging in the small field region  generates multi-scale expressions for the effective actions.  
 In section 2 we study the Green's functions (propagators),  critical points, and fluctuation covariances for certain multiscale Dirac operators that arise in this treatment. 
 This  follows the analysis of   Balaban, O'Carroll, and Shor \cite{BOS91} whose main technical tool is   a multiscale random walk expansion.   We start by reviewing their results.   Then we introduce a   modified propagator in Proposition \ref{sweet4},
 and  give a similar treatment.   In Proposition \ref{hughes} we use the random walk expansions to give 
  polymerized versions of these propagators, i.e.  we write them as  sums  of  terms localized in  polymers  (polymer  = a connected union of  blocks) . 
  In Proposition \ref{thrush1} we treat the determinant of the multiscale propagators  using the modified propagators  and show that it can be written as the exponential of  a
  sum of terms localized in polymers.    
  In section 3  we  consider the gauge field and study  the associated  multiscale  Green's functions, minimizers, and fluctuation covariances. 
This follows  the treatment of Balaban \cite{Bal84a}  - \cite{Bal85b}, \cite{Bal85d}  who   considers both axial and Landau gauges and the relation between them.
 He also develops     multiscale random walk expansions for the Landau gauge.   
We  first review and 
enlarge upon these results.   Then in Theorem \ref{REG} we prove a key  local regularity result for the Landau gauge  minimizers.  This  result is essential for using 
the gauge  fields as background fields in the fermion propagators. 
\bigskip

\textbf{Notation}:
\begin{enumerate}
\item
Throughout the paper   the convention is    that   $\one$ is a constant   independent of all parameters.  Also  
$C, \ga$  are  constants   ($C \geq 1,  \ga \leq 1$)    which may depend on $L$ and which    may change from line to line.

\item  Our basic fermion fields   are  Grassmann elements  $\Psi_{\al}(x)$ 
 and    $\bPsi_{\al}(x)$   indexed by  $x$ in some torus and  $1\leq   \al  \leq 4 $.     
 Integration over Grassmann variables uses the notation  
 \be 
   D\Psi  =  \prod_{x, \al} d(\Psi_{\al} (x) ) d(\bPsi_{\al} (x) )
 \ee

\item If $T$ is an operator on functions on  a lattice and $X$ is a subset of that lattice then we define the operator restricted to $X$ by  $T_X = 1_X T 1_X$.  If $T$ is an operator
on functions on bonds in the lattice the definition is the same, but $1_X$ restricts to bonds with at least one end in $X$.

\end{enumerate}

 \section{Block averaging for fermions}

 \subsection{global   averaging}  \label{2.1} 
 
 The renormalization group transformations depend on a block averaging operation  developed by Balaban, O'Carroll, and Schor \cite{BOS89}, \cite{BOS91}. 
 We start by reviewing the global version as in  \cite{Dim15b}, then move on to a multiscale version

Beginning  with a density  $\rho_0( A_0, \Psi_0)$  with   $A_0, \Psi_0$  defined      on  $\bbT^0_N $,    we  create   a sequence of densities  $\rho_k(  \cA,  \Psi_k)$
 defined for   $\cA$  on   $\tk$ and  $\Psi_k$    on      $ \bbT^0_{N-k}$.      They are defined recursively    
first by 
 \begin{equation}
\begin{split}  \label{kth}
 \tilde  \rho_{k+1} ( \cA,   \Psi_{k+1}) 
=  & \int    \de_G\Big( \Psi_{k+1} -  Q(\cA ) \Psi_k  \Big)  \rho_{k} (\cA,  \Psi_k)   D \Psi_k \\
 \end{split}
\end{equation}
where  $\Psi_{k+1}$   are  new Grassmann  variables    defined on the coarser lattice  $\bbT^1_{N-k}$.
 The   $\de_G$ is a Gaussian approximation to  
the delta function.  For a constant  $b = \one$  and $N_k$  it is defined by  
\be  \label{spring2}
\begin{split}
\de_G\Big( \Psi_{k+1} -  Q(A) \Psi_k   \Big)  
 =   &  N_k
   \exp  \B(  -    bL^{-1}   \B \langle   \bPsi_{k+1} -  Q(-A) \bPsi_k,    \Psi_{k+1} -  Q(A) \Psi_k  \B  \rangle   \B)       \\
\end{split}
\ee
   The averaging operator  $Q(\cA)$   has the form with $e_k=L^{\frac12k} e$
  \be    \label{dice2}
(Q(\cA) \Psi_k)(y)    =   L^{-3}   \sum_{x \in B(y)} e^{ ie_k \eta (\tau\cA)(y,x) }  \Psi_k(x)  
 \ee
Here    $B(y)$ is a cube with $L$ sites on a side centered on $y$ and 
\be \label{eleven}
(\tau\cA)(y,x) = 1/3!  \sum_{\pi} \cA(\Ga^{\pi} (y,x) ) 
\ee
  is an average over rectilinear paths from  $y$ to $x$.  For any such   path  $\Ga$ the expression $\cA(\Ga)=  \sum_{b \in \Ga} \cA(b)$ is an unweighted  sum over bonds   of length  $\eta  =L^{-k}$. 
The    constant    $ N_k $  is chosen so  $\int  D\Psi_{k+1}   \de_G  ( \Psi_{k+1}  )  =1$  and therefore   
\be     \label{bell}
 \int  \tilde   \rho_{k+1} (\cA,  \Psi_{k+1})\    D \Psi_{k+1}  =     \int     \rho_{k} (\cA,  \Psi_{k} )\   D \Psi_k
\ee

Next     one  scales  back to the unit lattice.   
 If   $\cA$ is a field  on   $\bbT^{-k-1}_{N-k-1}$   and  $ \Psi_{k+1}$ is a field   on $\bbT^{0}_{N-k-1}$   then
   then   
\be
\cA_{L}(b)  = L^{-1/2}   \cA( L^{-1}b) \hs  \Psi_{k+1,L}(x) =  L^{-1}\Psi_{k+1} (L^{-1}x)
\ee
are  fields  on  $\tk$   and  $\bbT^1_{N-k}$ respectively,   and we  define     
  \begin{equation}   \label{scaleddensity}
 \rho_{k+1 } ( \cA,  \Psi_{k+1})  =  \tilde  \rho_{k+1} (\cA_L, \Psi_{k+1,L}) L^{-8(s_N  - s_{N-k-1}) }   
 \end{equation}
 Here    $s_N =  L^{3N}$ is the number of sites in a 3 dimensional lattice with  $L^N$ sites on a side, and the 8 occurs since there
 are 8 fields at each site. 
Then one finds that 
 \begin{equation}  \label{preserve}
  \int       \rho_{k+1} (\cA,   \Psi_{k+1})   D \Psi_{k+1}
    =   L^{-8s_N} \int       \rho_{k} ( \cA_L,  \Psi_{k})    D \Psi_k 
\end{equation}
From this one can deduce  that for       $\cA$ on  $\tk$  and  $\Psi_k$ on   $\tz$ and  $\psi $ on   $\tk$
\be   \label{lincoln0}
   \int  \rho_{k } ( \cA,  \Psi_k)  D \Psi_k   =      \int   \rho_{0} (\cA_{ L^k}, \psi _{L^k} )   \  D \psi    
\ee
Thus  we are normalizing not to the integral of  $\rho_0$  but to a scaled  version.   If  $\rho_0$ 
came from  scaling up  a density  on $\bbT^{-N}_0$  this eventually returns us to an integral  over
this lattice.

The  individual   RG transformations can be composed into a single transformation.    Let  $Q_k(\cA) $
be  the $k$-fold   composition of   $ Q(\cA)$ defined on $\tk$ by
\be
Q_k(\cA) = Q(\cA) \circ \cdots \circ Q(\cA)
\ee
Then one can show    \cite{Dim15b} 
 that for     $\cA, \psi$  on   $\tk$  and  $\Psi_{k}$  on $\bbT^0_{N-k}$     
\be 
\label{two}
 \rho_{k} ( \cA,  \Psi_k)  
 =  \cN_k  \ \int     \exp  \B(  -  b_k      \B \langle   \bPsi_{k} -  Q_k(-\cA) \bpsi,    \Psi_{k} -  Q_k(\cA) \psi  \B  \rangle   \B)    
    \rho_0   (\cA_{L^k}, \psi _{L^k}) \ \  D  \psi    
\ee
where     
$b_k  =   b  (1-L^{-1})(1- L^{-k})^{-1}$
  and  $\cN_k$ is a normalizing  constant.

  Now suppose  $\rho_0$ is     a perturbation of the free  fermion   action:    
   \be 
    \rho_{0}(A,   \Psi_0)  = F_0( \Psi_0)    \exp \B(  - \blan  \bPsi_0,  ( \fD_{\cA} +\bar  m_0)    \Psi_0   \bran   \B)
\ee
Substitute this in (\ref{two}) and evaluate the integral by diagonalizing the quadratic form in $\psi$.  This involves the critical point for the form
which  comes at $\psi = \psi_k(\cA, \Psi_k)  \equiv \cH_k(\cA) \Psi_k$ where
\be  \label{route65}
\begin{split} 
         \cH_k(\cA )     =   &
\begin{cases}          b_k  S_k(\cA) Q^T_k(-\cA)   &    \textrm{   on   }  \Psi_k   \\
   b_k  S^T_k(\cA) Q^T_k(\cA)   &    \textrm{   on   }  \bPsi_k   \\
\end{cases}  \\
 \cS_k(\cA )  =&\B( \fD_{\cA} +\bar     m_k   +  b_k  P_k(\cA)  \B)^{-1} \hs   P_k(\cA) =Q_k^T(-\cA)Q_k(\cA) 
 \\ 
 \end{split}
 \ee
 Here $\fD_{\cA}$ and $P_k(\cA)$ are defined with the scaled coupling constant $e_k \equiv  L^{\frac12 k}e_0 = L^{-\frac12(N-k)}e$
 and the scaled fermion mass $\bar m_k = L^k \bar m_0 =L^{-(N-k)} \bar m$.
 Our expression  becomes 
 \be   \label{spinit}
   \rho_{k} (\cA,  \Psi_k)    =   \cN_k    \sZ_k(\cA)  F_k (  \psi_k(\cA) )   
      \exp  \Big( -    \blan \bPsi_k,    D_k(\cA)  \Psi_k\bran    \Big)
  \ee
where
\be  \label{route66}
\begin{split} 
 D_k(\cA)    = &      b_k -  b_k^2 Q_k(\cA)S_k(\cA)Q^T_k(-\cA)    \\
   F_k (\psi) 
    =  &   \sZ_k(\cA)  ^{-1}  \int      F_{0, L^{-k}}( \psi  +  \cW ) 
       \exp \B(   -    \blan  \bar \cW, \B( \fD_{\cA} +\bar     m_k   +  b_k  P_k(\cA)  \B)  \cW   \bran  \B)    D  \cW     \\
 \sZ_k(\cA)   =  & \int  \exp \B(   -    \blan  \bar \cW, \B( \fD_{\cA} +\bar     m_k   +  b_k  P_k(\cA)  \B)  \cW   \bran  \B)   D  \cW  = \det (\cS_k(\cA) )^{-1}
  \\
 \end{split}
 \ee
 and $ \bar \cW, \cW$ are new Grassmann variables defined on $\tk$.

If   we apply  the basic  RG transformation to the expression  (\ref{spinit}) we  get a new expression  for    $ \rho_{k+1} (\cA,  \Psi_{k+1})  $.
There is a new critical point at $\Psi_k =  \Psi_k^{crit}(\cA,\Psi_{k+1})= H_{k} (\cA) \Psi_{k+1}$ where  
 \be 
\begin{split}  \label{min1}
  H_{k} (\cA)     =  &     \begin{cases}      bL^{-1} \Ga_k (\cA)  Q^{T}(-\cA)  &   \textrm{  on  }   \Psi_{k+1}  \\
    bL^{-1}\Ga^T_k (\cA)  Q^{T}(\cA)  &  \textrm{  on  }   \bPsi_{k+1}     \\
    \end{cases} \\
    \Ga_k (\cA)  =  &  \B(D_{k}(\cA) +  bL^{-1}P(\cA) \B)^{-1}  \hs  P(\cA) = Q^T(-\cA) Q(\cA) \\
\end{split}      
\ee
Expanding around the critical point on identifies a new Gaussian integral  $\int  \cdots d \mu_{\Ga_k(\cA)}$ with a new normalization factor 
\be
\de Z_k(\cA) =   \int  \exp \B(   -    \blan  \bar W, \B( D_{k}(\cA) +  bL^{-1}P(\cA)  \B)  W   \bran  \B)   D W  = \det (\Ga_k(\cA) )^{-1}
\ee
One establishes the following scaling identities 
  \be    \label{z}
\begin{split} 
 [\psi_{k+1}(\cA, \Psi_{k+1})]_L = & \psi_k\B(\cA_L,  \Psi_k^{crit}(\cA_L,\Psi_{k+1,L}) \B) \\
\cN_{k+1}\sZ_{k+1} (\cA)   = &    L^{-8(s_N   -s_{N-k-1})}  \cN_k   N_{k}     \sZ_k(\cA_L) \de  \sZ_k(\cA_L)   \  \\
 F_{k+1} (\psi  ) 
    =  &   \int    F_k\B(  \psi _L      +    \cH_k(\cA_L )  W\B) d \mu_{\Ga_k(\cA_L) }(W)   \\
\end{split} 
\ee

 For future reference we note what this says about the free fermion partitions function
 \be \label{26}
  Z^f(N,0) 
\equiv  \int \exp \B(  - \blan  \bPsi,  ( \fD_0 +\bar  m_0)    \Psi   \bran   \B) \ D\psi
\ee
  This is $\int \rho_0 $ in the case $F_0=1$ and so is the same as  $\int \rho_k$ for any $k$.
  But if $F_0=1$ then $F_k=1$ and so by (\ref{spinit}) we have
  \be
  Z^f(N,0) 
=   \cN_k    \sZ_k(0)    \int  
      \exp  \B( -    \blan \bPsi_k,    D_k(0)  \Psi_k\bran    \B)  D \Psi_k =  \cN_k    \sZ_k(0)  \det (   D_k(0)  )
\ee

   \subsection{multiscale averaging}  \label{knots}
 We  are also interested in the case where the averaging is only done in a subset of the lattice.   In the first step this means replacing $\Psi_0$ on $\bbT_N^0$  by
 $\Psi_1$ on   $\bbT^1_N$ in some region  $\Om_1 \subset   \bbT^0_N$  (eventually a region where the gauge field is small).   
   Let   $M = L^m$ for some large integer  $m$.    We partition  either lattice into $LM$ cubes centered on
   $\bbT_N^{m+1}$,  and  take $\Om_1$ to be an arbitrary  union of such cubes. 
 Starting with a density $ \rho_0(A_0, \Psi_0) $ on $\bbT_N^0$ we define a new density for $\Psi_{1, \Om_1}$ on  $ \Om_1 \subset \bbT^1_N$ by
 \be  \label{ouch1}
  \begin{split}
&  \tilde    \rho_{1, \Om_1}(A_0,  \Psi_{1,  \Om_1  } ,  \Psi_{0, \Om^c_1} ) \\
  & =   N_{1, \Om_1}
   \int       \exp \left(   - bL^{-1} \blan \bPsi_{1}- Q(- A_0  ) \bPsi_0,   \Psi_{1}- Q(A_0 ) \Psi_{0}  \bran_{\Om_1}  \right) 
   \rho_0(A_0, \Psi_0)  D \Psi_{0, \Om_1}   \\
\end{split}  
 \ee  
 Here
$   N_{1, \Om_1}^{-1}  =    \int       \exp (   - bL^{-1} < \bPsi_{1},   \Psi_{1} >_{\Om_1}  )  D \Psi_{1,  \Om_1  }    $ 
 so that 
\be \label{lunchtime}
\int   \tilde    \rho_{1, \Om_1}(A_0,  \Psi_{1,  \Om_1  } ,  \Psi_{0, \Om^c_1} )  \  D \Psi_{1,  \Om_1  }   D \Psi_{0, \Om^c_1}  
=  \int     \rho_0(A_0, \Psi_0)  D \Psi_{0}  
\ee

  Next    scale  to  a density   defined for   $\Om_1$ a   union  of $M$ cubes centered on  $\bbT^m_{N-1}$,  fields          $\cA, \psi $   on  
 $\bbT^{-1}_{N-1}$,   and   $\Psi_{1, \Om_1}$   on  $\Om_1 \subset \bbT^0_{N-1}$.  We replace $\Om_1$ by $L \Om_1$,  $\cA$ by $\cA_L$,  $\Psi_{0, \Om_1^c}$ by $ [ \psi_{\Om^c_1}]_L$, and 
 $ \Psi_{1,  \Om_1  } $ by $ [\Psi_{1, \Om_1}]_ L $ and define
 \be  \label{ouch2}
      \rho_{1, \Om_1}(\cA,  \Psi_{1,  \Om_1  },  \psi_{ \Om^c_1}  ) =  \tilde    \rho_{1, L\Om_1}(\cA_L ,  [\Psi_{1, \Om_1}]_ L  , [ \psi_{\Om^c_1}]_L   )
      \si_{1, \Om_1}
 \ee
Here $\si_{1, \Om_1}$ are scaling factors which are somewhat arbitrary in this restricted setting.   We define them so that combined with (\ref{lunchtime}) we have 
\be
\int  \rho_{1, \Om_1}(\cA,  \Psi_{1,  \Om_1  },  \psi_{ \Om^c_1}  ) D \Psi_{1,  \Om_1}  D  \psi_{ \Om^c_1} 
 =    \int     \rho_0(\cA_L , \psi_{L})  D \psi  
\ee
 Specifically  $\si_{1, \Om_1} = L^{- 8 ( | \Om_1| - | \Om_1^{(1)}| )}$  where $\Om^{(1)}_1$ are the centers of $L$-cubes in $\Om_1$.

 We repeat this operation many times, successively averaging and then scaling down, each time in a smaller region. 
 After $k$ steps we will have a density  of the form  $ \rho_{k, \bom}  (\cA,   \Psi_{k, \bom}, \psi_{\Om_1^c})  $.
 Here $\bom$ is a sequence of regions in $\tk$ of the form 
\be  \label{esther1}
  \bom =  ( \Om_1,  \cdots,  \Om_k  )
\hs
\Om_1 \supset  \Om_2  \supset  \cdots  \supset   \Om_k
\ee   
where  $\Om_j$  is a union of  $L^{-(k-j)}M$ cubes  in $\tk$.   The  gauge field    $\cA$ is defined   on $\tk$   and       
\be \label{esther2}
 \Psi_{k, \bom} =  \B(     \Psi_{1,\de \Om_1}, \dots ,  \Psi_{k-1,\de \Om_{k-1}},   \Psi_{k, \Om_k} \B)   
\ee
Here 
\be 
   \de  \Om_j  = \Om_j - \Om_{j+1}
\ee
and   $\Psi_{j, \de \Om_j} $ is defined on
$   \de  \Om^{(j)}_j  = \Om^{(j)}_j - \Om^{(j)}_{j+1}$ where in general $\Om^{(j)}$ denotes the centers of $L^j$ cubes in $\Om$.
Thus   $\Psi_{j, \de \Om_j} $ is defined on   
$ \de  \Om^{(j)}_j  =    \de  \Om_j \cap  \bbT^{-(k-j)}_{N-k}
$. 
 The field $ \Psi_{k, \Om_k} $ is
 defined on the unit lattice  $   \Om^{(k)}_k  =   \Om_k \cap  \bbT^0_{N-k}$. and $\psi_{ \Om^c_1} $ is defined on $\Om_1^c\subset \tk$.

In the next   step   we  introduce  $\Om_{k+1} \subset \Om_k $,   a union of  $LM$ blocks in $\tk$
and and a new
field $\Psi_{k+1, \Om_{k+1}}$ defined on $\Om^{(k+1)}_{k+1}  = \Om_{k+1} \cap       \bbT^1_{N-k} $.
Then define 
\be 
\begin{split}
\bom^+  = &   (\bom,  \Om_{k+1}     ) =   ( \Om_1,  \cdots,  \Om_k,  \Om_{k+1}     )  \\
 \Psi_{k+1, \bom^+} = &  \B(   \Psi_{1,\de \Om_1}, \dots ,    \Psi_{k, \de  \Om_k},\Psi_{k+1, \Om_{k+1}} \B)   \\
\end{split}
\ee
The averaged  density is first
\be  \label{salubrious} 
\begin{split}
&  \tilde    \rho_{k+1, \bom^+}  (\cA,  \Psi_{k+1,\bom^+ }, \psi_{ \Om^c_1} )  =  N_{k+1, \Om_{k+1}}  \\
  &  \int 
 \exp \left(   - bL^{-1}  \blan \bPsi_{k+1}- Q(- \cA ) \bPsi_k,   \Psi_{k+1}- Q(\cA ) \Psi_k   \bran_{\Om_{k+1} }  \right) 
    \rho_{k, \bom}  (\cA,   \Psi_{k, \bom},\psi_{ \Om^c_1})   D \Psi_{k, \Om_{k+1}}    \\
\end{split}  
 \ee  
 where $N_{k+1, \Om_{k+1}}^{-1} =  \int 
 \exp (   - bL^{-1}  < \bPsi_{k+1},   \Psi_{k+1} >_{\Om_{k+1} } ) 
    D \Psi_{k+1, \Om_{k+1}} $ is  chosen to preserve the integral. 
Then      scale    defining   for $\bom^+$ in $\bbT^{-k-1}_{N-k-1}$ and associated 
  $\Psi_{k+1,\bom^+}$ and  $\cA$
 \be   \label{nutsy}    \rho_{k+1,    \bom^+  } (   \cA,  \Psi_{k+1,\bom^+}, \psi_{ \Om^c_1} )
 =  \  \tilde    \rho_{k+1, L\bom^+}   (\cA_L,  [\Psi_{k+1, \bom^+ }]_L , [\psi_{ \Om^c_1}]_L) \si_{k+1, \bom^+}
 \ee
where the scaling factors  $\si_{k+1, \bom^+}$ are  chosen  so that for each $k$
\be \label{rhombus}
\int    \rho_{k, \bom}  (\cA,  \Psi_{k, \bom}, \psi_{\Om^c_1}    )  D \Psi_{k, \bom} D \psi_{\Om^c_1} 
   = \int \rho_{0}  (\cA_{L^k},  \psi_{L^k} )   D \psi    
\ee

By composing the averaging operators  we obtain another expression for  
$\rho_{k, \bom}$.   First define  for $\psi$ on $\tk$ a field of type (\ref{esther2}): 
\be
\begin{split}
Q_{k, \bom}(\cA) \psi  = 
 \B(( Q_1(\cA  )  \psi)_{\de \Om_1},  \dots,   ( Q_{k-1}(\cA  )  \psi)_{\de \Om_{k-1}} , (Q_k(\cA)  \psi)_{\Om_k}   \B)        
\end{split}
\ee        
We also define
\be
  \bb^{(k)}  =    (  b^{(k)} _1,   \dots,   b^{(k)}_k  )    \hs   b^{(k)}_j  =   b_j L^{k-j} \ee        
Here $  b^{(k)} _j$ acts on $\de \Om_j$  for $j<k$  and  $  b^{(k)} _k =b_k$
acts on $\Om_k$.

\begin{lem}    \label{one} 
\be
\begin{split}
&   \rho_{k, \bom}  (\cA,  \Psi_{k, \bom}, \psi_{\Om^c_1}    )\\
&  =   \cN_{k, \bom}   \int 
 \exp \left(   - \blan  (\bPsi_{k,\bom}- Q_{k, \bom} (- \cA ) \bpsi), \bb^{(k)} \    (  \Psi_{k, \bom}- Q_{k, \bom} (\cA ) \psi )  \bran_{\Om_1} \right) 
    \rho_{0}  (\cA_{L^k},  \psi_{L^k} )   D \psi_{\Om_1}       \\
\end{split}  
 \ee  
 where    $  \cN_{k, \bom}^{-1}   =  \int   \exp (   - <  \bPsi_{k,\bom}, \bb^{(k)} ,      \Psi_{k, \bom} >_{\Om_1} )D \Psi_{k, \bom} $.
 \end{lem}  
\bigskip

\pr  The case $k=1$ can be obtained by combining (\ref{ouch1}) and (\ref{ouch2}) and scaling the integration variable.   Assume it is true for  $k$.     Then  
\be  \label{listerine00}
\begin{split}
&  \tilde    \rho_{k+1, \bom^+}  (\cA,  \Psi_{k+1,\bom^+ }, \psi_{ \Om^c_1} ) = N_{k+1, \Om_{k+1} }   \int 
 \exp \B(   - bL^{-1}  \blan \bPsi_{k+1}- Q(- \cA ) \bPsi_k,   \Psi_{k+1}- Q(\cA ) \Psi_k   \bran_{\Om_{k+1} }  \\
 &  - \blan  (\bPsi_{k,\bom}- Q_{ k,\bom} (- \cA ) \bpsi),  \bb^{(k)} \    (  \Psi_{k, \bom}- Q_{k, \bom} (\cA ) \psi )  \bran_{\Om_1}    \B) 
    \rho_{0}  (\cA_{L^k},  \psi_{L^k} )   d \Psi_{k, \Om_{k+1}}  D \psi_{\Om_1}    \\
\end{split}  
 \ee  
 The quadratic form in the exponential has form 
 \be \label{faro}
 bL^{-1}  \blan \bPsi_{k+1}- Q(- \cA ) \bPsi_k,   \Psi_{k+1}- Q(\cA ) \Psi_k   \bran_{\Om_{k+1} }  
   + b_k \blan  (\bPsi_{k}- Q_k (- \cA ) \bpsi),  \    (  \Psi_k- Q_k(\cA ) \psi )  \bran_{\Om_{k+1}}
 +  \cdots
 \ee  
 where the omitted terms do not depend on  $\Psi_{k, \Om_{k+1}}$.   We evaluate the integral over  $\Psi_{k, \Om_{k+1}}$ by
 expanding around the critical point of this function.   As in  \cite{Dim15b} this is  on $\Om_{k+1}$
 \be
\begin{split}
   \Psi^{\bullet}_{k} ( \Psi_{k+1},  \psi )
 =&  Q_{k} (\cA ) \psi      + \frac{ bL^{-1}} {b_k + bL^{-1} }  Q^T( -\cA ) \Psi_{k+1}  
     -  \frac{bL^{-1}}{  b_k + bL^{-1}  }   Q^T(-\cA)      Q_{k+1} (\cA ) \psi        \\
\end{split}  
\ee
and similarly for $ \bPsi^{\bullet}_{k} ( \Psi_{k+1},  \psi )$.
 Insert   $ \Psi_k   =   \Psi^{\bullet}_{k} + W$ and $  \bPsi_k   =   \bPsi^{\bullet}_{k} +\bar  W$ 
 into (\ref{faro}).  
 As in \cite{Dim15b} the term with no $W$'s is
 \be   \label{lazy}
\begin{split}
&  bL^{-1}  \blan \bPsi_{k+1}- Q(- \cA )     \Psi^{\bullet}_{k},   \Psi _{k+1}- Q(\cA )     \Psi^{\bullet}_{k} \bran_{\Om_{k+1} } 
+  b_k  \blan  (     \Psi^{\bullet}_{k}- Q_k (- \cA ) \bpsi),    (       \Psi^{\bullet}_{k}- Q_k (\cA ) \psi )  \bran_{\Om_{k+1} }  + \dots   \\
&=    b_{k+1} L^{-1}    \blan \bPsi_{k+1}- Q_{k+1}(- \cA ) \bpsi ,   \Psi _{k+1}- Q_{k+1}(\cA ) \psi    \bran_{\Om_{k+1} } + \dots 
  \\
\end{split}
\ee
Combined with the omitted terms  in $\Om^c_{k+1}$ this becomes
\be   
  \blan  (\bPsi_{k+1, \bom^+}- Q_{k+1, \bom^+} (- \cA ) \bpsi), \bb^+ \    (  \Psi_{ k+1,\bom^+}- Q_{k+1, \bom^+} (\cA ) \psi )  \bran_{\Om_1}
  \ee
   where  $ \bb^+  =   ( \bb^{(k)}, b_{k+1}L^{-1}  )   $.
The cross terms vanish  and the   term quadratic in  $W$ is  
 \be   <\bar  W, (b_k + bL^{-1}P(\cA))  W  >_{\Om_{k+1}} \ee
 
 Make these substitutions in    (\ref{listerine00})    
 and integrate over $W$ instead of  $\Psi_k$.   
 The integral over $W$  just gives a constant
 \footnote{In fact the constant is independent of $\cA$ as well as the fermi fields.  This is so since the integral over $W$ gives $\det ( (b_k + bL^{-1}P(\cA))$.
 This depends on $\cA$ only through $\tr P(\cA)$ which is independent  of $\cA$, see \cite{Dim15b}}
and so
 \be  \label{listerine0}
\begin{split}
&  \tilde    \rho_{k+1, \bom^+}  (\cA,  \Psi_{\bom^+ } ,\psi_{ \Om^c_1}) =  \const  \int \   D \psi _{\Om_1}  \\
& 
 \exp \B(   - \blan  (\bPsi_{k+1, \bom^+}- Q_{k+1, \bom^+} (- \cA ) \bpsi), \bb^+ \    (  \Psi_{ k+1,\bom^+}- Q_{k+1, \bom^+} (\cA ) \psi )  \bran_{\Om_1}   \B) 
    \rho_{0}  (\cA_{L^k},  \psi_{L^k} )   \\
\end{split}  
 \ee  
The result now follows by scaling as in (\ref{nutsy}).  We make the change of variables  replacing $\psi$ on $\tk$ by $\psi_L$ with $\psi$ on $\bbT^{-k-1}_{N-k-1}$.
We use   $ Q_{k+1, L\bom^+} (\cA_L ) \psi_L = [ Q_{k+1, \bom^+} (\cA ) \psi ]_L $ which involves a change in the coupling constant in the parallel translation from $e_k$ to $e_{k+1}$.
Also $L\bb^+$ is identified as  $ \bb^{(k+1)}$
This yields 
\be  \label{listerine}
\begin{split}
&    \rho_{k+1, \bom^+}  (\cA,  \Psi_{\bom^+ } ,\psi_{ \Om^c_1}) =  \const   \int     D \psi _{\Om_1} \\
& 
 \exp \B(   - \blan  (\bPsi_{k+1, \bom^+}- Q_{k+1, \bom^+} (- \cA ) \bpsi), \bb^{(k+1)} \  (  \Psi_{ k+1,\bom^+}- Q_{k+1, \bom^+} (\cA ) \psi )  \bran _{\Om_1}  \B) 
    \rho_{0}  (\cA_{L^{k+1}},  \psi_{L^{k+1}} )     \\
\end{split}  
 \ee  
But the constant must be $\cN_{k+1, \bom^+}$  so that (\ref{rhombus}) is satisfied.

\subsection{free flow}   \label{sudsy}

 Now specialize to the case of interest which is     
\be 
 \rho_{0}  (A_0,  \Psi_0 ) 
 =    \exp  \B( -   \blan  \bPsi_0,(\fD_{A_0}   + \bar   m_0) \Psi_0  \bran     \B)   F_0(\Psi_0)  D\Psi_0
\ee
Here we have included   arbitrary   function $  F_0(\Psi_0) $.      
Then   with $F_{0,L^{-k}}(\psi) = F_0(\psi_{L^k})$
\be  \label{lurch0}
\begin{split}
&  \rho_{k, \bom}  (\cA,  \Psi_{k,\bom},\psi_{ \Om^c_1}   )  = \cN_{k, \bom}  \int  \  D \psi_{\Om_1}  
  \\
 &   \exp \left(   - \blan  (\bPsi_{ k,  \bom}- Q_{ k,  \bom} (- \cA ) \bpsi),   \bb^{(k)}  (  \Psi_{ k,  \bom}- Q_{ k, \bom} (\cA ) \psi )  \bran_{\Om_1}
  -  \blan   \bpsi,(\fD_{ \cA } + \bar   m_k) \psi  \bran  \right)    F_{0,L^{-k}}(\psi)    \\
\end{split}  
 \ee  
 We expand around the critical point in $\psi$ for the quadratic form in the exponential  with $\psi_{\Om_1^c}$ fixed.   
 The critical point  $\psi^{crit}$ on $\Om_1$ satisfies
 \be   \label{sister}
\begin{split}
 Q_{k, \bom}^T(-\cA)  \bb^{(k)} \B(\Psi_{k, \bom}-Q_{k, \bom}(\cA)\psi^{\crit} \B)  
 -  (\fD_{\cA} + \bar m_k ) \psi^{\crit}  -  \fD_{\cA} \psi_{\Om_1^c}   = & 0 \\
 Q_k^T( \cA)  \bb^{(k)} \B(  \bPsi_{k,\bom} -Q_k(-\cA)\bpsi^{\crit} \B) 
 -     (\fD_{\cA} + \bar m_k ) ^T \bpsi^{\crit}  -  \fD^T_{\cA} \bpsi_{\Om_1^c}= & 0 \\
\end{split}
\ee
Define operators on $\Om_1 \subset \tk$
 \begin{equation}  \label{lurch} 
\begin{split}
  P_{  k, \bom}(\cA)  =  &  Q^T_{ k, \bom}(-\cA)   \bb^{(k)}\  Q_{ k, \bom}(\cA) \\
S_{k,\bom}(\cA) =& \B[ \fD_{\cA}+   \bar m_k  +  P_{ k, \bom}(\cA)  \B]_{\Om_1}^{-1}
\hs  \\
\end{split}
\end{equation}
Then the equations (\ref{sister}) are solved by $\psi^{\crit}=  \psi_{k,\bom}(\cA )  $ and $\bpsi^{\crit}=   \bpsi_{k, \bom}(\cA ) $ 
 which take the fixed values on $\Om_1^c$ and on $\Om_1$ are given by 
  \begin{equation}  \label{lurch2} 
\begin{split}
\psi_{k, \bom}(\cA)  =    \psi_{k, \bom}(\cA, \Psi_{k, \bom},   \psi_{ \Om^c_1} )  \equiv  &  S_{k,\bom}(\cA)
\B(Q^T_{k,\bom}(-\cA)  \bb^{(k)}\  \Psi_{k,\bom}    - \fD_{\cA}   \psi_{\Om_1^c} \B)   
\\
 \equiv  &  \cH_{k, \bom}(\cA) \Psi_{k,\bom}    -  S_{k,\bom}(\cA)\fD_{\cA}   \psi_{\Om_1^c}    \\
\bpsi_{k, \bom}(\cA)  =      \bpsi_{k, \bom}(\cA, \bPsi_{k, \bom},   \bpsi_{ \Om^c_1} )   \equiv  & S^T_{k,\bom}(\cA)
\left(Q^T_{k,\bom}(\cA)   \bb^{(k)}\  \overline   \Psi_{k,\bom}   -  \fD^T_{\cA}   \bpsi_{\Om_1^c} \right)     \\
 \equiv  &  \cH_{k, \bom}(\cA) \bPsi_{k,\bom}    -  S_{k,\bom}(\cA)\fD^T_{\cA}   \bpsi_{\Om_1^c}    \\
\end{split}
\end{equation}
Next   make the change of variables    
 $\psi =  \psi_{k, \bom}(\cA)  + \cW $, $\bpsi =  \bpsi_{k, \bom}(\cA)  + \bar  \cW$ on $\Om_1$  and integrate over $\cW$ instead of  $\psi$.  The cross terms
 in the exponential  vanish and we have   
\begin{equation}  \label{fable}
 \rho_{k, \bom}  (\cA,  \Psi_{k,\bom}, \psi_{ \Om^c_1}    ) =  \cN_{k, \bom} \  \sZ_{k,\bom}(\cA) \ \exp \B(-  \fS_{k, \bom} \B(\cA, \Psi_{k, \bom},  \psi_{k, \bom} (\cA) \B)\B) 
F_{k, \bom} \B(\psi_{k, \bom}(\cA) \B) 
\end{equation}
where  
\be   \label{loaner}
\begin{split}
 \fS_{k, \bom} \B(\cA, \Psi_{k, \bom},  \psi \B)
=&  \blan  (\bPsi_{k, \bom}- Q_{k, \bom} (- \cA ) \bpsi ),   \bb^{(k)}  (  \Psi_{k, \bom}- Q_{k, \bom} (\cA ) \psi )  \bran_{\Om_1 } 
+   \blan    \bpsi   ,   \B(\fD_{\cA}+ \bar   m_k \B)   \psi \bran \\
 F_{k, \bom} (\psi )  =  &    \int  F_{0,L^{-k}}(\psi)( \psi  + \cW )\  d\mu_{S_{k,\bom}(\cA)} 
(\cW)\\
\sZ_{k,\bom}(\cA)  =   &    \int 
  \exp \left(-  \blan \bar \cW,  \B[\fD_{\cA}+ \bar   m_k + P_{k, \bom}(\cA)  \B]_{\Om_1}  \cW   \bran \right) 
 D \cW  =  \det \B( S_{k,\bom}(\cA) \B)^{-1} \\
\end{split}
\ee
\bigskip

\begin{lem} \label{tootoo} With $\psi_{\Om_1^c} =0$
 \be \label{tootootoo}
 \fS_{k, \bom} \B(\cA, \Psi_{k, \bom},  \psi_{k, \bom} (\cA)  \B) = \blan \bPsi_{k, \bom}, D_{k, \bom} (\cA) \Psi_{k, \bom} \bran 
 \ee
 where 
 \be   \label{keykey}
  D_{k, \bom} (\cA) = \bb^{(k)}  - \bb^{(k)}Q_{k, \bom}(\cA) S_{k, \bom}(\cA) Q^T_{k, \bom}(-\cA)\bb^{(k)}
 \ee
 \end{lem} 

\pr  The expression can be written 
\be   \label{loaner1}
\begin{split}
& \fS_{k, \bom} \B(\cA, \Psi_{k, \bom},  \psi_{k, \bom} (\cA)  \B)
=  \blan  \bPsi_{k, \bom},   \bb^{(k)}   \Psi_{k, \bom})  \bran_{\Om_1 } -\blan  \bPsi_{k, \bom},   \bb^{(k)} Q_{k, \bom} (\cA )  \psi_{k, \bom} (\cA) )  \bran_{\Om_1 } 
 \\
- &   \blan Q_{k, \bom} (- \cA )  \bpsi_{k, \bom} (\cA)  ,   \bb^{(k)}   \Psi_{k, \bom} )  \bran_{\Om_1 }
+   \blan    \bpsi_{k, \bom} (\cA)   ,   \B(\fD_{\cA}+ \bar   m_k + P_{k, \bom}(\cA) \B)   \psi_{k, \bom} (\cA)  \bran \\
\end{split}
\ee
Inserting the expression  (\ref{lurch2}) for $\psi_{k, \bom}(\cA)$  the first two terms give the result. 
In the fourth term $ (\fD_{\cA} + \bar m_k+ P_{k, \bom}(\cA) ) \psi_{k, \bom}(\cA) = Q^T_{k, \bom}(-\cA)\bb^{(k)}\Psi_{k, \bom}$
and it exactly cancels the third term.

 \subsection{the next step}   \label{single}

If   we  start  with  the   expression  (\ref{fable})  for  $\rho_{k, \bom}$   and apply another renormalization transformation as in (\ref{salubrious}), (\ref{nutsy}) we 
we get another expression for   $\rho_{k+1, \bom^+ }$.   We work out some details of this transformation.  
We have  with  $\bom^+  =  ( \bom,  \Om_{k+1})$
\be   \label{manx}
\begin{split}
&   \tilde  \rho_{k+1, \bom^+  } (\cA,  \Psi_{k+1, \bom^+}, \psi_{\Om^c_1}) = \cN_{k, \bom} \  \sZ_{k, \bom} (\cA)  N_{k,\Om_{k+1}} \int  D\Psi_{k, \Om_{k+1}}   F_{k, \bom}  \B( \psi_{k, \bom}(\cA)\B)     \\  
&   \exp \B(-  bL^{-1}
\blan   \bPsi_{k+1}-Q(-\cA)\bPsi_k, \Psi_{k+1}-Q(\cA)\Psi_k \bran_{\Om_{k+1}}  -    \fS_{k, \bom} \B(\cA, \Psi_{k, \bom},  \psi_{k, \bom} (\cA) \B) \B)  \\
\end{split}
\ee 
We want to evaluate the integral by expanding the quadratic form in the exponent around its critical points. 
Let $\Psi^{crit}_{k, \bom^+}, \bPsi^{crit}_{k, \bom^+}$ be the critical points  in    $\Psi_{k, \bom}, \bPsi_{k, \bom}$    with 
values in $\Om^c_{k+1}$ fixed.  So we are computing critical points   in $\Psi_{k, \Om_{k+1}},  \bPsi_{k, \Om_{k+1}}$.       These can be computed by a generalization of the formula (\ref{min1}).  However we will not use this, and instead develop a more
local expression.

We introduce the operators on $\Om_1 \subset \tk$
\be
\begin{split}
P^0_{k+1,\bom^+}(\cA) = & Q^T_{k+1, \bom^+}(- \cA) \bb^+ Q_{k+1, \bom^+}( \cA) \\
S^0_{k+1,\bom^+}(\cA) = & \B[  \fD_{ \cA }+   \bar m_k  +   P^0_{ k+1, \bom^+}(\cA)  \B]_{\Om_1}^{-1}\\
\end{split}
\ee
These scale to  $P_{k+1,\bom^+}(\cA), S_{k+1,\bom^+}(\cA) $ respectively.   We also define the field $  \psi^0_{k+1, \bom^+} (\cA)$ which
is $\psi$ on $\Om^c_1$ and  on $\Om_1$ is 
  \be
  \psi^0_{k+1, \bom^+} (\cA) =   \psi^0_{k+1, \bom^+} \B(\cA, \Psi_{k+1, \bom^+}, \psi_{\Om^c_1} \B) 
  =  S^0_{k+1,\bom^+}(\cA) \B(Q^T_{k+1,\bom^+} (-\cA)\bb^+ \  \Psi_{k+1, \bom^+} - \fD_{\cA} \psi_{\Om^c_1}\B)
  \ee
(This  scales to  $ \psi_{k+1, \bom^+}(\cA) $, see below).

\begin{lem} 
\be  \label{55}
\begin{split}
  \Psi^{\crit} _{k, \bom^+}(\cA)  = &   \Psi^{\bullet} \B( \Psi_{k+1},  \psi^0_{k+1, \bom^+}(\cA) \B) \hs \textrm{ on } \Om_{k+1} \\
  \psi^0_{k+1, \bom^+} (\cA)  = &   \psi_{k, \bom}  \B( \cA,   \Psi^{\crit} _{k, \bom^+}(\cA), \psi_{\Om^c_1} \B)  \hs \textrm{ on } \Om_{1} \\
\end{split}
\ee
and similarly for the conjugate fields.
\end{lem}
\bigskip

\pr Let
\be
\begin{split}
 &J( \Psi_{k+1, \Om_{k+ 1}}, \Psi_{k, \bom},    \psi )   =  bL^{-1} 
\blan   \bPsi_{k+1}-Q(-\cA)\bPsi_k, \Psi_{k+1}-Q(\cA)\Psi_k \bran_{\Om_{k+1}}  \\
+ &\blan  (\bPsi_{k, \bom}- Q_{k, \bom} (- \cA ) \bpsi),   \bb  (  \Psi_{k, \bom}- Q_{k, \bom} (\cA ) \psi  )  \bran_{\Om_1 } 
+    \blan    \bpsi   ,   \B(\fD( \cA )+ \bar   m_k \B)   \psi \bran   \\
\end{split}
\ee
The quadratic form in the exponent in (\ref{manx}) is this expression  evaluated at its critical point    $\psi = \psi_{k, \bom}(\cA),  \bpsi = \bpsi_{k, \bom}(\cA)$.
We can  find an expression for  its critical points  $\Psi^{crit}_{k, \bom^+}, \bPsi^{crit}_{k, \bom^+}$ by finding critical points  of  $J( \Psi_{k+1, \Om_{k+ 1}}, \Psi_{k, \bom},    \psi ) $
in   $\Psi_{k} , \bPsi_{k}$  on $\Om_{k+1}$ and $\psi, \bpsi$ on $\Om_1$ simultaneously.  Denoting these by  $\Psi^{crit}_k, \bPsi^{crit}_k$ and $\psi^{crit}, \bpsi^{crit}$
we find the equations
\be   \label{limber}
\begin{split}
\frac{ \pa J}{ \pa \bPsi_k } = & - bL^{-1} Q^T(-\cA)\B( \Psi_{k+1}-Q(\cA)\Psi^{crit} _k\B)  + 
 b_k\B( \Psi^{crit}_{k}-Q(\cA)\psi^{crit} \B) =  0 \\
 \frac{ \pa J}{ \pa \bpsi } = & - Q_{k, \bom}^T(-\cA) \bb\B(\Psi^{crit}_{k, \bom^+}-Q_{k, \bom}(\cA)\psi^{crit}\B)  
 +  (\fD_{\cA} + \bar m_k ) \psi^{crit} + \fD_{\cA }\psi_{\Om_1^c} =  0 \\
\frac{ \pa J}{ \pa \Psi_k } = & -bL^{-1} Q^T(\cA) \B(\bPsi_{k+1}-Q(-\cA)\bPsi^{crit}_k\B)   
+b_k \B( \bPsi^{crit}_{k}-Q(-\cA)\bpsi^{crit} \B) =  0 \\
\frac{ \pa J}{ \pa \psi }   = & -Q_{k, \bom}^T( \cA) \bb\B( \bPsi^{crit}_{k, \bom^+}-Q_{k,\bom}(-\cA)\bpsi^{crit}\B) 
 +    (\fD_{\cA} + \bar m_k ) ^T \bpsi^{crit} +  \fD^T_{\cA }\bpsi_{\Om_1^c} =  0 \\
\end{split}
\ee 
where    $\Psi^{crit}_{k, \bom^+}$ is  $\Psi_{k, \bom}$ with $\Psi_k$ replaced by $\Psi_k^{crit}$ on $\Om_{k+1}$.
The first and second equations have the solutions
\be   \label{longo2}
 \begin{split}
   \Psi^{crit}_k  = &  \Psi^{\bullet} ( \Psi_{k+1}, \psi^{crit}) \ \ \   \hs \textrm{ on } \Om_{k+1}\\ 
 \psi^{crit}   = &  \psi_{k, \bom} (\cA,  \Psi^{crit}_{k, \bom^+}, \psi_{\Om_1^c})  \hs \textrm{ on } \Om_{1} \\
 \end{split}
 \ee

The second equation in (\ref{limber}) can also be written
 \be \label{equine}
  \B(\fD_{\cA} + \bar m_k + P_{k, \bom}(\cA)\B) \psi^{crit} =    Q_{k, \bom}^T(-\cA)\bb\ \Psi^{crit}_{k, \bom^+} - \fD_{\cA }\psi_{\Om_1^c}
 \ee
 We claim that this is the same as 
 \be \label{equine2}
  \B(\fD_{\cA} + \bar m_k + P^0_{k+1, \bom^+}(\cA) \B) \psi^{crit} =  Q_{k+1, \bom^+}^T( -\cA )\bb^+ \ \Psi_{k+1, \bom^+}  - \fD_{\cA }\psi_{\Om_1^c}
   \ee
Indeed in $\Om^c_{k+1}$ we have   $ P_{k, \bom}(\cA) =  P^0_{k+1, \bom^+}(\cA)$ and $Q_{k, \bom}^T(-\cA)\bb\  \Psi^{crit}_{k, \bom^+}= 
 Q_{k+1, \bom^+}^T( -\cA )\bb^+ \ \Psi_{k+1, \bom^+}  $ so they agree.  On the other hand  in $\Om_{k+1}$ we use the expression (\ref{longo2}) for 
 $   \Psi^{crit}_k$ and obtain
 \be
 \begin{split}   Q_{k, \bom}^T(-\cA) \bb \  \Psi^{crit}_{k, \bom^+}= &
b_k Q_{k}^T(-\cA)\Psi^{crit}_{k} \\
     =  & b_k  P_{k} (\cA ) \psi ^{crit}     + \frac{b_k bL^{-1}} {b_k + bL^{-1} }  Q_{k+1}^T( -\cA ) \Psi_{k+1}  
     -  \frac{b_kbL^{-1}}{ ( b_k + bL^{-1})   }       P_{k+1} (\cA ) \psi^{crit}  \\
     =  & b_k  P_{k} (\cA ) \psi ^{crit}       + b_{k+1}L^{-1}   Q_{k+1}^T( -\cA ) \Psi_{k+1}  
     -  b_{k+1}L^{-1}P_{k+1} (\cA ) \psi^{crit}  \\
      =  & P_{k, \bom} (\cA ) \psi ^{crit}     +    Q_{k+1, \bom^+}^T( -\cA ) \bb^+ \Psi_{k+1, \bom^+}  
     - P^0_{k+1, \bom^+} (\cA ) \psi^{crit}  \\
\end{split}
\ee
Substitute this on the right side of (\ref{equine}) and get (\ref{equine2}).  So (\ref{equine}) and (\ref{equine2}) are equivalent as claimed.

  The equation (\ref{equine2}) has the solution
$\psi^{crit}  =  \psi^0_{k+1, \bom^+}(\cA) $.    With this identification the identities (\ref{longo2}) become
\be   \label{longo3}
 \begin{split}
   \Psi^{crit}_k  = &  \Psi^{\bullet} ( \Psi_{k+1}, \psi^0_{k+1, \bom^+ }(\cA) )  \hs \textrm{ on } \Om_{k+1}\\ 
\psi^0_{k+1, \bom^+}(\cA)  = &  \psi_{k, \bom}(\cA,  \Psi^{crit}_{k, \bom^+}, \psi_{\Om_1^c})\ \ \ \  \ \hs \textrm{ on } \Om_{1} \\
 \end{split}
 \ee
This is the same as (\ref{55}) since $\Psi^{crit}_{k, \bom^+} = \Psi^{crit}_k$ on $\Om_{k+1}$.
This completes the proof.
\bigskip

Now in (\ref{manx}) we expand around the critical point  by the transformation on $\Om_{k+1}$ (i.e. on the unit lattice $\Om^{(k)}_{k+1}$)
\be  \label{diag}
\Psi_k =  \Psi^{\crit}_{k, \bom^+}(\cA) + W   \hs   \bPsi_k =\bPsi^{\crit}_{k, \bom^+}(\cA) + \bar  W
\ee
By (\ref{55}) this also entails that
 \be \label{diag2}
 \psi_{k, \bom} (\cA)=   \psi^0_{k+1, \bom^+}(\cA) + \cW_{k, \bom}(\cA)   \hs    \bpsi_{k, \bom} (\cA)=   \bpsi^0_{k+1, \bom^+}(\cA) + \bar  \cW_{k, \bom}(\cA)   
 \ee
where 
\be
  \cW_{k, \bom}(\cA)  =  \psi_{k, \bom}(\cA, W, 0) =   \cH_{k, \bom}(\cA) W  =     b_k  S_{k, \bom}(\cA)  Q^T_{k}(-\cA)  W
\ee
We  introduce
 \be
\begin{split}
& \fS^0_{k+1, \bom^+} \B(\cA, \Psi_{k+1, \bom^+}, \psi \B)\\
= 
 &\blan  ( \bPsi_{k+1, \bom^+}-Q_{k+1, \bom^+}(-\cA)\bpsi),\bb^+ \ ( \Psi_{k+1, \bom^+}-Q_{k+1, \bom^+}(\cA)\psi  )\bran_{\Om_1}  +  \blan \bpsi , (\fD_{\cA} + \bar m_k ) \psi \bran   \\
\end{split}
\ee
(This scales to $ \fS_{k+1, \bom^+}(\cA, \Psi_{k+1, \bom^+}, \psi )$, see below). 

\begin{lem}  \label{summer} 
Under the translations (\ref{diag}), (\ref{diag2})  the quadratic form  in (\ref{manx}) 
\be
  \label{skyblue1}
\begin{split}  
& bL^{-1}
\blan   \bPsi_{k+1}-Q(-\cA)\bPsi_k, \Psi_{k+1}-Q(\cA)\Psi_k \bran_{\Om_{k+1}} 
 + \fS_{k, \bom}\B (\cA,  \Psi_{k,\bom},    \psi_{k, \bom}(\cA)\B )     \\
\end{split}  
\ee
becomes   
\be  \label{skyblue2}
\begin{split}  
 &  \fS^0_{k+1, \bom^+} \B( \cA,   \Psi_{k+1, \bom^+} ,    \psi^0_{k+1, \bom_{k+1} }(\cA) \B)
+   \blan  \bar W,   \B[ D_{k, \bom} (\cA ) + bL^{-1} P(\cA)  \B ]_{\Om_{k+1}  } W   \bran   \\
\end{split}  
\ee
\end{lem}  
\bigskip

\pr  Since we are at the critical point the cross terms vanish.   The terms in  $\bPsi^{\crit}_{k, \bom^+}(\cA) $ and $  \bpsi^0_{k+1, \bom^+}(\cA)$
are
\be
\begin{split}
 & bL^{-1} 
\blan   \bPsi_{k+1}-Q(-\cA)\bPsi^{\crit}_{k, \bom^+}(\cA), \Psi_{k+1}-Q(\cA)\Psi^{\crit}_{k, \bom^+}(\cA) \bran_{\Om_{k+1}} 
 +    \fS_{k, \bom}\B(\cA, \Psi^{\crit}_{k, \bom^+}(\cA), \psi^0_{k+1, \bom^+}(\cA) \B)   \\
  = & 
  \fS^0_{k+1, \bom^+} \B( \cA,   \Psi_{k+1, \bom^+} ,    \psi^0_{k+1, \bom_{k+1} }(\cA) \B)   \\
\end{split}
\ee
Here we have used   (\ref{lazy}) in $\Om_{k+1}$ and $\Psi^{\crit}_{k, \bom^+}= \Psi_{k, \bom} = \Psi_{k+1, \bom^+}$ in $\Om_1 - \Om_{k+1}$.
The terms in $W, \cW_k(\cA)$ are identified  by lemma  \ref{tootoo}  as 
\be  \label{mope}
\begin{split}
&  bL^{-1} 
 \blan  Q(-\cA)W,  Q(\cA)W \bran  +    \fS_k\B(\cA, W, \cW_{k, \bom}(\cA)\B)  \\   
  =  &   bL^{-1}  <\bar  W, P(\cA) W > +  \blan  \bar W,  [D_{k, \bom} (\cA )]_{\Om_{k+1} }  W  \bran. \\
\end{split}
\ee
  This completes the proof.
\bigskip

Now in  (\ref{manx})  we make the change of variables  and  integrate over the new Grassmann variables 
   $\bar W, W$  on $\Om_{k+1}$  instead of  $\Psi_k, \bPsi_k$
   This gives 
     \be   \label{manx2}
\begin{split}
&   \tilde  \rho_{k+1, \bom^+ } (\cA,  \Psi_{k+1, \bom^+}) =  \cN_{k, \bom} \  \sZ_{k, \bom} (\cA)  N_{k,\Om_{k+1}}
 \exp \left(-  \fS^0_{k+1, \bom^+} \B( \cA,   \Psi_{k+1, \bom^+} ,    \psi^0_{k+1, \bom^+ }(\cA) \B)    \right) 
 \\  
&   \int     
 F_{k, \bom} \B(  \psi^0_{k+1, \bom^+}(\cA)+  \cW_{k, \bom^+}(\cA) \B) 
  \exp \left(   -     \blan\bar  W,  \B[ D_{k, \bom}(\cA)  +  bL^{-1}   P(\cA)\B]_{\Om_{k+1} }W\bran   \right)  \ D W     \\
\end{split}
\ee 
Next   identify the  Gaussian integral $\int [ \cdots ]   d \mu_{\Ga_{k, \bom^+}(\cA)}$   with covariance
\be
\Ga_{k, \bom^+}(\cA) =   \B[ D_{k, \bom}(\cA)  +  bL^{-1}   P(\cA)\B]^{-1}_{\Om_{k+1}}
\ee
and normalization factor
\be \label{kitten}
\begin{split}
\de  \sZ _{k, \bom^+}(\cA)   =  &  \int   \exp \B(   -     \blan\bar  W,  \B[ D_{k, \bom}(\cA)  +  bL^{-1}   P(\cA)\B]_{\Om_{k+1}} W\bran       \B) \ DW   
= \det\B( \Ga_{k, \bom^+} (\cA)\B)^{-1}
 \\
\end{split}
\ee
Then  (\ref{manx2}) is rewritten as 
  \be   \label{manx3}
\begin{split}
&   \tilde  \rho_{k+1, \bom^+ } (\cA,  \Psi_{k+1, \bom^+}) =  (\cN_{k, \bom}   \sZ_{k, \bom} (\cA))(  \de  \sZ_{k, \bom^+}(\cA)  N_{k,\Om_{k+1}} ) 
 \\  
&   \exp \left(-  \fS^0_{k+1, \bom^+} \B( \cA,   \Psi_{k+1, \bom^+} ,    \psi^0_{k+1, \bom^+ }(\cA) \B)    \right)  \int     
 F_{k, \bom} \B(  \psi^0_{k+1, \bom^+}(\cA)+  \cW_{k, \bom^+}(\cA) \B) 
  \  d \mu_{\Ga_{k, \bom^+}(\cA)} (W)   \\
\end{split}
\ee 
  Next we  scale by (\ref{nutsy}).
  We have
  \be 
  \begin{split}
  \psi^0_{k+1, L \bom^+}\B( \cA_L, [ \Psi_{k+1, \bom^+} ]_L, [\psi_{\Om_1^c}]_L\B)
  = &  \B[  \psi_{k+1,  \bom^+}( \cA) \B]_L \\
   \fS^0_{k+1, L \bom^+}\B(\cA_L, [ \Psi_{k+1,\bom^+}]_L,   [  \psi_{k+1,  \bom^+}( \cA) ]_L\B)  
  =  &   \fS_{k+1, \bom^+}\B(\cA, \Psi_{k+1, \bom^+}, \psi_{k+1, \bom^+}(\cA)  \B)  \\
  \end{split}
 \ee
 and we get 
 \be   \label{kingly}
 \begin{split}
 & \rho_{k+1, \bom^+} (\cA,  \Psi_{k+1, \bom^+})  
   =    (\cN_{k, L\bom}   \sZ_{k, L\bom} (\cA_L))(  \de  \sZ_{k,L \bom^+}(\cA_L)  N_{k,L\Om_{k+1}} ) \si_{k+1, \bom^+}   \\
  &     \exp  \B( -      \fS_{k+1, \bom^+}(\cA, \Psi_{k+1, \bom^+}, \psi_{k+1, \bom^+}(\cA)  )   \B) \int   F_{k,L \bom} \B(  [ \psi_{k+1, \bom^+}(\cA)]_L + \cW_k(\cA_L) \B)   d \mu_{\Ga_{k, L \bom^+}(\cA_L)} (W)\\
\end{split}   
 \ee
Taking  $F_0 =1$  we have  $F_{k, \bom}  = 1 $.    Comparing
this   with   (\ref{fable}) for  $k+1$ we find 
\be    \label{z1}
\cN_{k+1, \bom^+}\sZ_{k+1, \bom^+} (\cA)   =    (\cN_{k, L\bom}   \sZ_{k, L\bom} (\cA_L))(  \de  \sZ_{k, L\bom^+}(\cA_L)  N_{k,L\Om_{k+1}} ) \si_{k+1, \bom^+} 
\ee
and so 
\be   \label{kingly2}
 \begin{split}
 & \rho_{k+1, \bom^+} (\cA,  \Psi_{k+1, \bom^+})  
   =  \cN_{k+1, \bom^+} \sZ_{k+1, \bom^+} (\cA)  \\
  &     \exp  \B( -      \fS_{k+1, \bom^+}(\cA, \Psi_{k+1, \bom^+}, \psi_{k+1, \bom^+}(\cA)  )   \B) \int   F_{k, L\bom} \B(  [ \psi_{k+1, \bom^+}(\cA)]_L + \cW_k(\cA_L) \B)   d \mu_{\Ga_{k, \bom^+}(\cA_L)} (W)\\
\end{split}   
 \ee
This  integral  is a basic fluctuation integral of a type we investigate further for specific  $F$.

\bigskip 

\subsection{random walk  expansion}  \label{study it}

In the analysis of the renormalization group flow one needs detailed control over the Green's functions (propagators).  This is achieved with a random walk
expansion which we now explain for the Dirac Green's function
\be   
S_{k, \bom} (\cA)   = \B [  \fD_{\cA}     + \bar  m_k   +  P_{k, \bom} (\cA  ) \B]^{-1}_{\Om_1} 
\ee
We still have  in $\tk$      that    $\bom =   (\Om_1,   \dots,    \Om_k )$,    that        $\Om_1 \supset  \Om_2  \supset  \cdots  \supset  \Om_k$,    and  that    $\Om_j$ is  a union  of   $L^{-(k-j)}M$ cubes.   We  assume  also the separation condition       
\be  \label{spacing} 
    d(  \Om_{j+1}, \Om_j^c )  \geq   6  L^{- (k-j) } M  
\ee 
However the possibility  that   $\Om_j = \Om_{j+1} = \tk$ is not excluded.  

To state the result we need a number of  definitions. 
\begin{itemize}
\item
Let  $\pi_j( \de \Om_j) $   be  all    $L^{-(k-j)}M$ cubes  in  $\de\Om_j$ and  $\pi(\bom)   =  \cup  \pi_j(\de \Om_j) $.

\item If  $ \sq \in \pi_j(\de \Om_j)$    then  $\tilde \square$ is the enlargement defined by adding a layer  
of  $L^{-(k-j) }M$ cubes.   Similarly we define an $n$-fold enlargement $\tilde \sq^n$.   If  $X$ is a union of such $\sq$ then $\tilde X$
is the union of the $\tilde \sq$, etc.

\item   $  \fG_{k,\bom} $ is  all  complex   gauge fields  $  \cA  $   on   $ \tk $   such that        
 for   $b \in \square  \in \pi_j(  \de \Om_j)$
\be 
 | \cA(b)  | \leq       L^{\frac12   (k-j) }  e_j^{-1 + \ep}
 \ee

\item 
 $ \tilde   \fG_{k,\bom} $ is  all  gauge fields of the form    $  \cA=  \cA_0  + \cA_1$
where
\begin{itemize}
\item    $ \cA_0 $ is real and for  each    $\square \in \pi_j(  \de \Om_j) $     is gauge equivalent  in $\tilde    \square ^{5}$ to a field $\cA' \in   \fG _{k,\bom}$
\item  
 $\cA_1 \in  \fG  _{k, \bom}$ is complex. 
 \end{itemize}

\item
If  $y \in \de \Om_j$ define 
\be  \label{tacky}
   \De_y =   \textrm{ the }  L^{-(k-j)}  \textrm{ cube in  }   \de \Om_{j}  \textrm{ centered on } y \in \de \Om_j^{(j)}
\ee 
We also define an enlargement $\tilde   \De_y$  by adding a layer of  $ L^{-(k-j)}$ cubes around $\De_y$, and let $\zeta_y$
be a smooth partition of unity with $\supp\  \zeta_y \subset \tilde   \De_y$.

\item
Associated with the sequence  $\bom$  is   a   scaled distance .  It is defined for  $(y,y')  \in [ \bom] = \cup_{j=1}^k  \de \Om_j^{(j)}$  by
\begin{equation}   \label{stingray}
d_{\bom}(y,y')   =  \inf_{\ga:    y \to  y'}  \sum_{j=1}^k  L^{k-j}  \ell(\ga  \cap \de   \Om_j )
\end{equation}   
with  $\de \Om_k  =  \Om_k$.
The infimum is over paths   $\ga$    joining   $y,y'$    in the lattice  $\bbT^{-k}_{\sN-k}$  such that  in    $ \de  \Om_j$ the path
   $\ga$  consists of   $L^{-(k-j)}$ links   in       $\de \Om_j^{(j)}$.     The    factor  $L^{k-j}$  in $d_{\bom}(y,y')   $  means we count
these     links      as unit length.   The $d_{\bom}(y,y')  $  satisfy  (lemma 2.1 in  \cite{Bal84b} )  
\be  \label{multisum} 
  \sum_{y'} e^{  -  \ga  d_{\bom} (y,y') } =  \sum_j \sum_{y' \in \de \Om_{j}^{(j)}}  e^{  -  \ga  d_{\bom} (y,y') }  \leq  C
\ee

\item
A covariant H\"{o}lder derivative is defined by on functions on $\tk$ for $0<\al<1$ and  $|x-y| <1$
\be \label{holder}
(\de_{\al, \cA} f)(x,y) = \frac{e^{ie_k \eta \cA(\Ga_{xy}) } f(y) - f (x) }{ |x-y|^{\al}} \hs \eta =L^{-k}
\ee
where $\Ga_{xy}$ is a path from $x$ to $y$.

\end{itemize}

The main result      due to Balaban,  O'Carrol,  Schor      \cite{BOS91} is the following; see also \cite{Dim04}.  

 \begin{lem}   \cite{BOS91}  \label{sweet3}    Let   $M$  be sufficiently large (depending on  $L$),  
  and  $e_k$  sufficiently small (depending on  $L,M$),   and let   $\cA \in  \tilde   \fG _{k,\bom}$.   
Then there is a      random walk  expansion  
 \begin{equation}  \label{spoof}
 S_{k, \bom}(\cA) =    \sum_{\om }  S_{k,\bom, \om}(\cA)
 \end{equation}
where     $\om$ is a sequence of cubes    $   \om   =  (\square_0, \square_1, \dots,   \square_n )$
 in  $\pi(\bom) $  
such that  $\square_i, \square_{i+1}$ are equal or  nearest neighbors.  Each term is analytic in $\cA  \in \tilde \fG _{k, \bom} $ and
the expansion     converges to a function analytic in $\cA  \in \tilde \fG _{k, \bom} $   which satisfies  
for    $y \in \de \Om^{(j)}_j,  y' \in \de \Om^{(j')}_{j'} $
  \be   \label{sycamore4}
  \begin{split}
|1_{\De_{y}} S_{k, \bom}( \cA)1_{\De_{y'}} f |  \leq     &  C  L^{-(k-j')}   e^{  -\ga  d_{\bom} (y,y') } \|f\|_{\infty}   
 \\
 L^{-\al(k-j)}   \| \de_{\al, \cA }    \zeta_{y}     S_{k, \bom}( \cA  )1_{\De_{y'}} f \|_{\infty}   \leq     &  C L^{-(k-j')}   e^{  -\ga  d_{\bom} (y,y') } \|f\|_{\infty}    \\
   \end{split}
\ee
 \end{lem}
\bigskip

\rems
\begin{enumerate}
\item
An advantage of this type of estimate is that it is preserved under composition of operators.   Furthermore 
using (\ref{multisum}) one can sum over $\De_{y' }$ and 
 deduce for $x \in \de \Om_j$ (or at least one of $x,y \in \de \Om_j$)
\be   \label{sycamore5}
|(S_{k, \bom} ( \cA) f)(x) |,\ \    L^{-\al(k-j)} | (\de_{\al, \cA }       S_{k, \bom} ( \cA  ) f )(x,y) |   \leq      C  \sup_{j'}   L^{-(k-j')} \sup_{x' \in \de \Om_{j'}}  |f(x') |    
\ee
 The  right side is of course less than  $C \| f\|_{\infty}$.
\item
In the estimates (\ref{sycamore4})  factors $L^{(k-j) }$ and  $L^{(k-j') }$ can be exchanged as desired.   This is so because the ratio  is  bounded by 
$L^{|j-j'|}$.  If $|j-j'| \leq 1$ this is bounded by a constant.   If $|j-j'| \geq 2$ then  the separation condition (\ref{spacing}) means that $y,y'$ must be far apart.  Indeed
 by lemma 2.1 in \cite{Bal84b}
\be  \label{lobster}
 L^{|j-j'|} \leq  e^{ \cO(M^{-1}) d_{\bom}(y,y') }
 \ee
 and for $M$ large this can be compensated by an adjustment in $\ga$.
\item
We sketch some details of the proof since  we will need to refer to it.  The formulation is a little different  because we started by scaling up to a unit lattice, but the 
essentials are all the same.  
The proof depends constructing local inverses for   $ \fD_{\cA}     + \bar  m_k   +  P_{k, \bom} (\cA  )$. This is the content of
the following lemma will be discussed further.  
\end{enumerate}

\begin{lem}     \cite{BOS91}     \label{sweet2}    Under the same hypotheses   for  any  $\square   \in \pi_j(  \de \Om_j) $  
there is    an  operator    $S_{k, \bom}(\square,   \cA)$ on functions on a domain  $ \Om_1(\sq) \cap \Om_1$  with     $ \tilde \sq^4 \subset \Om_1(\sq) \subset \tilde \sq^5$
such that  $ S_{k, \bom}(    \square,   \cA    )  $ only depends on $\cA$ in $\Om_1(\sq)  \cap \Om_1$ and
\be  
 \label{twotwo}   \B(  \B[   \fD_{\cA}     + \bar  m_k   +  P_{k, \bom} (\cA  )  \B]_{\Om_1}    S_{k,\bom}(\square, \cA)  f \B) (x )   =  f(x)  \hs      x  \in   \tilde  \square
\ee 
and      
\be    \label{orca}
\begin{split}
| 1_{\De_{y}}  S_{k, \bom}(    \square,   \cA    )    1_{\De_{y'}}  f  | 
  \leq  &     C    L^{-(k-j')}  e^{- \ga d_{\bom}(y,y')  }  \| f \|_{\infty}   
\\
 \| \de_{\al, \cA }    \zeta_{y}     S_{k, \bom}(    \square,   \cA    )  1_{\De_{y'}} f \|_{\infty} 
   \leq     &  C L^{-(1 - \al)(k-j')}    e^{  -\ga  d_{\bom} (y,y') } \|f\|_{\infty}    \\
   \end{split}
   \ee
\end{lem} 
\bigskip

\pr   (of lemma \ref{sweet3} assuming lemma  \ref{sweet2}) 
Let    $h^2_{\square}$  be  a partition of unity  indexed by cubes   $\square \subset  \pi(\bom) $       with  $\sum_{\square} h_{\square}^2 =1$  and   $\supp\  h_{\square}$ well inside  $ \tilde \square$.  We  define a parametrix  using the operators of  lemma \ref{sweet2} to be the operator on $\Om_1$
\be 
 S_{k, \bom} ^*(\cA)  = \sum_{\square}  h_{\square}  S_{k, \bom} (   \square, \cA) h_{\square} 
  \ee
  where the sum is over $\sq$ in an neighborhood  of $\Om_1$. 
Then  by (\ref{twotwo})
\be     
\begin{split}
  & \B[\fD_{ \cA}   + \bar  m_k   +    a_k    P_{k, \bom} (\cA  ) \B]_{\Om _1}  S_{k, \bom} ^*(\cA) \\
 = &   \sum_{\square}  \B[\fD_{ \cA}   + \bar  m_k   +     P_{k, \bom} (\cA  ) \B]_{\Om _1}   h_{\square}   S_{k, \bom} (   \square, \cA) h_{\square} \\
 =   &  \sum_{\square}  h_{\square}  \B[\fD_{ \cA}   + \bar  m_k   +  P_{k, \bom} (\cA  ) \B]_{\Om _1} S_{k, \bom} (   \square, \cA) h_{\square}+   \sum_{\square}  
   \B[ \fD_{ \cA}   + \bar  m_k   +    P_{k, \bom} (\cA ),h_{\sq}\B]_{\Om _1} S_{k, \bom} (   \square, \cA) h_{\square} \\
=    &  I  -  \sum_{\square}   K_{\square}(\cA)  S_{k, \bom} ( \square, \cA) h_{\square}  \equiv  I -K \\
  \end{split}
  \ee
  where $  K_{\square }(\cA)   =  K_{\square, \bom }(\cA)  $ is   given by 
  \be  
    K_{\square} (\cA)  =  \B[  [ h_{\sq},  \fD_{ \cA}   +\bar  m_k   +     P_{k, \bom}  (\cA) ] \B]_{\Om_1} 
     \ee
 Then   
  \be
   S_{k, \bom} (\cA)  =   S_{k, \bom} ^*(\cA) ( I - K)^{-1}   =    S_{k, \bom} ^*(\cA) \sum_{n=0}^{\infty}  K^n
  \ee    
  provided the series   converges.  
  This can be written   as    the random walk expansion 
   \begin{equation}  \label{g1}
 S_{k, \bom} (\cA) =    \sum_{\om }  S_{k,\bom, \om}(\cA)
 \end{equation}
where   for $\om= (\sq_0, \sq_1, \dots, \sq_n)$ 
\be \label{lumpy}  
S_{k, \bom, \om } (\cA)=\B( h_{\square_0} S_{k, \bom} (  \square_0,\cA )h_{\square_0}\B)
 \B(K_{\square_1}(\cA) S_{k, \bom} (   \square_1,\cA)h_{\sq_1}\B)\cdots    \B(K_{\square_n}(\cA) S_{k, \bom} ( \square_n,\cA)h_{\square_n}\B)
\ee
 Note  that    $ S_{k,\bom, \om}(\cA)$  only depends on  $\cA$ in  $ \bigcup_{i=0}^n  \tilde \square^5_i $.

The functions   $\{ h_{\square}\}$ can be chosen so that  for  $\square \subset   \pi_j(  \de \Om_j) $
\be
| \pa h_{\square} | \leq   \one L^{k-j} M^{-1}
\ee
Then $K_{\sq}(\cA)$ can be expressed in terms of $\pa h_{\sq}$ and combined with (\ref{orca}) one can show
\be  
    \label{spitfire0}
     |1_{\De_y}  K_{{\square}}(\cA) S_{k, \bom}  (\cA, \square)1_{\De_{y'}}f|   \leq   C M^{-1}   e^{  -  \ga  d_{\bom} (y,y') }  \|f \|_{\infty}
\ee
Now insert  multiscale  localization functions $1_{\De_y}$  as in (\ref{tacky})  between the various factors in (\ref{lumpy}) so that
\be \label{lumpy2}  
\begin{split}
&S_{k, \bom, \om } (\cA)\\
&= \sum_{y_1, \dots, y_n} \B( h_{\square_0} S_{k, \bom} (  \square_0,\cA )h_{\square_0} \B)1_{\De_{y_1}}
\B( K_{\square_1}(\cA) S_{k, \bom} (   \square_1,\cA)h_{\sq_1} \B)\cdots   1_{\De_{y_n}}  \B(K_{\square_n}(\cA) S_{k, \bom} ( \square_n,\cA)h_{\square_n} \B)\\
\end{split}
\ee
Note that only terms with $y_j \in \tilde \sq_{j-1} \cap \tilde \sq_j \neq \emptyset$ contribute.
 Estimate the resulting expression using  (\ref{spitfire0})
and for  the first factor $ S_{k, \bom} (  \square_0,\cA ) $ using  (\ref{orca}).   Then estimate the sum over localizations using  (\ref{multisum}) and an associated convolution
inequality also from lemma 2.1 in \cite{Bal84b}.
One finds  with a new $C, \ga$  that   if  $|\om|  = n$ then  
\be   
\begin{split}
    \label{night}
  |1_{\De_{y}}S_{k,  \bom,    \om}(\cA)1_{\De_{y'}} f|  \leq   &  C  L^{-(k-j)} (CM^{-1} )^n e^{  -  \ga  d_{\bom} (y,y') }  \|f \|_{\infty} \\
     \| \de_{\al, \cA }    \zeta_{y}   S_{k,  \bom,    \om}(\cA)1_{\De_{y'}} f \|_{\infty}  \leq   &  C  L^{-(1- \al)(k-j)} (CM^{-1} )^n e^{  -  \ga  d_{\bom} (y,y') }  \|f \|_{\infty}\\
 \end{split}
 \ee
This is sufficient to establish the convergence  of  the expansion for $M$  sufficiently  large,   since the number of paths with a fixed length $n$ is bounded
by  $\one^n$.   Replacing some factors $L^{-(k-j)}$ by   $L^{-(k-j')}$  the bounds  (\ref{sycamore4})   on $S_{k,  \bom}(\cA)$   follow. \bigskip

\pr (of lemma \ref{sweet2})
\bigskip

\noindent
\textbf{part I: }    
  First some preliminary results for global operators and $\cA =0$. 
   The scaling identity   $   S^0_{k+1}(0) f_L =  L [ S_{k+1}(0) f]_L$  means the kernels are related by
  \be S_{k+1}(0,x,x')  = L^2 S^0_{k+1}(0, Lx, Lx')
  \ee
  Combine this with  the identity  $ S^0_{k+1}(0)    =  S_k(0)   +   \cH_k(0)  \Ga_{k} (0)  \cH^T_k(0)   $  which is the global version of (\ref{oshkosh2}) from  appendix \ref{A}
Then  iterate the result to get the relation 
  \be \label{telescope}
    S_k  ( 0; x,x')    =   \sum_{j=0}^{k-1}    L^{2(k-j)}  \B( \cH_j(0) \Ga_j(0) \cH_j^T(0)  \B) (L^{k-j} x,  L^{k-j} x' )  
\ee
The convention here is that $\cH_0(0) = I$ and $\Ga_0(0) = S^0_1(0)$.
Using some Fourier analysis one can show that    $\cH_j(0), \de_{\al}\cH_j(0)$   and  $ \Ga_j(0)$  have  exponential decay  \cite{BOS89}.  Then from  the representation
(\ref{telescope}) 
 one has
\be \label{slush}
  | S_k  ( 0; x,x')   | \leq C d(x,x')^{-2} e^{-\ga d(x,x') } \hs  
   |\de_{\al} S_k  ( 0; x,x')   | \leq C d(x,x')^{-2- \al} e^{-\ga d(x,x') } 
   \ee
   with the convention that $d(x,x) = L^{-k}$.
  It follows that  if $\De_x$ is an $L^{-(k-j)}$ cube and $\De_{x'}$ is an $L^{-(k-j')}$ cube
  \be   \label{slugish} 
\begin{split}
\|  1_{\De_x}  S_{k}(0)   1_{\De_{x'}}  f\|_{\infty}   \leq   &   C  L^{-(k-j')} e^{- \ga d(x,x')  }\\
\|  \de_{\al }    \zeta_{x}     S_{k}( 0 )  1_{\De_{x'}} f\|_{\infty}   \leq   &  C  L^{-(1-\al)(k-j')} e^{- \ga d(x,x')  } \\
\end{split}
\ee
For the first inequality the  idea is to   estimate the integral over  $\De_{x'}$ by
$ \int_{\De_{x'}}  d(x,y)^{-2} dy \leq \one  L^{-(k-j')}$  and for the second to use$ \int_{\De_{x'}}  d(x,y)^{-2- \al} dy \leq \one  L^{-(1- \al)(k-j')}$.
See \cite{BOS91},   \cite{Dim04} for more details.

Let $Y$ a union of $M$ cubes in the unit lattice $\tz$ and consider  the restricted operator
 \be
 \Ga_{k,Y}(0) = \B[ D_k(  0  )   +  bL^{-1} P(0)\B]_Y^{-1}
 \ee
  In   \cite{BOS91}   it is shown  by some
Fourier analysis  and taking advantage of  the  Wilson form of the free fermion action that    
\be 
     \| \Ga_{k, \sq}( 0) f \|_2    \leq    C \|  f\|_2   
 \ee  
 Furthermore   from the representation (\ref{route66})  of $D_k(0)$  and the exponential decay of $S_k(0)$ one can show  
\be    \label{lunkhead}
   | \B( D_k(  0  )   +  bL^{-1} P(0)\B) (x,x')  |  \leq     Ce^{-\ga d(x,x')  }
\ee
Then it follows by a lemma of Balaban on unit lattice operators  \cite{Bal83b}, \cite{BOS91}    that  
\be    \label{base1}  
  |    \Ga_{k,Y}(0; x,x')|    \leq    Ce^{-\ga d(x,x')  }
\ee

If   $Y$ is a union of $LM$-cubes in $\tk$ ,  we also need to consider
\be   \label{prince}
S_{k,Y}  (0)   =  \B(   \fD_{0}     + \bar  m_k   + b_k [ P_k(0 )]_{Y^c}    +   b_{k+1}L^{-1} [ P_{k+1}(0 )]_{Y}     \B )^{-1}
\ee
This is related   to    $S_{k}  (0) $
by  
\be  
S_{k,Y}  (0 )   =  S_k (0 )   + \cH_k(0)\Ga_{k,Y} (0) \cH^T_k(0)
\ee
which is (\ref{oshkosh2}) in the case that $\bom$ is a sequence of full tori and $\Om_{k+1}= Y$. 
We  deduce from  (\ref{base1}) and  (\ref{slugish})  that the bounds (\ref{slugish}) also hold for $S_{k,Y}  (0 )  $.
 \bigskip

\noindent
\textbf{part II:} Now to the proof itself.  Let     $\square_0  \in  \pi_j( \Om_j )$.   We define on $\tk$
\be
S_{k, \bom}(\sq_0, 0) =  S_{k, \bom(\sq_0)}(0) 
\equiv    \B[  \fD_{0}     + \bar  m_k  +   P_{k, \bom(\sq_0)} (0) \B]_{\Om_1 \cap \Om_1(\sq_0)}^{-1} 
   \ee
where the averaging operator  $P_{k, \bom(\sq_0)} (0)$  is  based on  a sequence  $\bom(\sq_0)$  localized  around $\sq_0$.
We would like the averaging operators to also be compatible with the original sequence $\bom$ which is potentially awkward.
We settle for enforcing compatibility only close to $\sq_0$. With this in mind we take
   \be  
\bom( \square_0 )  =  \B ( \Om_1(\square_0),   \dots,   \Om_{j}(\square_0), \Om_{j+1}(\sq_0) \B ) 
  \ee   
Here $ \Om_j(\square_0) = \square^{\sim 4}_0 $ and for $i = j-1,j-2, \dots, 1$  we define  sucessively larger sets   $\Om_i(\square_0)$  to be a  union of   $L^{-(k-i)}M$  cubes 
forming a cube centered on $\sq_0$ and satisfying  the minimal separation condition condition.  
Then  we   add a region  $\Om_{j+1}  (\square_0)   = \tilde  \sq_0^3   \cap \Om_{j+1} $. 
Thus we have in increasing size
\be
\Om_{j+1}  (\square_0)   = \tilde  \sq_0^3   \cap \Om_{j+1}  \hs   \Om_j(\square_0) = \square^{\sim 4}_0 \hs   d(  \Om_{i+1}, \Om_i^c )  =   6  L^{- (k-i) } M
\ee
Note that     $\Om_1(\square_0) \subset  \tilde \square^5_0  $.

To get estimates on  $S_{k, \bom(\sq_0)}(0) $ we again develop a random walk expansion.
Form a   parametrix  by 
\be
   S^*  (\square_0)   =     \sum_{\square \in  \pi( \bom(\square_0))}   h_{\square}   S_{\square}   h_{\square}  
\ee
where     $h^2_{\square  } $ is  a  partition of unity indexed by $\pi( \bom(\square_0))$    with  $\supp \ h_{\square}  \subset   \tilde    \square$ and the  operator 
$S_{\sq}$ is defined on $ \tk$  as follows.    For a cube  $\sq$ in  $ \pi( \bom(\square_0)) $ we again 
have two cases which are 
\begin{enumerate}
\item  $ \sq \in \pi_i(\de \Om_i(\sq_0) ) $ and 
  $\tilde  \sq^3 \subset \de \Om_i(\sq_0)$ for some $i$
\item   $\tilde \sq^3$ intersects both $   \Om_{i}(\sq_0), \Om_{i+1}(\sq_0)$ for some $i$
\end{enumerate}
Corresponding to these two cases we define
 \be   \label{stingerstinger}
      S_{\sq}  =
\begin{cases}   \B (\fD_0  + \bar m_k +  b^{(k)}_iP_i(0) \B)^{-1}    & \hs  \textrm{ case 1}  \\
        \B( \fD_0  + \bar m_k +   b^{(k)}_i [P_i(0)]_{Y^c}   +    b^{(k)}_{i+1} [P_{i+1}(0)]_Y     \B)^{-1}  
      &       \hs    \textrm{ case 2 }  \\
 \end{cases}
\ee 
where in the second case
\be 
    Y \equiv    \tilde  \sq^3 \cap  \Om_{i+1}(\square_0) 
    \ee
We  further modify this in case  $\sq_0$ is  an $L^{-(k-1)}M$ cube in  $\pi_1 (\de \Om_1) $  and $\tilde  \sq^3_0 \cap \Om_1^c \neq \emptyset$.   In that case $\bom(\sq_0) = \Om_1(\sq_0) =\tilde \sq_0^4$
and we just define
\be
S_{\sq}  = \B[ \fD_0  + \bar m_k +  b^{(k)}_1P_1(0)         \B]^{-1} _{\Om_1}  
\ee
In all cases since $\supp \  h_{\sq} \subset \tilde \sq \subset  \Om_1(\sq_0)$ the $ h_{\square}   S_{\square}   h_{\square}  $ is an operator on  $ \Om_1 \cap \Om_1(\sq_0)$
and so is $S^*$.   The operator $S_{\sq}$  has averaging operators $b^{(k)}_iP_i(0)$ in each  $\de \Om_i(\sq_0)$. This agrees with  $   P_{k, \bom(\sq_0)} (0)$ and so it
satisfies the key identity 
\be  
 \label{bump1}    \B(   \B[  \fD_{0}     + \bar  m_k  +   P_{k, \bom(\sq_0)} (0) \B]_{\Om_1 \cap \Om_1(\sq_0)}  S_{\square}  f \B) (x )   =  f(x)  \hs      x  \in   \tilde \square
\ee

For estimates we have:

\begin{lem} \label{sweet5}  
Let $ \sq_0 \in \pi(\bom)$ and $\sq \in   \pi(\bom(\square_0))$ and let $y,y'$ and $\De_y,\De_{y'}$ be as in (\ref{tacky}).
\begin{enumerate}
\item     $S_{\sq}$ satisfies for $ \tilde \sq \cap \De_y \neq \emptyset$ and  $ \tilde \sq \cap \De_{y'} \neq \emptyset$
\be        \label{loaf}
\begin{split}
| 1_{\De_y}  S_{\sq}       1_{\De_{y'}}  f  |   \leq    &  C    L^{-(k-j')}  e^{- \ga d_{\bom} (y,y')  }  \| f \|_{\infty}  \\
| 1_{\De_y} \de_{\al}  S_{\sq}       1_{\De_{y'}}  f  |   \leq   &  C    L^{-(1- \al)(k-j')}  e^{- \ga d_{\bom} (y,y')  }  \| f \|_{\infty}  \\
\end{split}
\ee
\item There is  a random walk expansion  for $ S_{k, \bom(\sq_0)} (0) $ based on $S_{\sq}$ which 
yields the bounds  
 \be        \label{orca2}
 \begin{split}
| 1_{\De_{y}}   S_{k, \bom(\sq_0)} (0)   1_{\De_{y'}}  f  |   \leq   &  C    L^{-(k-j')}  e^{- \ga d_{\bom}(y,y')  }  \| f \|_{\infty}  \\
| 1_{\De_{y}} \de_{\al}  S_{k, \bom(\sq_0)} (0)     1_{\De_{y'}}  f  |   \leq   &  C    L^{-(1- \al)(k-j')}  e^{- \ga d_{\bom}(y,y')  }  \| f \|_{\infty}  \\
\end{split}
\ee
\end{enumerate}
\end{lem}
\bigskip

\pr At first suppose that  $ \tilde    \sq^5_0 \subset \de \Om_j$.  Then $ \Om_{j+1}(\sq_0) = \emptyset$ and all $\sq \in \pi(\bom(\sq_0))$ are also
in $\de \Om_j$.   
Suppose further that $\sq$ is in case 1 so   $S_{\sq} =  ( \fD_0  + \bar m_k +  b^{(k)}_iP_i(0) ) ^{-1}   $.    There is a mismatch here
 between the lattice $\tk$ and
the averaging operator $P_i(0)$ which usually occurs on $\bbT^{-i}_{N-i}$. 
From (\ref{slush}) and a variation of   (\ref{slugish}) we do have the estimate on $\bbT^{-i}_{N-i}$  for  $x,x' \in \bbT^0_{N-i} $ unit cubes   $\De_x, \De_{x'}$ 
\be
       \label{halfloaf}
| 1_{\De_x} \B ( \fD_0  + \bar m_i +  b_iP_i(0) \B)^{-1}         1_{\De_{x'}}  f  |   =| 1_{\De_x} S_i(0)  1_{\De_{x'}}  f  |  \leq     C    e^{- \ga  d(x,x')  }  \| f \|_{\infty}  
\ee
We also have    the scaling relation  
\be  
 \B (\fD_0  + \bar m_k +  b^{(k)}_iP_i(0) \B)^{-1}  f_{L^{-(k-i)}} = L^{-(k-i)}
  \B ( ( \fD_0  + \bar m_i +  b_iP_i(0) )^{-1}  f \B)_{L^{-(k-i)}} 
  \ee 
and then (\ref{halfloaf}) scales down to the estimate   on $\bbT^{-k}_{N-k}$  for $x,x' \in \bbT^{-(k-i)}_{N-k} $ and $ L^{-(k-i) }$ cubes  $\De_x, \De_{x'}$ 
\be
  \label{halfloaf2}
| 1_{\De_{x}}   \B ( \fD_0  + \bar m_k +  b_iP_i(0) \B)^{-1}         1_{\De_{x'}}  f  |   \leq       C    L^{-(k-i)}  e^{- \ga L^{(k-i)}d(x,x')  }  \| f \|_{\infty}   
\ee

Now consider  cubes $\De_{y},\De_{y'} $ intersecting $\tilde \sq$ as required by the lemma.  Then  $\De_{y},\De_{y'} \subset \de \Om_j$
and hence are $L^{-(k-j)}$ cubes, 
The previous bound  and $L^{-(k-i)} \leq CL^{-(k-j)}$    yields the estimate
\be
| 1_{\De_{y}}   \B ( \fD_0  + \bar m_k +  b_iP_i(0) \B)^{-1}         1_{\De_{y'}}  f  |   \leq   C L^{-(k-j)} 
\sup_{x \in \De_y}   \sum_{x' \in \De_{y'}}   C   e^{- \ga L^{(k-i)}d(x,x')  }  \| f \|_{\infty}  
\ee
On the one hand we could take   $ L^{(k-i)}d(x,x')   \geq  L^{(k-j)}d(x,x') \geq  L^{k-j} d(y,y') -1 $
and  bound the exponential  by $\one  e^{- \ga L^{(k-j)}d(y,y') } $.  On the other hand we have
$\sum_{x'}  e^{- \ga L^{(k-i)}d(x,x')  }  \leq    C  $.
Splitting the exponent we use both estimates and obtain with a new $\ga$ 
\be
\begin{split}
  \label{halfloaf3}
 | 1_{\De_{y}}  S_{\sq}        1_{\De_{y'}}  f  |  =| 1_{\De_{y}}   \B ( \fD_0  + \bar m_k +  b_iP_i(0) \B) ^{-1}         1_{\De_{y'}}  f  |  
\leq  &   C    L^{-(k-j)} e^{- \ga L^{(k-j)}d(y,y') }  \| f \|_{\infty}   \\
\end{split}
\ee
Since  $L^{k-i}d(y,y')  =  d_{\bom}(y,y') $  here this gives the desired result. 
The estimate on the  H\"{o}lder   derivative is similar.

 Suppose again  $ \tilde    \sq^5_0 \subset \de \Om_j$,  but now  $\sq$ is in case 2. 
    So $\tilde \sq^3$ intersects both $   \Om_{i}(\sq_0), \Om_{i+1}(\sq_0)$ and
 $S_{\sq} =   (\fD_0  + \bar m_k +   b^{(k)}_i [P_i(0)]_{Y^c}   +    b^{(k)}_{i+1} [P_{i+1}(0)]_Y     )^{-1}$. 
     The relevant scaling relation is 
\be  
\begin{split}
& \B (\fD_0  + \bar m_k +   b^{(k)}_i [P_i(0)]_{Y^c}   +    b^{(k)}_{i+1} [P_{i+1}(0)]_Y     \B)^{-1}  f_{L^{-(k-i)}} \\
  & = L^{-(k-i)}
  \B (  \B(\fD_0  + \bar m_i +   b_i [P_i(0)]_{L^{k-i}Y^c}   +    b_{i+1} L^{-1}  [P_{i+1}(0)]_{L^{k-i}Y } \B)^{-1}   f \B)_{L^{-(k-i)}} \\
 \end{split}
  \ee 
The latter is  an operator of the form  (\ref{prince}) and  has the bounds  (\ref{slugish}).  Arguing as before we find that the bound  (\ref{halfloaf3})
is replaced by
\be
  \label{halfloaf4}
| 1_{\De_{y}}   \B (\fD_0  + \bar m_k +   b^{(k)}_i [P_i(0)]_{Y^c}   +    b^{(k)}_{i+1} [P_{i+1}(0)]_Y     \B)^{-1}        1_{\De_{y'}}  f  |   \leq       C    L^{-(k-j)}  e^{- \ga L^{(k-j)}d(y,y')  }  \| f \|_{\infty}   
\ee
which is the desired bound.

\bigskip

Now we relax the condition that    $ \tilde    \sq^5_0 \subset \de \Om_j$ and allow that   $ \tilde    \sq^5_0 $ intersect both $\Om_j$ and $\Om_{j+1}$.  The cube $\sq_0$
itself may be in either  $\Om_j$ and $\Om_{j+1}$.  For $\sq$ in case 1  we still have the bound (\ref{halfloaf3}) but now it is not exactly what is required since either $\De_y$ or $\De_{y'}$
may be in $ \Om_{j+1}$ and hence a larger $L^{-(k-j-1)}$ cube.  This requires some adjustments in the bound.
Suppose for example that $\De_y$ is in $\Om_j $ and  $\De_{y'}$ is  in $\Om_{j+1}$. Then from (\ref{halfloaf3})
\be 
| 1_{\De_{y}}   \B ( \fD_0  + \bar m_k +  b_iP_i(0) \B) ^{-1}         1_{\De_{y'}}  f  |  
\leq     C    L^{-(k-j)}   \sum_{z \in \De_{y'}}  e^{- \ga L^{(k-j)}d(y,z) }  \| f \|_{\infty}  
\ee
where the sum is over $z$ on $\bbT^{-(k-j)}_{N-k}$.  We can replace the  $ L^{-(k-j)}  $ in front by the larger $ L^{-(k-j-1)} $.   
Now on  the one hand in the exponent  we  can  take $L^{(k-j)}d(y,z) \geq  L^{(k-j)}d(y,y') - 1$ and then argue that $ L^{(k-j)}d(y,y') \geq d_{\bom}(y,y')$ as follows.  Consider a minimal  path between $\Ga_{y,y'}$  from $y$ to $y'$ whose length is $d(y,y')$. 
Let $y''$ be   the intermediate point  where the path crosses from $ \Om_j$ to $\Om_{j+1}$. Then 
\be 
\begin{split}
L^{k-j}d(y,y') = & L^{k-j}\ell( \Ga_{y,y'}))
 =   L^{k-j} \ell( \Ga_{y,y''})) + L^{k-j}\ell( \Ga_{y'',y'}))\\ 
 \geq &  L^{(k-j)} \ell( \Ga_{y,y''})) + L^{k-j-1}\ell( \Ga_{y'',y'}))
 \geq    d_{\bom} (y,y') \\
\end{split}
\ee
On  the other hand we could just take    $\sum_{z \in \De_{y'}}  e^{- \ga L^{(k-j)}d(y,z) }  \leq C$. Splitting the exponent we do both and obtain with a new $\ga$ 
\be 
| 1_{\De_{y}}   \B ( \fD_0  + \bar m_k +  b_iP_i(0) \B) ^{-1}         1_{\De_{y'}}  f  |  
\leq     C    L^{-(k-j-1 )}   e^{- \ga d_{\bom}(y,y') }  \| f \|_{\infty}  
\ee
This is the required bound in this case.   The instance $\sq$ in case 2 is treated similarly.
\bigskip

For the case $\tilde \sq^3 \cap \Om_1^c \neq \emptyset$ and  $S_{\sq} =  [ \fD_0  + \bar m_k +  b^{(k)}_1P_1(0)    ]^{-1} _{\Om_1} $
instead of (\ref{halfloaf}) we have on the lattice  $\bbT^{-1}_{N-1} $ and for  unit cubes 
\be
       \label{halfloaf5}
| 1_{\De_x} \B [\fD_0  + \bar m_1 +  b_1P_1(0) \B] ^{-1}_{L^{k-1}\Om _1}         1_{\De_{x'}}  f  |   \leq     C   e^{- \ga d(x,x')  }  \| f \|_{\infty}  
\ee
Indeed the operator scales up to the unit lattice operator   $[ \fD_0 + m_0 + b_1L^{-1}P(0) ]^{-1}_{L^k \Om_1} = \Ga_{k,  L^k \Om_1}(0) $ and the bound follows from (\ref{base1}).
Then we scale down to the operator $S_{\sq}$ on  $\tk$ by  
\be  
 \B [ \fD_0  + \bar m_k +  b^{(k)}_1P_1(0) \B]^{-1}_{\Om_1}  f_{L^{-(k-1)}} = L^{-(k-1)}
  \B [ \B[ \fD_0  + \bar m_1 +  b_1P_1(0) \B]^{-1}_{L^{k-1}  \Om_1} f \B] _{L^{-(k-1)}} 
  \ee 
and this again leads to the bound (\ref{loaf}) in this case.  This completes the proof of part 1. 

\bigskip

For part 2   the random walk expansion is generated as in lemma \ref{sweet3}.  We  compute on $\Om_1  \cap \Om_1(\sq_0)$
\be     \label{kingdom}
\begin{split}
&  \B[\fD_{ 0}   + m_k   +      P_{k, \bom(\sq_0)} (0  ) \B]_{\Om _1 \cap \Om_1(\sq_0 ) }  S^*(\sq_0)   \\
=& \sum_{\sq } \B[\fD_{ 0}   + m_k   +       P_{k, \bom(\sq_0)} (0  ) \B]_{\Om _1 }   h_{\square}   S_{\square}   h_{\square}  \\
=& \sum_{\sq}    h_{\square}   S_{\square}  \B[\fD_{ 0}   + m_k   +        P_{k, \bom(\sq_0)} (0  ) \B]_{\Om _1 } h_{\square}  
- \sum_{\sq}  K_{\square}    S_{\square}   h_{\square} 
\\
=  &   I  -  \sum_{\square}   K_{\square}   S_{\square}   h_{\square}  \\
\end{split}
\ee
Here in the first step we drop the restriction of the operator to $\Om_1(\sq_0)$ on the right which 
is allowed since $\supp \  h_{\sq} \subset \Om_1(\sq_0)$.  In the second step we define
 $  K_{\sq}   =  K_{\sq , \bom(\sq_0) }  $  to be the operator on $\Om _1 \cap \Om_1(\sq_0 )$
  \be  
    K_{\sq }   =  \B[  [ h_{\sq},  \fD_{ 0}   +\bar  m_k   +     P_{k, \bom(\sq_0)}  (0) ] \B]_{\Om _1  } 
     \ee
 In the last step we use the identity (\ref{bump1}). 
 From (\ref{kingdom}) we generate    the random walk expansion $ S_{k, \bom(\sq_0)} (0) =    \sum_{\om }  S_{k,\bom(\sq_0) , \om}(0) $
where   
\be  S_{k, \bom(\sq_0), \om } (0)=  (h_{\square_0}   S_{\sq_0}  h_{\square_0})
 (K_{\square_1} S_{\sq_1} h_{\sq_1} ) \cdots    (K_{\square_n}  S_{\sq_n} h_{\square_n})
\ee
As  before it leads to the estimates (\ref{orca2}). This completes the proof of lemma \ref{sweet5}.
\bigskip

  \noindent
\textbf{part III:}  We have the basic estimates on  $  S_{k, \bom}(\sq_0, 0) =  S_{k, \bom(\sq_0)}(0) $
and now  show that it satisfies the remaining condition  of lemma \ref{sweet2}. 
This is the   identity (\ref{twotwo})  which  says that  for $x \in \tilde \sq_0 $ 
and either $\tilde \sq_0 \subset \de \Om_j$ or $\tilde \sq_0$ intersecting both $\de \Om_j, \de \Om_{j+1}$
\be  \label{suntory}
\B(  \B[ \fD_{0}     + \bar  m_k  +   P_{k, \bom} (0)\B] _{\Om_1} S_{k, \bom(\sq_0)} (0)   f \B)(x)  = f(x)
\ee
This follows   since both    $ [P_{k, \bom} (0)] _{\Om_1} $ and  $ [ P_{k, \bom(\sq_0)} (0)] _{\Om_1} $  have averaging operator  $b^{(k)}_j P_j(0)$ in $\de \Om_j$
and $b^{(k)}_{j+1} P_{j+1}(0)$ in  $\de \Om_{j+1}$. (This is where we use the condition  $\Om_{j+1}  (\square_0)   = \square^{\sim 3}_0   \cap \Om_{j+1} $). 
Thus  we can replace   $ [P_{k, \bom} (0)] _{\Om_1} $ by $ [ P_{k, \bom(\sq_0)} (0)] _{\Om_1} $ in (\ref{suntory}).   
Then  since $ S_{k, \bom(\sq_0)} (0)$ maps to functions on $\Om_1 \cap \Om_1(\sq_0)$
we can  replace $[ \cdots] _{\Om_1} $  by $ [ \cdots ] _{\Om_1 \cap \Om_1(\sq)} $. The result follows since $ S_{k, \bom(\sq_0)} (0) $ is the inverse operator. 
This completes the proof  of lemma \ref{sweet2} for the case $\cA=0$. 
\bigskip

\noindent
\textbf{part IV:}  
 Lemma \ref{sweet2}   for  $\cA \in  \tilde   \fG _{k,\bom}$ can be obtained using gauge transformations and expansions around $\cA =0$.
See \cite{BOS91} for details.  
\bigskip

\rems \

\noindent
 (1.) 
 The bounds on $\cS_{k, \bom}$  lead to bounds on  $ \cH_{k, \bom} ( \cA)f  = S_{k, \bom} Q^T_{k, \bom}(- \cA)\bb^{(k)}f $ for $f=\{ f_{j, \de \Om_j} \}$ on $[\bom ]$.   The 
 $\bb^{(k)}$ supplies a factor $L^{(k-j')}$  on $\de \Om_{j'}$ which is cancels the $L^{-(k-j')}$ in (\ref{sycamore5}).  This yields 
 on $\de \Om_j$
 \be  \label{slavic}
| \cH_{k, \bom} ( \cA) f|,  \ \   L^{-\al (k-j)} |  \de_{\al, \cA }  \cH_{k, \bom} ( \cA) f|    \leq       C  \| f  \|_{\infty} 
 \ee
Or using the sharper bound (\ref{sycamore4}) as well as (\ref{lobster})  we have  on $\de \Om_j$
 \be 
L^{-(k-j)}| \cH_{k, \bom} ( \cA)  f |, \  L^{-(1+\al)  (k-j)} |  \de_{\al, \cA }  \cH_{k, \bom} ( \cA) f|      \leq       C   \sup_{j'} L^{-(k-j')} \|f_{j', \de \Om_{j'}} \|_{\infty}
 \ee
  \bigskip
 
\noindent
(2.)   We   can     introduce weakening parameters  $\{ s_{\square} \}$  indexed by $\sq \in \pi( \bom)$ with values  $0 \leq  s_{\square} \leq  1$.   
For  $\om = ( \sq_0, \sq_1, \dots, \sq_n)$ the operator $S_{k, \bom, \om}$ only connects points and depends on $\cA$ in 
\be 
 X_{\om}   \equiv  \bigcup_{i=0}^n  \tilde \square^5_i 
\ee
We weaken this term  with a factor  
\be  
s_{\om}  =  \prod_{\square \subset X_{\om}}  s_{\square}  
 \ee
 and define
 \be  \label{again}
   S_{k, \bom} (s, \cA) =    \sum_{\om} s_{\om}  S_{k,\bom,  \om}(\cA)
\ee
If  $s_{\square}$  is  small  then the coupling through  $\square$ is reduced.  
If  $\om = \sq$ is a single cube   then  $|\om| =0$ and  $X_{\om} = \emptyset$ and in  this case  we define  $s_{\om} = 1$.  
Hence the  $   S_{k, \bom}(s,\cA)$  interpolate between
$S_{k, \bom}(1,\cA)=  S_{k, \bom}(\cA)  $   and the strictly local  operator  $  S_{k, \bom}(0,\cA) = S^*_{k, \bom}(\cA)$.

All the previous results  hold as well for    the  $ S_{k, \bom}(s,\cA)$.  
   In fact we can allow     complex  $s_{\square}$   satisfying   $ |s_{\square}|  \leq  M^{\al_0}$ for $\al_0 <1$.  This still leaves a convergence factor 
   of $M^{-(1- \al_0)}$ which suffices for $M$ sufficiently large.     The weakened propagator  $S_{k, \bom} (s,\cA)$ also gives a weakened operator  $\cH_{k, \bom}(s, \cA)$  which satisfies
   all the above bounds.
\bigskip

\subsection{another random walk expansion}

For our treatment of fermion determinants we will need a variation of the previous treatment. Now instead of  
$S_{k,  \bom}(\cA)$ we consider  for $y \geq 0$ the operator on $\Om_1 \subset \tk$
\be  \label{kingsly}
   S_{k, \bom^+,y}(\cA) = \B[ \fD_{\cA}     + \bar m _k  +  [P_{\bom}(\cA)]_{\Om^c_{k+1} }
    +[ \al_k  P_k(\cA)  + \beta_k  P_{k+1} (\cA)]_{\Om_{k+1} } \B]^{-1}_{\Om_1}  
  \ee   
where
\be
 \al_k =    \frac{b_k  i\ga_3y  }{b_k + i\ga_3y}  
    \hs    \beta_k  = \frac{b_k^2 bL^{-1} }{( b_k + bL^{-1}+ i\ga_3y) (b_k + i\ga_3y )}
 \ee
This   interpolates between $S^0_{k+1, \bom^+}(\cA)$ at  $ y=0$ and    $S_{k,  \bom}(\cA)$ at  $y = \infty$.

\begin{lem}  \label{sweet4}  Lemma \ref{sweet3} and lemma \ref{sweet2} hold for  $ S_{k, \bom^+,y}(\cA)$ with bounds uniform in $y \geq 0$.  
\end{lem}
\bigskip

\pr  \textbf{part I}: 
Follow the proof of lemma \ref{sweet2} and start with   the global version of $S_{k,y}(0)$ at $\cA =0$.  This  
is related to $ \Ga_{k,y}(0)   =[    D_{k} (0)  +  bL^{-1} P(0) + i \ga_3 y]^{-1} $.  More generally we consider
\be 
\Ga_{k,y,Y}(0)   = \B[    D_{k} (0)  +  bL^{-1} P(0) + i \ga_3 y\B]^{-1}_Y = \ga_3\B [    (D_{k} (0)  +  bL^{-1} P(0))\ga_3 + i  y\B]^{-1}_Y
\ee
We first claim that
\be    \label{base}
  |    \Ga_{k,y,Y } (0; x,x')|    \leq    Ce^{-\ga d(x,x')  }
\ee
If $y$ large we make the expansion
\be   
\begin{split}
 \Ga_{k,y,Y} (0)    =   &    [  D_k( 0 )   +  bL^{-1} P( 0)  + i \ga_3y    ]_Y  ^{-1}     \\
=   &  ( i \ga_3 y ) ^{-1}    \B [  I   +    [ D_k( 0 )   +  bL^{-1} P( 0)]_Y    ( i \ga_3 y ) ^{-1}    \B ]  ^{-1}   \\ 
=   &  ( i \ga_3 y ) ^{-1}  \sum_{n=0}^{\infty}   (-1)^n   \B(  [ D_k( 0 )   +  bL^{-1} P( 0) ]_Y    ( i \ga_3 y ) ^{-1}   \B)^n   \\ 
\end{split}
\ee
The  bound  (\ref{lunkhead}) says   $ | ( D_k(  0  )   +  bL^{-1} P(0)) (x,x')  |  \leq     Ce^{-\ga d(x,x')  }$. 
Thus if $y$ is sufficiently large,  say $y \geq C_0$  for some $C_0$, the series converges and the bound (\ref{base})
holds.  On the other hand suppose $y \leq C_0$.  Take $Y=\sq$ a single cube.
 Since  $[( D_{k} (0)  +  bL^{-1} P(0)) \ga_3]_{\sq} $ is
self-adjoint  we have
 \be 
    \|\Ga_{k,y,\sq }(0)   f \|_2    \leq    C \| \B[   \B(D_{k} (0)  +  bL^{-1} P(0)\B)\ga_3 \B]^{-1}_{\sq}  f  \|_2    \leq  C   \| \Ga_{k, \sq}(0)  f \|_2    \leq   C \|  f\|_2   
 \ee  
  Also by  (\ref{lunkhead}) again and the bound on $y$ 
\be
  \B |  \B( D_k(  0  )   +  bL^{-1} P(0)  + i \ga_3y  \B)(x,x')   \B|  \leq    C e^{-\ga d(x,x')  }
\ee
Then   (\ref{base})  follows from the last two bounds and  Balaban's  lemma  on unit lattice operators  \cite{Bal83b}, \cite{BOS91}.   
\bigskip

   Now we use the  identity     (the global version of (\ref{oshkosh1}) in  appendix  \ref{A} )   
   \be  
  S_{k,y}(0)    =  S_k(0)   +   \cH_k(0)  \Ga_{k,y} (0)  \cH^T_k(0)   
 \ee
combined with the expansion  (\ref{telescope}) for $S_k(0)$
to get
\be  \label{telescope2}
    S_{k,y}  ( 0; x,x')    = \B( \cH_k(0)  \Ga_{k,y} (0)  \cH^T_k(0)\B)(x,x') +     \sum_{j=0}^{k-1}    L^{2(k-j)}  \B( \cH_j(0) \Ga_j(0) \cH_j^T(0)  \B) (L^{k-j} x,  L^{k-j} x' )  
\ee
This is estimated  as before and establishes the bounds (\ref{slugish}) for $S_{k,y}$.   
We also need to consider on $\tk$
\be   \label{prince2}
S_{k,y, Y}  (0)   =  \B(   \fD_{0}     + \bar  m_k   + b_k[ P_k(0 ) ]_{Y^c}    +  [ \al_k P_k(0) + \beta_k P_{k+1}(0)]_{Y}     \B )^{-1}
\ee
This is related   to    $S_{k,y}  (0) $
by  
\be  
S_{k,y, Y}  (0 )   =  S_{k,y} (0 )   + \cH_k (0)\Ga_{k,y,Y} (0)\cH_k^T(0)
\ee
which is a special case of (\ref{oshkosh1}).  This yields   the bounds (\ref{slugish}) for $S_{k,y, Y}  (0 ) $.
This concludes the modifications of part I.
\bigskip

\noindent
\textbf{part II} : 
Continuing with the modifications of  the proof of lemma \ref{sweet2} we first consider $\cA=0$ and look for  operators satisfying for $x \in \tilde \sq_0$
\be  
 \label{33}   \B(  \B[   \fD_{0}     + \bar  m_k   + [ P_{k, \bom} (0 )]_{\Om^c_{k+1} } + [ \al_k P_k(0) + \beta_k P_{k+1}(0)]_{\Om_{k+1}} \B]_{\Om_{k+1}}    S_{k,\bom}(\sq_0, 0)  f \B) (x )   =  f(x) \ee 
Our previous choice of $S_{k,\bom}(\sq_0, 0)$ still works away from $\Om_{k+1}$ so we only need to consider the case of $\sq_0 \in \pi_k( \Om_k)$ in or close to $\Om_{k+1}$.
Formerly  we would  have taken   $  [  \fD_{0}     + \bar  m_k  +   P_{k, \bom(\sq_0)} (0) ]_{ \Om_1(\sq_0)}^{-1} 
  $ with $\bom(\sq_0) = ( \Om_1(\sq_0),\dots, \Om_k(\sq_0) )$.  Now we modify this  by adding $\Om_{k+1}(\sq_0) =  \tilde \sq_0^3 \cap \Om_{k+1}$ and defining 
\be
S_{k, \bom}(\sq_0, 0)=   \B[  \fD_{0}     + \bar  m_k  + [  P_{k, \bom(\sq_0)} (0)]_{\Om^c_{k+1}(\sq_0)}  +  [\al_k P_k(0) + \beta_k P_{k+1}(0)]_{\Om_{k+1}(\sq_0)}  \B]_{ \Om_1(\sq_0)}^{-1} 
   \ee
This has averaging operator $\al_k P_k(0) + \beta_k P_{k+1}(0)$ in  $\sq_0 \cap \Om_{k+1}$ and $P_k(0)$ in $\sq_0 \cap \Om^c_{k+1}$  and so satisfies (\ref{33}).

Again we need a random walk expansion for $S_{k, \bom}(\sq_0, 0)$ which means we need operators   $S_\sq$ for $\sq \in \pi(\bom(\sq_0) )$  such that 
for $x \in \tilde \sq $
\be  \label{orca7}
\B( \B[  \fD_{0}     + \bar  m_k  + [  P_{k, \bom(\sq_0)} (0)]_{\Om^c_{k+1}(\sq_0)}  +  [\al_k P_k(0) + \beta_k P_{k+1}(0)]_{\Om_{k+1}(\sq_0)}  \B]_{ \Om_1(\sq_0)}
S_{\sq} f\B) (x) = f(x) 
\ee
We make the same choice as in (\ref{stingerstinger}) with the  addition
 \be   \label{stingerstingerstinger}
      S_{\sq}  =
\begin{cases}   \B (\fD_0  + \bar m_k + \al_k P_k(0) + \beta_k P_{k+1}(0\B)^{-1}    & \hs  \textrm{ case 1}  \\
        \B( \fD_0  + \bar m_k +  b_k[  P_k (0)]_{Y^c}  +  [\al_k P_k(0) + \beta_k P_{k+1}(0)]_Y  \B)^{-1}  
      &       \hs  \textrm{ case 2}  \\
 \end{cases}
\ee 
where case 1 is $ \tilde \sq^3 \subset \Om_{k+1}(\sq_0)$ and case 2 is $\tilde \sq ^3$ intersects both $ \Om_{k+1}(\sq_0)$ and $\Om_k(\sq_0)$ and in this case
   $ Y  \equiv  \tilde \sq ^3 \cap \Om_{k+1}(\sq_0) $.
These choices  satisfies (\ref{orca7}).  As noted we have good estimates on such operators and the rest of the proof  proceeds as in lemma \ref{sweet2} and lemma \ref{sweet3}.

\subsection{polymer propagators} 

  A  \textit{polymer}       $ X $   is a   connected union of  $M$-cubes  with the convention that two cubes are connected if they have a face in common.
 The set of all polymers in $\tk$  or   $\tz$  is denoted $\cD_k$.  For a polymer $X$ a  polymer function  $E(X, \cA)$  of a gauge field $\cA$ on $\tk$  is a function  
 which only depends on $\cA$ in $X$.     We are generally interested in functions of the form  $ E(\cA) = \sum_X E(X, \cA)$.
 The sum  will converge since  $E(X, \cA)$ generally has decay in the size of $X$ of the form $\cO(e^{-\ka d_M(X)}$). Here   
\be
  M d_M(X)  =    \textrm{  length of a shortest continuum tree   intersecting every  }  M  \textrm{  cube   in }  X   
 \ee  
 and the positive constant  $\ka = \one$ is large enough so that if $\square$ is a single $M$ cube 
 \be 
 \sum_{X \in \cD_k, X \supset \sq }   e^{ -  \ka   d_M(X  )}  \leq \one
 \ee

 We also consider multiscale polymers
  As in sections \ref{knots}  and \ref{knots2}  suppose we are given a decreasing sequence of small field  
 regions  $\bom  =   (\Om_1,  \Om_2,  \dots,   \Om_k) $   where  $\Om_j$ is a  union of  $L^{-(k-j)}M$ cubes and  $\de \Om_j = \Om_j - \Om_{j+1}$. Multiscale polymers $Y$ are now connected unions of cubes   in $\tk$ with the restriction that a cube from $\de \Om_j$ be
an $L^{-(k-j)}M$ cube.  The set of all multiscale polymers in denoted $\cD_{k, \bom}$.   A multi-scale polymer function $E(Y, \cA)$ only depends on $\cA$ in $Y$ and   typically decays like $e^{-\ka |Y|_{\bom}}$
where  $|Y|_{\bom}$ is  the number of cubes in $Y$, so  
\be
|Y|_{\bom} = \sum_{j=1}^k | Y \cap \de \Om_j |_{L^{-(k-j}M}
\ee

We   localize  Dirac   propagators (= Green's functions) depending on a background gauge field   by giving  polymer  expansions:

\begin{lem}  \label{hughes}
Under the hypotheses of lemma \ref{sweet3} 
\begin{enumerate}
\item   $S_{k, \bom} (\cA )  =  \sum_{Y \in \cD_{k, \bom} }  \hat  S_{k, \bom}  (Y, \cA  ) $  where  $ \hat  S_{k, \bom}  (Y, \cA , x,x' ) $  has support in $x,x' \in Y$ and  depends on $\cA$ only in $Y$  and for any constant
 $\hat \kappa = \one$ we have for $M$
sufficiently large 
\be \label{arty1}
    | 1_{\De_y}\hat S_{k, \bom} (Y, \cA  )1_{\De_{y'}} f|  \leq   C L^{-(k-j) } e^{-\ga d_{\bom}(y, y')} e^{ - \hat  \ka  |Y|_{\bom} } \|f\|_{\infty}   
\ee
\item   $S_{k, \bom} (\cA )  =  \sum_{X \in \cD_{k} }   S_{k, \bom}  (X, \cA  ) $  where  $   S_{k, \bom}  (X,  \cA, x,x'   ) $ has support in $x,x' \in X$,  depends on $\cA$ only in $X$,  and for any constant
 $  \kappa = \one$ we have for $M$
sufficiently large 
\be \label{arty2}
    | 1_{\De_y} S_{k, \bom} (X, \cA  )1_{\De_{y'}} f|  \leq   C L^{-(k-j) } e^{-\ga d_{\bom}(y, y')} e^{ -  \ka  d_M(X) } \|f\|_{\infty}   
\ee
\item
The same  results  hold  for $S_{k, \bom^+,y} $ as defined in (\ref{kingsly})
\end{enumerate}
\end{lem} 
\bigskip

\pr (1.)  We have  $S_{k, \bom} (\cA )  = \sum_{\om}  S_{k, \bom, \om}(\cA) $ where
   for a multiscale     walk   $\om = (\sq_0, \sq_1, \dots,  \sq_n)$  the term  $S_{k, \bom, \om}(\cA) $ is localized in $X_{\om} = \cup_i \tilde \sq^5_i   $.
   Let $\bar X_{\om}$ be the smallest element of $\cD_{k, \bom}$ containing $X_{\om}$.
   We  define   
   \be
\hat  S_{k, \bom}   (Y, \cA  )  = \sum_{\om:  \bar X_{\om}  = Y  }   S_{k, \bom,  \om} (\cA) 
  \ee
    From (\ref{night}) we have that   $S_{k, \bom,  \om} (\cA)$  satisfies the bound (\ref{arty1}) with $(CM^{-1})^n$ rather than
  $ e^{ - \hat  \ka  |Y|_{\bom} } $.
  However if $\sq$ is an $L^{-(k-j)}M$ cube in $\de \Om_j$ then $\overline{\sq^{\sim 5}}$ may intersect $\de \Om_{j+1}$ or  $\de \Om_{j-1}$
  Allowing for this we have $|\overline{\sq^{\sim 5}}|_{\bom} \leq  L^3|\sq^{\sim 5}|_{L^{-(k-j}M} = 11 L^3 $
  Then 
  \be
 |\bar X_{\om}|_{\bom}   =   \B|\bigcup_{\sq \in \om} \overline{\sq^{\sim 5}}\B |_{\bom} \leq \sum_{\sq \in \om}  | \overline{\sq^{\sim 5}} |_{\bom}
 \leq 11 L^3 n  
   \ee 
   Thus for $M$ sufficiently large  we can take   $(CM^{-1})^{n/2} \leq     e^{ - \hat  \ka  | \bar X_{\om}|_{\bom} }    =    e^{ - \hat  \ka  |Y|_{\bom} } $.
  The  remaining factor $(CM^{-1})^{n/2} $ is enough  for  convergence of the sum over $\om$   as before and we have the result. 
  \bigskip
  
  \noindent
  (2.)  Let $\bar Y \in \cD_k$ be the union of all $M$ cubes intersecting $Y \in \cD_{k, \bom}$  and define
   \be
 S_{k, \bom}   (X, \cA  )  = \sum_{\bar Y = X  } \hat  S_{k, \bom} (Y, \cA) 
  \ee
But $|Y|_{\bom} \geq |\bar Y|_M = |X|_M \geq d_M(X) $ so  in the estimate on   $\hat  S_{k, \bom} (Y, \cA) $ we can use part of
$e^{ - \hat  \ka  |Y|_{\bom} } $ to extract a factor $e^{- \ka d_M(X) }$.  In the sum over $Y$ we must have  $\De_y \subset Y$.
For an $L^{-(k-j)} $ cube   $\De_y  \subset  \de \Om_j$ let $\sq_y$ be the   $L^{-(k-j)} M$ cube containing it.  Then for $\hat \ka - \ka$ large enough  the sum over 
$Y$ is controlled by    (see Appendix in \cite{Dim13b})
\be
\sum_{Y \supset \De_y}  e^{ - ( \hat  \ka - \ka)  |Y|_{\bom} }
  \leq \sum_{Y \supset \sq_y }  e^{ - ( \hat  \ka - \ka)  |Y|_{\bom} }
 \leq  1
\ee

 \bigskip

 \noindent
  (3.) The proof is entirely similar.

\subsection{polymer determinants}  

We develop a polymer expansion for the fermion  determinant, generalizing results in \cite{Bal96}, \cite{Dim15b}. 
Recalling that     $\bom^+  = (\bom, \Om_{k+1}  ) $ 
 consider the fermion determinant which is     from (\ref{kitten})
    \be    
      \de \sZ_{k, \bom^+} (\cA)      =   \det  \B(   \B [ D_{k, \bom} (\cA) + bL^{-1}  P(\cA)  \B] _{\Om_{k+1}}   \B)  
 \ee   
    where  by (\ref{keykey})    \be   \B[ D_{k, \bom} (\cA)  \B ] _{\Om_{k+1}}    =       b_k - b_k^2  Q_{k} (\cA) S_{k, \bom}  (\cA )Q^T_{k}(- \cA)    
  \ee

 \begin{lem}   \label{thrush1}   For  $\cA  \in \tilde  \fG _{k, \bom}$
 \be 
 \begin{split}
   \de \sZ_{k, \bom^+} (\cA)      
   =   &  \exp  \B(   \B(    (1 - L^{-3} )     \log  b_k       +  L^{-3}   \log  (     b_k + bL^{-1} )   \B)4  | \Om^{(k)}_{k+1}|  +
    \sum_{X \cap \Om_{k+1} \neq \emptyset } E^d_{k, \bom^+}(X, \cA  )  \B) \\
\end{split}  
\ee
where    the sum is over $X \in \cD_k$ and
\be 
| E^d_{k, \bom^+} (X, \cA  ) |  \leq    CM^3   e^{ -  \ka  d_M(X  )}  
\ee
\end{lem}
\bigskip

\pr  The proof follows \cite{Dim15b} where the global case is treated.  If   $X \subset   \Om_{k+1}$ then  $ E^d_{k, \bom^+} (X, \cA  )$  is   independent of  $\bom^+$   and  identical with the global quantity.

The 
  $ [ D_{k, \bom} (\cA) + bL^{-1} P(\cA)  ] _{\Om_{k+1}}   $  on $\Om_{k+1}^{(k)}$ is  not self-adjoint,   but   $ [( D_{k, \bom} (\cA) + bL^{-1} P(\cA))\ga_3  ] _{\Om_{k+1}}   $ is self-adjoint   and  they have the same  determinant so      
\be   
   \de \sZ_{k, \bom^+} (\cA)    
= \exp \B( \Tr  \log \B[  (D_{k, \bom} (\cA) + bL^{-1}  P(\cA))  \ga_3 \B]_{\Om_{k+1}} \B)
\ee  
We  use  a representation  of  the logarithm   from   \cite{Dim15b}  which is for any $R_0>0$
\be     \label{singing}  
\begin{split}
 & \log   \B[  (D_{k, \bom} (\cA) + bL^{-1}  P(\cA))  \ga_3 \B]_{\Om_{k+1}}  \\ 
  =&   
   \B[ D_{k, \bom}(\cA) + bL^{-1}P(\cA) \B]_{\Om_{k+1}}  \int_{R_0}^{\infty} \frac{dy}{y}   \Ga_{k,\bom^+,y}(\cA)  
-  i  \ga_3   \int_{0}^{R_0} dy   \Ga_{k, \bom^+, y}(\cA)  + \log  R_0     + \frac{   i\pi  }{2}    \\
\end{split} 
\ee
where $ \Ga_{k, \bom^+, y}(\cA)=[ D_{k, \bom}(\cA) + bL^{-1}P(\cA) + i \ga_3 ]^{-1}_{\Om_{k+1}}$. 
In appendix \ref{A} we establish the identity
\be
  \Ga_{k, \bom^+,y }(\cA)    =   
\B[   \sB_{k,y}(\cA)   +   b_k^2  \sB_{k,y}(\cA) Q_k(\cA)   S_{k, \bom^+,  y}(\cA)Q_k^T(-\cA) \sB_{k,y}(\cA)  \B]_{\Om_{k+1} }
\ee
where
\be
\sB_{k,y}(\cA)  =    \frac{1}{b_k+ i\ga_3y}  (I -  P(\cA) )   +   \frac{1}{ b_k + bL^{-1}+ i\ga_3y }  P(\cA) 
\ee
Since $\Ga_{k, \bom^+, y}(\cA)  = \cO(y^{-1})$ we can take the limit $R_0 \to \infty$ in (\ref{singing}). The first term goes to zero.  The divergent part of 
the second term is canceled by the  $\log  R_0$.  These limits are discussed in more detail in \cite{Dim15b}.
One finds    
\be     
\begin{split}
&    \de \sZ_{k, \bom^+} (\cA)      =  \exp \B(        \log  b_k  \Tr   [ I - P(\cA) ]_{\Om_{k+ 1} }    +     \log  (     b_k + bL^{-1} ) \Tr  [ P(\cA)  ]_{\Om_{k+ 1} }   \\
&- i  b_k^2  \int_0^{\infty   }   \  \Tr \B(   \B[ \ga_3\sB_{k,y}(\cA)  Q_k (\cA)   S_{k, \bom^+,y }(\cA)Q_k^T(-\cA) \sB_{k,y}(\cA) \B) \B]_{\Om_{k+1} }   \    dy   \B)  
\\
\end{split}
\ee
The  trace of the projection operator is
\be   
\Tr[  P(\cA) ]_{\Om^{(k)}_{k+ 1} } = \Tr[Q^T(-\cA) Q(\cA) ]_{\Om^{(k)}_{k+ 1} }  =\Tr[  Q(\cA)Q^T(-\cA)  ]_{\Om^{(k+1)}_{k+ 1} }  = 4|{\Om^{(k+1)}_{k+ 1} }| = 4 L^{-3} | \Om^{(k)}_{k+1}|  
\ee
Also  insert  the polymer expansion $  S_{k, \bom^+ ,y }(\cA)= \sum_{X \in \cD_k} S_{k, \bom^+ ,y }(X, \cA)$ of lemma \ref{hughes}    which yields   
\be     
   \de \sZ_{k, \bom^+} (\cA)      =  \exp \B(   \B(    (1 - L^{-3} )     \log  b_k       +  L^{-3}   \log  (     b_k + bL^{-1} )   \B) 4 | \Om^{(k)}_{k+1}| 
     +   \sum_{X  \cap \Om_{k+1} \neq \emptyset }         E^d_{k, \bom^+  } (X, \cA ) \B)  
\ee 
where     
\be   E^d_{k, \bom^+  } (X, \cA )   = 
     i  b_k^2  \int_0^{\infty   }   \  \Tr \B(   \B[ \ga_3 \sB_{k,y}(\cA)  Q_k (\cA)   S_{k, \bom^+,y}(X, \cA)Q_k^T(-\cA) \sB_{k,y}(\cA)  \B) \B]_{\Om_{k+1} }     \B)  \    dy 
\ee
The trace here  is estimated by (\ref{arty2}) for $S_{k, \bom^+,y}(X, \cA)$ (with $\ka \to \ka+1$) by 
\be  \label{lake}
\begin{split}
  | \Tr( \cdots )  |  
= & \B| \sum_{x \in  X \cap \Om^{(k)}_{k+1} }
\blan  Q^T_k(-\cA) \sB^T_{k,y}(\cA) \ga_3 \de_x,     S_{k, \bom^+,y}(X, \cA) Q^T_k(-\cA)\sB_{k,y}(\cA) \de_x \bran \B|
\\                        
\leq  &  C   e^{ -  (\ka + 1)   d_M(X  )}  
 \sum_{  x \in  X \cap \Om^{(k)}_{k+1}   }  \| Q^T_k(-\cA)\sB^T_{k,y}(\cA) \ga_3 \de_x\|_1 \| Q^T_k(-\cA) \sB_k(\cA,y) \de_x\|_{\infty} \\
\leq &  C M^3  e^{ -  \ka   d_M(X  )}  
(1+ |y|^2|)^{-1}\\
\end{split}
\ee
Here we used $\sB_k(\cA,y) = \cO(y^{-1})$ as $y \to \infty$  and
\be 
 | X \cap \Om^{(k)}_{k+1}   |  \leq \Vol ( X ) = M^3 |X|_M \leq \one M^3 (d_M(X) +1) \leq \one M^3 e^{d_M(X)}
\ee
The estimate (\ref{lake}) yields the estimate on $   E^d_{k, \bom^+  } (X, \cA )$.
\bigskip

\section{Block averaging for gauge fields  }

For the renormalization transformations for  gauge fields  we      follow the  treatment originated by   Balaban       
 \cite{Bal84a}, \cite{Bal84b},  \cite{Bal85b},      and   Balaban, Imbrie, and Jaffe  
 \cite{BIJ85},  \cite{BIJ88}.   See also     \cite{Dim15},  \cite{Dim15a}, \cite{Dim15b}. 

\subsection{block averaging}

(a.) We define   block averaging operators on gauge fields, see  \cite{BIJ85}  \cite{Imb86} for more details.
For a  field $A$ on    $\bbT^0_{N-k}$       define  an  averaged field  $\cQ A$    on   oriented bonds   in      $\bbT^{1}_{N-k} $ by     (for reverse oriented bonds take minus this)  
\be   
\begin{split}
(\cQ A) (y,  y + L e_{\mu} )  
= &L^{-4}  \sum_{x \in B(y)  }    A(  \Ga_{x,  x +  L e_{\mu} } ) \\
\end{split}
\ee
where   $  \Ga_{x,  x +  L e_{\mu} }$ is the straight line between the indicated points,  $A(\Ga)= \sum_{b \in \Ga} A(b)$, and $B(y)$
is a cube with $L$-sites on a side centered on $y \in   \bbT^{1}_{N-k}$.

 Similarly  $\cQ$ can be defined on any lattice  (with  no weight factors)
 and we define compositions by   $ \cQ_k  =   \cQ   \circ \cdots \circ    \cQ  $  ($k$ times).
 In particular it maps functions $\cA$ on (bonds in) $\tk$ to functions $\cQ_k \cA $ on (bonds in) $\tz$
 and is given by 
 \be   \label{moonshine}
(  \cQ_k \cA) (y,  y +  e_{\mu} )  = L^{-4k} \sum_{x \in B_k(y) }\cA  ( \Ga_{x,  x +   e_{\mu}})=
 \int_{|x -y|  < \frac12  }  L^{-k}   \cA  ( \Ga_{x,  x +   e_{\mu}})\ dx
\ee
This has the adjoint mapping $A$ on (bonds in) $\tz$ to $\cQ_k^T A$ on (bonds in) $\tk$ of length $\eta = L^{-k}$
\be \label{lunge}
(\cQ_k^T A)(b)
=       L^{-1}  \sum_{x, \mu:  \Ga(x, x + Le_{\mu} ) \ni b}                  A ([x],  [x] +  e_{\mu} ) 
\ee
where $[x]$ is the unique $ y \in \tz$ such that $ x \in B_k(y)$.
If $Q$  is the averaging operator on scalars  we have the identities
$   dQ = \cQ d  $.  
For example for a scalar $f$
\be 
\begin{split}
(dQf) (y, y+Le_{\mu}) = &  L^{-1}\B( Qf (y+Le_{\mu}) -Qf (y) \B) = L^{-4} \sum_{x \in B(y)} f\B((x+L e_{\mu}) -f(x) \B) \\
= & L^{-4} \sum_{x \in B(y)} df\B ( \Ga_{x, x +Le_{\mu}} \B) = (\cQ df ) (y, y+Le_{\mu})
\end{split}
\ee
More generally
 \be  \label{king1}
   dQ_k = \cQ_k d      
  \ee

\noindent
(b.) We also define an operator  $\cQ^{(2)}$ on functions on plaquettes (squares)  as follows.
 For any  $x \in \bbT^0_{N-k}$ let    $p$ denoted a unit  plaquette  designated by its corners $ [ x, x+ e_{\mu}, x+ e_{\mu} + e_{\nu}, x + e_{\nu}  ] $
 and let  $P_x$
be the $L$-plaquette
\be
  P_x   =   [ x, x+ Le_{\mu}, x+ Le_{\mu} + Le_{\nu}, x + Le_{\nu}  ]
\ee
For  $F$ defined on plaquettes in $\bbT^0_{N-k}$  we define $\cQ^{(2)} F $
on plaquettes in $ \bbT^{1}_{N-k} $ by
\be   
  (\cQ^{(2)} F )(P_y)   = L^{-5} \sum_{x \in B(y)}  F(P_x)  \hs F(P_x) = \sum_{p  \subset  P_x} F(p)
\ee
where $y \in \bbT_{N-k}^1$.  
 We  have the identity $dQ = \cQ d  $ and more generally 
 \be  \label{king2}
    d \cQ_k  = \cQ_k^{(2)} d   
  \ee
  
  \noindent
 (c.)  Also define  a  surface    averaging operator    $\cQ^s$ on  a  unit  lattice  by
     \be   
   ( \cQ^s  \cA )(y, y+ Le_{\mu})   =   L^{-2}  \sum_{b'  \in B^s(y,  y+Le_{\mu} )  } \cA(b')
   \ee
 where   $B^s(y,  y+Le_{\mu} ) $ is the set of unit  surface  bonds joining the cubes     $B(y)$  and  $B(y + L e_{\mu})$.
 This satisfies  $\cQ \cQ^{s,T} = I$ and   $\cQ^s  \cQ^{s,T} = L $ and   $  d Q^{T} f=  \cQ^{s,T}df $.

  We also have the $k$-fold composition  $\cQ^s_k=\cQ^s \circ \cdots \cQ^s$ mapping  $\cA$ on $\tk$  to $\cQ^s_k\cA$ on $\tz$ given  by 
  \be
  ( \cQ^s _k \cA )(y, y+  e_{\mu})   =   L^{-2k}  \sum_{b  \in B^s_k(y,  y +e_{\mu} )  } \cA(b)
  \ee
  where $B^s_k(y, y+ e_{\mu})$ is surface bonds in $\tk$ joining $B_k(y)$ and $B_k(y + e_{\mu})$.
   This has the adjoint
  \be
   ( \cQ^{s,T} _k A )(b)  
  = \begin{cases}  L^{k} A (y, y+  e_{\mu}) 
  & \hs b \in B^s _k(y, y+  e_{\mu}) \\
  & \hs \textrm{ otherwise } \\
  \end{cases} 
  \ee
  and  satisfies the identities 
  \be \label{only}
   \cQ_k \cQ^{s,T}_k = I \hs \cQ^s_k  \cQ^{s,T}_k = L^k  \hs   d Q^{T}_k =  \cQ^{s,T}_kd
   \ee

 \noindent
 (d.) Further define an edge averaging operator  on functions on
 plaquettes by 
 \be
  ( \cQ^e  F )(p')   =   L^{-1}  \sum_{p  \in B^e(p')   } F(p)
\ee
Here if   $ p' $ is a plaquette in   $\bbT^{1}_{N-k} $ with corners $[y, y + Le_{\mu}, y+ Le_{\mu} + L e_{\nu}, y+ L e_{\nu}]$
then $B^e(p')$ are  the  $p$ are plaquettes $p$ with corners  $[x, x + e_{\mu}, x+ e_{\mu} + e_{\nu}, x+ e_{\nu}]$
   in  $\bbT^0_{N-k}$  such that $x \in B(y), x+ e_{\mu} \in B(y + Le_{\mu})$, etc.
 This satisfies $ \cQ^e  \cQ^{e,T} = L^2 $ and $  d \cQ^{s,T}A =  \cQ^{e,T}dA$.

 We also have the $k$-fold composition  $\cQ^e_k=\cQ^e \circ \cdots \cQ^e$ mapping  $\cF$ on (plaquettes in ) $\tk$  to $\cQ^e_k\cF$ on
  (plaquettes in )  $\tz$ given  by 
  \be
  ( \cQ^e_k \cF )(p')   =   L^{-k}  \sum_{p  \in B_k^e(p' )  } \cF(p)
  \ee
  Here if $p'$ is the unit  plaquette  $[y, y + e_{\mu}, y+ e_{\mu} +  e_{\nu}, y+  e_{\nu}]$ then with $\eta = L^{-k}$ the set 
 $ B^e_k(p')$ are the plaquettes $[ x, x + \eta e_{\mu}, x + \eta'e_{\mu} + \eta e_{\nu},x + \eta e_{\nu})]$ in $\tk$  such that
 $x \in B_k(y)$, $ x+ \eta  e_{\mu} \in B_k(y+ e_{\mu} )$, etc.
    This has the adjoint
  \be \label{tingle}
   ( \cQ^{e,T} _k F )(p)  
  = \begin{cases}  L^{2k} F (p') 
  & \hs p \in B^e_k (p') \\
  & \hs \textrm{ otherwise } \\
  \end{cases} 
  \ee
  and  satisfies the identities 
  \be
    \cQ^e_k  \cQ^{e,T}_k = L^{2k}  \hs   d Q^{s,T}_k =  \cQ^{e,T}_k d
   \ee

\subsection{global axial gauge averaging}   

We now explain the renormalization group transformation for gauge fields,  starting with a review of the global version in  \cite{Dim15b}.  
Starting  with a   density  $\rho_0 (A_0)  $  defined for  $A_0$ on (bonds in )   $\bbT^0_N$  we define  a sequence of   densities  $ \rho_{k} (A_{k}) $
defined for functions  $A_k$  on $\tz$ as follows. 
Given $\rho_k(A_k)$ and      $A_{k+1}$ on     $ \bbT^1_{N-k}$  first   define      
\be    \label{bumble0} 
\tilde   \rho_{k+1} (A_{k+1})   = \int   \de ( A_{k+1} - \cQ A_k)  \de   (\tau A_k) \rho_k(A_k)  \  DA_k   
\ee
Here we have   introduced  an axial gauge fixing  delta   function   $ \de   (\tau A_k)$
with 
 \be 
\de  (\tau A_k)    =  \prod_{y  \in   \bbT^{1}_{N-k}  }  \prod_{x  \in B(y),  x \neq  y}    \de \B((\tau   A_k)(y, x) \B)   
\ee
where $(\tau A_k)(y,x)$ is defined in  (\ref{eleven}).       In making this definition we are not  just coarse graining, but also introducing a gauge fixing function step by step.

Then  we      define   $   \rho_{k+1}(A_{k+1}) $  for     $A_{k+1}$ on    $ \bbT^0_{N-k-1}$        
by 
\be  
\label{scaleddensity2}
    \rho_{k+1}(A_{k+1}  )  =     \tilde  \rho_{k+1} (A_{k+1,L})  L^{\frac12 (b_N- b_{N-k-1})   -\frac12 (s_N- s_{N-k-1}) }
\ee
Here   $b_n =  3L^{3N}$ is the number of bonds in a three dimensional   toroidal lattice with $L^N$ sites on a side, and again
$s_N = L^{3N}$ is the number of sites.  (So the scaling factor could be written  $ L^{s_N- s_{N-k-1} }) $.

The result of the iteration can  be computed explicitly as
\be  \label{fourfour} 
\begin{split}
  \rho_k (A_k)  =     &    \int    \de (A_k -   \cQ_k  \cA )  \de  (\tau_k  \cA)  \rho_{0,L^{-k}} ( \cA)    D\cA  \\
  \end{split}  
\ee
where now   $\cA$ is defined on bonds in     $\tk$.    Then       $ \cQ_k \cA$ is  defined in (\ref{moonshine}) and and the   gauge fixing function is   now   
 \be 
  \de  (\tau_k  \cA)   \equiv        \prod_{j=0}^{k-1}    \de  (\tau  \cQ_j \cA )   
 \ee

Now suppose we start with $\rho_0(A_0) =  F_0( A_0) \exp( - \frac12 \|d A_0 \|^2)$ for an arbitrary function $F_0$.  Then 
\be  \label{fourfourfour} 
\begin{split}
  \rho_k (A_k)  =     &    \int    \de (A_k -   \cQ_k  \cA )  \de  (\tau_k  \cA)  F_{0,L^{-k}} ( \cA)   \exp \B( - \frac12 \|d \cA \|^2\B)  D\cA  \\
  \end{split}  
\ee
 Define $\cA^{\sx}_k $ on $\tk$ by
\be
\cA^{\sx}_k  = \cA^{\sx}_k (A_k) =       \cH^{\sx}_kA_k  = \textrm{ minimizer of } \| d \cA \|^2  \textrm{ subject to } \cQ_k \cA = A_k, \tau_k \cA =0
\ee
and make  the change of variables   $\cA  =  \cA^{\sx}_k    + \cZ$ in the integral.    The term $1/2  \| d \cA \|^2 $ splits     and  one finds 
\be  \label{total1} 
      \rho_k(A_{k})   =    \sZ_k      F_k (   \cA^{\sx}_k )     \exp \B(   - \frac12   \|  \cA^{\sx}_k   \|^2    \B)   
 \ee
where 
\be  \label{total2}
\begin{split}
 F_k (\cA  )   
   = & \sZ_k^{-1}    \int    \de (  \cQ_k \cZ  )  \de (\tau_k \cZ)    F_{0,L^{-k}}  (\cA+ \cZ)    \exp   \B( - \frac 12 \|  d \cZ \|^2    \B)  D \cZ
    \\   
\sZ_{k}   =   &   \int    \de (  \cQ_k \cZ  )   \de (\tau_k \cZ)     \exp  \B ( - \frac 12 \|  d \cZ \|^2   \B )  D \cZ \\
\end{split}
\ee
It is also useful to study the next step going from $\rho_k$ to $\rho_{k+1}$; see \cite{Dim15}, \cite{Dim15b} or  the multiscale version which we now take up.

\subsection{multiscale axial gauge averaging} \label{knots2}

As in section  \ref{knots}   we consider   a  decreasing  sequence of small field regions   $\bom  =  (\Om_1,  \dots  \Om_k) $   in    $\tk$. 
We  use the same  notation $\Om_j$ for a set of bonds such that at least one end is   in $\Om_j$.    We define again    $\de \Om_j   =  \Om_j  -  \Om_{j+1}$ 
and  $\de \Om^{(j)}_j   =  \Om^{(j)}_j  -  \Om^{(j)}_{j+1}$  in $\bbT^{-(k-j)}_{N-k}$ where for example $  \Om^{(j)}_j $ are bonds joining the  centers of
$L^j$ cubes  with at least one end in  $\Om_j$. 
Associated  with  the  sequence  we have  a sequence of functions  
\be     
    A_{k, \bom  }  
   =    (A_{0, \Om_1^c},   A_{1, \de  \Om_1},     \dots,   A_{k-1, \de \Om_{k-1}},  A_{k, \Om_k})
\ee
where  $A_{0, \Om_1^c}$ is defined on $\Om_1^c \subset \tk$,  $ A_{j, \de \Om_j }$ is defined on   $ \de \Om_j^{(j)} $,  and $A_{k, \Om_k}$ is defined on $\Om_k^{(k)} \subset \bbT^0_{N-k} $.

Starting with  $\rho_0(A_0)$  we define a sequence  of  densities    $\rho_{k, \bom} (A_{k, \bom})$  as follows. 
  Given the density   $\rho_{k, \bom}$
 add    a new union of $LM$ cubes  $\Om_{k+1} \subset  \Om_k$ and define $   \bom^+   =  ( \bom,  \Om_{k+1}  )   $
  and $A_{k+1, \bom^+} = (A_{k, \bom}, A_{k+1, \Om_{k+1} } )$ where $A_{k+1, \Om_{k+1}} $
   is defined on $  \Om^{(k+1)}_{k+1} \subset \bbT^1_{N-k} $.
Instead of   (\ref{bumble0})  
we define
  \be    \label{bumble2} 
\tilde   \rho_{k+1, \bom^+} (A_{ k+1,\bom^+})  
  =    \int   \de_{\Om_{k+1}} ( A_{k+1} - \cQ A_{k})  \de_{ \Om_{k+1} } ( \tau A_{k}  )  \rho_{k, \bom} (A_{k, \bom} )  DA_{k, \Om_{k+1}}   
\ee
where the integral is over $A_{k, \Om_{k+1}}   $ on $\Om^{(k)}_{k+1}$.  Note that $\cQ A_{k}$ depends on $A_k(b)$ for bonds outside
$\Om_{k+1}$, but we only integrate over  $A_k(b)$ for bonds in $\Om_{k+1}$. The  axial gauge fixing is
\be  
\de_{ \Om_{k+1} } ( \tau A_{k}  )= \prod_{y \in  \Om^{(k+1)}_{k+1}}  \prod_{x \in B(y), x \neq y  }  \de  \B( (\tau A_k  ) (y,x)  \B) 
\ee
Then we scale again and define for $  \bom^+$ in $\bbT^{-k-1}_{N-k-1}$ with associated $A_{k+1, \bom^+} $ 
\be  
\label{scaleddensity3}
    \rho_{k+1, \bom^+}(A_{k+1, \bom^+}  )  =   \tilde  \rho_{k+1, L \bom^+} \B( [A_{k+1, \bom^+}]_L \B)  \si'_{k+1, \Om_{k+1} } 
\ee
with scaling factors  $ \si'_{k+1, \Om_{k+1} } $ still to be chosen.

 When we compose these operations  the result will be  expressed in terms of an  averaging  operator $ \cQ_{k,\bom}  $ defined on $\cA$ on $\tk$ by 
 \be    
 \cQ_{k,\bom}  \cA = \B(\cA_{\Om^c_1}, \cQ_{1, \de \Om_1 } \cA, \dots ,   \cQ_{k-1, \de \Om_{k-1} } \cA,  \cQ_{k, \Om_k} \cA \B) 
 \ee
 This is a field of type $A_{k, \bom}$ and we define
 the delta function
 \be
   \de \B(A_{k, \bom} -   \cQ_{k, \bom}   \cA \B)  \equiv \prod_{j=0}^k \de_{\de \Om_j} \B(A_{j} - \cQ_j \cA\B)
 \ee
 Here   $\de \Om_k \equiv \Om_k$ and $\de \Om_0 = \Om_1^c$. The convention is  $\cQ_0 = I$ so the $j=0$ delta function
 is  $\de_{\Om^c_1} (A_0 - \cA)$ and just sets $A_0 =\cA$ on $\Om^c_1$.
  We also define a   hierarchical gauge fixing  function 
 \be
    \de    (  \tau_{k,\bom } \cA)    =      \prod_{j=1}^{k}    \de_{  \Om_j   }  (\tau  \cQ_{j-1} \cA  )   
\ee

 \begin{lem} \label{stumble} For       $\cA:    \tk \to \bbR$  one can choose the scaling factors   $ \si'_{k+1, \Om_{k+1} } $ so that
\be 
 \label{bumble3}
  \rho_{k, \bom} (A_{k,  \bom})  =      \int    \de \B(A_{k, \bom} -   \cQ_{k, \bom}   \cA \B)   \de     \B(  \tau_{k, \bom } \cA \B)      \rho_{0,L^{-k}} ( \cA) \   D\cA
    \ee 
\end{lem}
 \bigskip

 \pr   The proof is by induction.   Assuming it is true for  $k$  we have     
  \be    
  \begin{split}
&\tilde   \rho_{k+1, \bom^+} (A_{k+1, \bom^+})  =    \\
    &  \int   \de_{\Om_{k+1}} ( A_{k+1} - \cQ A_{k})  \de_{\Om_{k+1}}  ( \tau A_k  )    \de  \B(A_{k,\bom } -   \cQ_{k, \bom}   \cA  \B)  
      \de     \B(  \tau_{k,  \bom } \cA \B)    \rho_{0,L^{-k}} ( \cA) \     \ DA_{k, \Om_{k+1}}\ D\cA   \\
\end{split} 
\ee
 But    since $A_k = \cQ_k \cA$  on $\Om_k$ and hence $\Om_{k+1}$
   \be  \label{doubt1}
   \de_{\Om_{k+1}}  ( \tau A_k  )      \de    \B (  \tau_{k,  \bom } \cA  \B)=  \de_{\Om_{k+1}}  ( \tau  \cQ_k \cA   )      \de     \B(  \tau_{k,  \bom } \cA \B)     
      =    \de    \B (  \tau_{ k+1,\bom^+  } \cA \B)   \ee
Also 
 \be \label{doubt2}
 \int   \de_{\Om_{k+1}} ( A_{k+1} - \cQ A_k)    \de  \B(A_{k, \bom} -   \cQ_{k,\bom}   \cA  \B)    DA_{k, \Om_{k+1}}   
  =      \de \B (A_{k+1, \bom^+} -   \cQ_{ k+1,\bom^+ }   \cA \B ) 
 \ee  
 Thus we have 
  \be    \label{bumble} 
 \tilde   \rho_{k+1, \bom^+} (A_{k+1,  \bom^+})    
  =     \  \int       \de \B(A_{k+1, \bom^+} -   \cQ_{k+1,  \bom^+ }   \cA \B)  \de \B( \tau_{k+1, \bom^+} \cA\B)    \rho_{0,L^{-k}} ( \cA) \   D\cA  
\ee
Then scale by (\ref{scaleddensity3}) and   change variables from $\cA$ to $\cA_L$ using 
    $\cQ_{k+1,  L \bom^+ }  \cA_L  =  (\cQ_{k+1, \bom^+}  \cA)_L$ and    $\tau_{k+1, L \bom^+}   \cA_L  =  (\tau_{k+1, \bom^+}   \cA)_L$.
    This introduces scaling factors and we choose    $ \si'_{k+1, \Om_{k+1} } $ to exactly cancel these.  Hence
\be
\begin{split}
&   \rho_{k+1, \bom^+} (A_{ k+1, \bom^+}) 
 = 
 \int       \de \B(A_{k+1, \bom^+} -   \cQ_{k+1,  \bom^+ }   \cA \B)  \de \B( \tau_{k+1, \bom^+} \cA\B)    \rho_{0,L^{-k-1}} ( \cA) \   D\cA    \\
\end{split}
\ee 
This  completes the proof.
\bigskip

\rem   In addition to the multiscale averaging operator $\cQ_{k, \bom}$ on 1-forms (functions on bonds), there are multiscale averaging 
operators $Q_{k, \bom}$ on 0-forms (scalars) and $\cQ^{(2)}_{k, \bom}$  on 2-forms (functions on plaquettes).   These satisfy identities 
generalizing (\ref{king1}), (\ref{king2}) namely
\be  \label{sweetlou}
d Q_{k, \bom} f = \cQ_{k, \bom} df  \hs    d \cQ_{k, \bom} \cA = \cQ^{(2)}_{k, \bom} d \cA 
\ee
In the second case we may have to evaluate  $  d \cQ_{k, \bom} \cA$ on a plaquette in  $\bbT^{-(k-j)}_{N-k}$ with some points in
$\de \Om_j$ and some in $\de  \Om_{j-1}$.   But in $\de \Om_{j-1}$ we have $\cQ_{k, \bom} \cA = \cQ_{j-1} \cA$ defined on the finer
lattice $\bbT^{-(k-j+1)}_{N-k}$.  In this case the convention is that  $  d \cQ_{k, \bom} \cA$ on    $\bbT^{-(k-j)}_{N-k}$  bonds in $\de \Om_{j-1}$
is interpreted as  $\cQ ( \cQ_{j-1} \cA ) = \cQ_j \cA$.

    \subsection{free flow} \label{freeflow2}
  
 Now suppose that we start with a perturbation of the  the free action
 \be   
    \rho_0(A_0)   =         \exp\B( - \frac12 \|d  A_0   \|^2\B)    F_{0} (A_0)   
\ee
for some bounded function $F_0$.    Then
  \be 
 \label{nougat1} 
  \rho_{k, \bom} (A_{k,  \bom})  =    \int    \de  \B(A_{ k,\bom} -   \cQ_{k, \bom}  \cA \B)  \de   \B (\tau_{k,\bom} \cA \B)   \exp\B( - \frac12 \|d  \cA   \|^2\B)
  F_{0, L^{-k}} (\cA)   \   D\cA 
    \ee 
One can show that  the quadratic form  $\|d \cA \|^2  $ is positive definite on the constrained surface so the integral exists.
We evaluate the integral by expanding around the minimizer  $\cA^{\sx}_{k,  \bom} $ on  $\tk$   defined  by
\be \label{undertake}
\cA^{\sx}_{k,  \bom} 
  = \textrm{
  minimizer       of  }      \|d \cA \|^2   \textrm{   subject to  }      \cQ_{k,\bom} \cA = A_{k, \bom} ,  \    \tau_{k,\bom}  \cA  =0   
    \ee
 Now for $\cA$ satisfying the constraints define $\cZ$ by  $\cA =    \cA^{\sx}_{k,  \bom} + \cZ $.  The new field $\cZ$ vanishes on $\Om_1^c$ since $\cA, \cA^{\sx}_{k,  \bom}$ have the same fixed value there. 
    The cross term in 
   $\frac12 \| d \cA\|^2$ vanishes and we have
      \be  \label{undo}
     \frac12    \|  d  \cA   \|^2   =        \frac12    \|  d  \cA^{\sx} _{k, \bom}  \|^2
        + \frac12    \|  d  \cZ  \|^2  
  \ee
  Changing to an integral over $\cZ$ yields
    \be 
 \label{nougat2} 
  \rho_{k, \bom} (A_{k, \bom})  =     \sZ_{k, \bom}  \exp  \B(  -     \frac12    \|  d  \cA^{\sx} _{k, \bom}  \|^2   \B)\   F_{k, \bom}  (\cA^{\sx}_{k, \bom} )   
   \ee 
 where
 \be  \label{wander}
 \begin{split}
  F_{k, \bom}  (\cA )  
  = &    \sZ_{k, \bom}^{-1}   \int   
    \de \B(  \cQ_{k,\bom}   \cZ \B)  \de   \B(\tau_{k,\bom} \cZ \B)   \exp\B( - \frac12 \|d \cZ \|^2\B)   F_{0, L^{-k}} (   \cA  + \cZ  ) 
  \   D\cZ   \\
 \sZ_{k, \bom}
  = &      \int     
    \de \B(  \cQ_{k,\bom}   \cZ \B)  \de   \B(\tau_{k,\bom} \cZ \B)   \exp\B( - \frac12 \|d \cZ \|^2\B)
  \   D\cZ   \\
 \end{split}
\ee    
 
 We discuss the minimizer    in more detail. 
 Define an axial gauge Green's function     $  \cG^{\sx} _{k, \bom}  $on  $ \tk$   by      
\be  \label{gint0}
 \cG^{\sx}_{k, \bom} J   =      \textrm{    minimizer  of }      \frac12  \| d \cA \|^2 -   \blan \cA , J \bran     \textrm{  subject to   }    
        \cQ_{k,\bom} \cA=0  \    \tau_{k,\bom}  \cA  =0    
   \ee 
 This vanishes on $\Om_1^c$ or if $J$ has support in $\Om^c_1$, so it is essentially an operator on $\Om_1$.
 We do not have an explicit expression for  $\cG^{\sx} _{k, \bom}  $ but do have the representation 
 \be \label{gint}
 \exp\B(  \frac12\blan J,    \cG^{\sx}_{k, \bom} J \bran \B)
   =  \sZ_{k, \bom}^{-1}  \int \exp \B( -\frac12  \| d \cZ \|^2 + <\cZ , J> \B)  \de  (  \cQ_{k, \bom}  \cZ )  \de    (\tau_{k,\bom} \cZ) D \cZ
  \ee 
   Because of the constraint this is not entirely  routine.     The details are discussed   Proposition A.3 in \cite{BIJ85}.    They also show that   that  if $\cZ$ is in the constrained surface
   then $  \cG^{\sx}_{k, \bom}\ \de d \ \cZ =\cZ$.   
   
 \begin{lem}  The minimizer $\cA^{\sx}_{k, \bom} $  is given by 
    \be    \label{citrus1}
    \begin{split}
\cA^{\sx}_{k, \bom}   = &   \cH^{\sx}_{k, \bom}   A_{k, \bom}  \equiv
    \B(   \cQ_{k,\bom}^{s,T} -   \cG^{\sx} _{k, \bom}    \de   d     \cQ_{k, \bom}^{s,T} \B)A_{k, \bom}  \\
  \\  
 \end{split}
 \ee
 \end{lem}

\pr   By (\ref{only}) one solution of    $ \cQ_{k,\bom} \cA = A_{k,\bom}$  and $\tau_{k,\bom}  \cA  =0$     is   $\cA =    \cQ^{s,T}_{k,\bom} A_{k,\bom}$
with the convention that   $ \cQ^{s,T}_{k,\bom}  =I$ on $\Om_1^c$.   Thus  the general solution of the constraints in 
(\ref{undertake})  are    functions  $ \cA =  \cQ^{s,T}_{k,\bom} A_{k,\bom} + \cA'$
where    $ \cQ_{k,\bom} \cA' = 0 ,  \    \tau_{k,\bom}  \cA'  =0    $.  With these constraints  we seek to minimize in $\cA'$ 
\be 
\frac12 \| d \B( \cQ^{s,T}_{k,\bom} A_{k,\bom} + \cA' \B)\|^2
=    \frac12  \| d \cA' \|^2 + \blan \cA', \de d  \cQ^{s,T}_{k,\bom} A_{k,\bom}  \bran    
  + \dots
\ee
where $\de = d^T$ and the omitted terms do not depend on $\cA'$.
 The solution is 
 \be 
  \cA'=   -   \cG^{\sx} _{k, \bom}     \de   d     \cQ_{k, \bom}^{s,T}A_{k, \bom}  \ee
which gives the result. 
\bigskip  

\rem
   As noted earlier  the cross term in (\ref{undo})  must vanish.  But it instructive to see how this
  comes about.  We have
   \be 
  \begin{split}
  \blan  d\cA^{\sx} _{k, \bom}, d \cZ \bran   = &   \blan d  \B(I - \cG^{\sx} _{k, \bom} \de d \B)   \cQ_{k, \bom}^{s,T}A_{k, \bom},  d \cZ \bran  
   =    \blan    \cQ_{k, \bom}^{s,T}A_{k, \bom}   ,    \B(I - \de d \cG^{\sx} _{k, \bom}\B)   \de  d \cZ \bran =0 \\  
    \end{split}
  \ee 
The last step follows since $\cZ$ satisfies the constraints and  so  $\cG^{\sx} _{k, \bom}  \de d \cZ =\cZ$ as noted above.
 \bigskip

\subsection{the next step}    \label{nextstep}
     Suppose we are starting with the expression  (\ref{nougat2}) for  $ \rho_{k,  \bom}(A_{k, \bom}) $.   In the next step     we  have
\be  \label{six}
\begin{split}
& \tilde \rho_{k+1, \bom^+} (A_{k+1,\bom^+}) 
 =     \sZ_{k, \bom}  \\
 &   \int       \de_{\Om_{k+1}}(A_{k+1} -    \cQ A_{k}  )\   \de_{\Om_{k+1} }(  \tau  A_k)  \    
    \exp  \B(   -\frac12    \|  d  \cA^{\sx} _{k, \bom}  \|^2    \B)    F_{k, \bom} (\cA^{\sx}_{k, \bom} )    dA_{k,\Om_{k+1}}  \\
 \end{split}
\ee
To evaluate the integral we define
\be
\begin{split}
   A^{\min}_{k, \bom^+}  
   = &
  \textrm{ minimizer of  }        \|    d\cA^{\sx}_{k, \bom}  \|^2    \textrm{ in }  A_{k, \bom}    \textrm{ subject to }   \\
 & \cQ  A_k = A_{k+1}, \tau A_k = 0  \textrm{ on } \Om_{k+1}
  \textrm{ with } A_{k, \bom} \textrm{ fixed on } \Om^c_{k+1}  \\
      \end{split}
       \ee      
  Expand around the minimizer  by
  \be    \label{ant1}
   A_{k, \bom}    = A^{\min}_{k, \bom^+}     +  Z  
\ee 
Then  $Z$ has support  on  the unit lattice $\Om^{(k)}_{k+1}$. 
This   the transformation   induces   
\be   \label{ant2}
   \cA^{\sx}  _{k, \bom}    =    \cA^{0,\sx}_{k+1, \bom^+}   +  \cZ^{\sx}_{k, \bom}
\ee
where   
\be   
  \cA^{0,\sx}_{k+1, \bom^+}    =        \cA^{\sx}_{k, \bom}  (  A^{\min}_{k, \bom^+ } )
     \hs 
\cZ^{\sx}_{k, \bom}  \  =        \cA^{\sx}_{k, \bom} ( Z)     = \ \    \cH^{\sx}  _{k, \bom} Z 
\ee 
Note   that since   $A_{k, \bom}=\cQ_{k,\bom}  \cA^{\sx}_{k, \bom}   $  
we   have  
  \be 
   A^{\min}_{k,  \bom^+ }   =   \cQ_{k,\bom}   \cA^{0,\sx}_{k+1, \bom^+}   
   \ee
 This  is a useful representation of    $A^{\min}_{k,  \bom^+ } $  on  $\Om_{k+1}$ since as we will see   $ \cA^{0, \sx}_{k+1, \bom^+}$  is just   $ \cA^{\sx}_{k+1, \bom^+} $  before scaling.
Under the transformation  (\ref{ant2})     the cross terms vanish and so 
\be  \label{english}
\frac12 \|  d  \cA^{\sx} _{k, \bom} \|^2   =    \frac12 \|  d  \cA^{0,\sx}_{k+1, \bom^+}   \|^2  +
\frac12  \| d\cZ^{\sx}_{k, \bom}  \|^2  
 \ee 
 We   can also write    
 \be   
\frac12  \|d  \cZ^{\sx}_{k, \bom}  \|^2    =  \frac12  \blan  Z,   [ \De_{k, \bom}]_{\Om_{k+1}}  Z \bran   \hs    \De_{k, \bom}  =  \cH^{\sx, T}  _{k, \bom}  \de d      \cH^{\sx}  _{k, \bom}
 \ee
 Making the change of variables (\ref{ant1}), (\ref{ant2}) in (\ref{six}) we have 
\be  \label{spotless} 
\begin{split}
&\tilde \rho_{k+1, \bom^+} (A_{k+1,\bom^+}) \ 
 =    \  \sZ_{k, \bom}   \  \de  \sZ_{k, \bom^+}   
   \exp \B(     - \frac12 \|  d  \cA^{0,\sx}_{k+1, \bom^+}   \|^2        \B)  
   F^*_{k+1, \bom^+} (  \cA^{0,\sx}_{k+1, \bom^+} )   \\
 &   F^*_{k+1, \bom^+ } (  \cA   )  
    =  \de   \sZ_{k, \bom^+}   ^{-1}   \int    F_{k, \bom}\B(  \cA     +   \cZ^{\sx}_{k, \bom}     \B) 
       \de_{\Om_{k+1} } (  \cQ   Z  )   \de_{\Om_{k+1}} (  \tau Z)  \      \exp \B(   - \frac12  \blan  Z,   [ \De_{k, \bom}]_{\Om_{k+1}}  Z \bran     \B)  D Z \\
&  \de   \sZ_{k, \bom^+}   
    =  \int         \de_{\Om_{k+1} } (  \cQ   Z  )   \de_{\Om_{k+1}} (  \tau Z)  \      \exp \B(   - \frac12  \blan  Z,   [ \De_{k, \bom}]_{\Om_{k+1}} Z \bran     \B)  D Z \\   
\end{split}
 \ee
 Then  scale  as in (\ref{scaleddensity3}) and obtain
 \be \label{wink}
 \begin{split}
  \rho_{k+1, \bom^+} (A_{k+1,\bom^+}) \ 
 =  &    \si'_{k+1, \Om_{k+1} }  \  \sZ_{k,L \bom}   \  \de  \sZ_{k,L \bom^+}   
   \exp \B(     - \frac12 \| d\cA'_L\|^2        \B)  
   F^*_{k+1, L\bom^+} ( \cA'_L ) \\
  \end{split}
 \ee
 where  $\cA'$ on $\bbT_{N-k-1}^{-k-1}$ is  temporarily  defined by 
$
  \cA'_L   =   \cA^{0,\sx}_{k+1, \bom^+}   \textrm { at } [A_{k+1, \bom^+ }]_L 
$.
 
 We generate some identities by comparing  (\ref{wink})  with  the expression   (\ref{nougat2}) for  $ \rho_{k+1, \bom^+} (A_{k+1,\bom^+})$. 
   If $F_0 =1$ then $F_{k+1, \bom^+} =1 $ and $F^*_{k+1, \bom^+} =1$
 and setting the fields equal to zero we find that
 \be
   \sZ_{k+1, \bom^+}  =     \si'_{k+1, \Om_{k+1} }   \sZ_{k, L\bom}   \  \de  \sZ_{k, L\bom^+}   
 \ee
 It also  follows also that  $ \exp (     - \frac12 \| d \cA^{\sx}_{k+1, \bom^+}\|^2        ) = \exp (     - \frac12 \|d \cA'_L\|^2   )$. 
 Now for general $F$ the identity becomes   
 \be      \label{intrigue}
F_{k+1, \bom^+} \B( \cA^{\sx}_{k+1, \bom^+}\B)=  F^*_{k+1, L\bom^+} (   \cA'_L)   
 \ee
  Now suppose  $F_0(A_0) = <A_0, J> $.  Then  in (\ref{wander})  the integral over $\cZ$ vanishes  so $F_{k}(\cA) =< \cA_{L^{k}}, J>$.
   Similarly  $F_{k+1}^* (\cA) = F_{k}(\cA) =< \cA_{L^{k}}, J>$. 
    So (\ref{intrigue}) becomes   
  $
< [ \cA^{\sx}_{k+1, \bom^+}]_{L^{k+1}}, J > =      <  \cA'_{L^{k+1}}, J > 
 $.
 Hence  $\cA' = \cA^{\sx}_{k+1, \bom^+}$ or
 \be  
 [ \cA^{\sx}_{k+1, \bom^+}]_L \equiv     \cA^{0,\sx}_{k+1, \bom^+}   \textrm { at } [A_{k+1, \bom^+ }]_L 
\ee
So 
\be
 \rho_{k+1, \bom^+} (A_{k+1,\bom^+}) \ 
 =     \sZ_{k+1, \bom^+}    \exp \B(     - \frac12 \| d\cA^{\sx}_{k+1,  \bom^+ } \|^2        \B)  
   F^*_{k+1, \bom^+} \B( [ \cA^{\sx}_{k+1,  \bom^+ }]_L \B) \\
\ee

\bigskip

\noindent
\textbf{The fluctuation integral:} In (\ref{spotless})  we have integrals of the form for $Z$ on $\Om^{(k)}_{k+1}$
\be 
\de   \sZ_{k, \bom^+}   ^{-1} \int   f(Z)
       \de (  \cQ   Z  )   \de (  \tau Z)  \      \exp \B(   - \frac12  \blan  Z,   [ \De_{k, \bom}]_{\Om_{k+1}}  Z \bran     \B)  D Z
\ee
One can show that the. quadratic form $<  Z,   [ \De_{k, \bom}]_{\Om_{k+1}}  Z >$ is positive definite on the subspace $\cQ Z =0, \tau Z =0$
with a lower bound depending only on $L$ \cite{Bal84b}.
To evaluate such an integral  we 
 parametrize the  subspace   $\cQ Z =0, \tau Z =0$ as in \cite{Dim15b}.   We take    
 $Z  =  C \tilde Z  $ where  $\tilde Z$ is a pair   $ \tilde Z = ( \tilde Z_1, \tilde  Z_2)$.   The field  $\tilde Z_1$ is defined on bonds  within  each block $B(y)$ and satisfies 
   $ \tilde Z_1  \in \ker \tau $.  The field    $\tilde Z_2$   is    defined on bonds joining   $B(y), B(y')$ denoted $B(y,y')$,  but not
the central bond on each face denoted  $b(y,y')$.    The mapping   $Z  =  C \tilde Z  $  is the identity  on all bonds except the central bonds and assigns a
value to the central bonds so that   $ \cQ   Z =0,   \tau Z=0$. 
Then   the integral  can be written
\be   \label{springgarden4}
\de \sZ_{k, \bom^+}   ^{-1}    \int     f( C \tilde Z )  
     \exp \B(  - \frac12   \blan C  \tilde  Z , [ \De_{k, \bom}  ]_{\Om_{k+1}}  C \tilde  Z   \bran     \B) 
    \    D    \tilde   Z   \ee
If we define
\be 
C_{k, \bom^+}   =  \B(  C^T     [ \De_{k, \bom}  ]_{\Om_{k+1}}  C \B) ^{-1}
\ee
then this    integral can be expressed with  the Gaussian measure  $\mu_{C_{k, \bom^+ } }$  with covariance  $C_{k, \bom^+ } $   as      
\be    \label{sometimes2} 
      \int          f( C \tilde Z ) \    d \mu_{ C_{k, \bom^+ } }   ( \tilde  Z )    
\ee
This can also be written as
\be    \label{sometimes3} 
      \int          f(  Z ) \    d \mu_{ C'_{k, \bom^+ } }   (   Z )    
\ee
where now 
\be
C'_{k, \bom^+ }  =   C C_{k, \bom^+ } C^T  = C \B(  C^T     [ \De_{k, \bom}  ]_{\Om_{k+1}}  C \B) ^{-1}C^T
\ee
and the integral is now over $Z$ on $\Om^{(1)}_{k+1} $.
\bigskip

\subsection{general gauges and minimizers}
\label{generalgauges}

Until now we have to working in the axial gauge which is good for positivity properties but poor for ultraviolet properties.  Here we discuss the
axial gauge in more general terms and   introduce the Landau gauge which has better ultraviolet properties.

Gauge  transformations $\cA \to \cA + d \la$ form a group and generate orbits (gauge equivalence classes).
We are also  interested in imposing averaging  conditions $\cQ_{k, \bom} \cA = A_{k, \bom}$.  This is a condition on orbits if we make the restriction
$Q_{k, \bom} \la = 0$  since by (\ref{sweetlou})  $\cQ_{k, \bom} d\la =  dQ_{k, \bom} \la = 0$.   Gauge transformations with $Q_{k, \bom} \la = 0$
are called \textit{restricted} and the group of restriction gauge transformations define restricted orbits.  A choice of gauge is a choice of 
of a point on each restricted orbit.  Axial and Landau are two instances 

\subsubsection{axial gauge}

 The (symmetrized) axial gauge condition is that on $ \Om_j$  we have  $\tau (\cQ_{j-1} \cA) = 0$ for $j=1, \dots , k$.
On $\de  \Om_j$  we get conditions  $\tau (\cQ_{i-1} \cA) = 0$ for $i=1, \dots, j$.
This can be written as follows. For   $x_0,x_1, \dots, x_j$  with   $x_i \in \de \Om^{(i)}_j $ 
and   $x_{i-1} \subset B(x_i)$  the condition is
\be  \label{axial}
   \frac{1}{3!} \sum_{\pi}(\cQ_{i-1} \cA)\B( \Ga^{\pi} (x_i, x_{i-1}) \B)=0  \hs i=1, \dots, j 
\ee
where  the rectilinear path $\Ga(y,x)$  changes
coordinates from $y$ to $x$ in standard order, and $\Ga^{\pi}(y,x)$ permutes the order.    We could also consider the unsymmetrized axial gauge condition where
the average over $\pi$ is omitted and we just require $(\cQ_{i-1} \cA)\B( \Ga (x_i, x_{i-1})=0$.

Note that there is no  axial  condition on $\cQ_j\cA =A_j$ on $\de \Om_j$ which is fixed on a restricted orbit. 
\bigskip

  \begin{lem} \label{sumpter}  { \ }
  
  \begin{enumerate}
  \item For any $\cA$ there is a unique restricted $\la$ so that $\cA - d \la$ is axial.
  \item  Every restricted orbit has a unique  axial gauge representative
  \end{enumerate} 
\end{lem}
\bigskip

\pr  (1.) The statement is true either in either symmetrized or unsymmetrized case.  We give the proof in the latter case.
The claim is   that there is a $\la$ on $\tk$ such that $\cA' =   \cA - d \la$
satisfies the axial condition. 
Then $\la$ must satisfy 
 \be
(\cQ_{i-1} \cA)( \Ga(x_i, x_{i-1}) ) =(\cQ_{i-1} d \la )( \Ga(x_i, x_{i-1}) )
\ee
But
\be
(\cQ_{i-1} d \la )( \Ga(x_i, x_{i-1}) ) = (d Q_{i-1} \la )( \Ga(x_i, x_{i-1}) )= (Q_{i-1}\la)(x_{i-1}) - (Q_{i-1}\la)(x_{i})
\ee
Hence the condition becomes for $i=1, \dots, j$ on $\de \Om_j$ 
\be
(Q_{i-1}\la)(x_{i-1}) - (Q_{i-1}\la)(x_{i}) =  (\cQ_{i-1} \cA)( \Ga(x_i, x_{i-1}) ) \equiv \mu_i (x_{i-1}) 
\ee

In  the case $i = j$  this  says $(Q_{j-1}\la)(x_{j-1}) - (Q_{j-1}\la)(x_{j})  = \mu_j(x_{j-1}) $  on $\de \Om_j$.   We also have  the restricted gauge condition $Q_j \la = 0$ on $\de \Om_j$.  This gives
two equations for  $Q_{j-1}\la$ which are
\be \label{sundry}
\begin{split}
(Q_{j-1}\la)(x_{j-1}) - ( Q_{j-1}\la)(x_{j})  = &   \mu_j( x_{j-1})  \hs x_{j-1}  \neq x_j \\
(QQ_{j-1} \la )(x_j) = & 0 \\
\end{split}
\ee
This has the form $T( Q_{j-1} \la ) = \tilde \mu_j $  where    $\cQ_{j-1} \la$ and $\tilde \mu_j$  are functions on  $\de \Om_j^{(j-1)}$. 
The  operator $T$   is a bijection   since it preserves the dimension and  has kernel zero.  Thus  (\ref{sundry})
uniquely determines $Q_{j-1} \la$.  

Next for $i=j-1$ the equation is 
$(Q_{j-2}\la)(x_{j-2}) - (Q_{j-2}\la)(x_{j-1}) = \mu_{j-1}(x_{j-2})$  on $\de \Om_j$. 
 This gives
two equations for  $Q_{j-2}\la$ which are
\be
\begin{split}
(Q_{j-2}\la)(x_{j-2}) - (Q_{j-2}\la)(x_{j-1}) = & \mu_{j-1 }(x_{j-2}) \hs x_{j-2}  \neq x_{j-1} \\
(QQ_{j-2} \la )(x_{j-1}) = & (Q_{j-1} \la)(x_{j-1}) \\
\end{split}
\ee
and as before these uniquely determine $Q_{j-2} \la$  on $\de \Om^{(j-2)}_j$. 

 Continuing in this fashion we eventually find a 
unique $\la$ on $\de \Om_j$ satisfying all the conditions.

(2.)  There is at least one axial representative by (1.).   Suppose $ \cA$ and $\cA'$ are both axial and $\cA = \cA' - d \la$ with  $Q_{k, \bom} \la=0$.  As above we then have   on $\de \Om_j$
\be
0 = (\cQ_{i-1} d \la )( \Ga(x_i, x_{i-1}) ) = (Q_{i-1}\la)(x_{i-1}) - (Q_{i-1}\la)(x_{i}) 
\ee
Together with the condition $Q_j \la =0$ on $\de \Om_j$ this implies $\la =0$ as above.    Hence $\cA = \cA'$.

\subsubsection{Landau gauge}
Landau gauge in the continuum is defined by setting the divergence of $\cA$ to   zero. On the lattice this would be $\de \cA \equiv d^T\cA=0$.   We cannot accomplish this with a restricted
gauge transformation and instead 
ask that $\| \de   \cA \|^2$ be a minimum following \cite{Bal84b}.
Starting with any 
$\cA$ we look for  $\cA^{\la_0} = \cA - d \la_0$ with  the restricted gauge function
 $\la_0$   chosen to minimize $\|  \de \cA^{\la_0} \|^2 = \| \de \cA - \De \la_0 \|^2 $.
If $\la_0$ is such a minimum with  $Q_{k, \bom} \la_0 =0$ and  $Q_{k, \bom} \la_1 = 0$ then  $\| \de \cA - \De (\la_0 + t \la_1)\|^2 $
has a vanishing derivative at $t=0$ and so $<  \De \la_1, (\de \cA - \De \la_0 ) >=0$.   This says that $R_{k, \bom}  (\de \cA - \De \la_0 ) =0$
where $R_{k, \bom}$ is the projection onto $\De \ker Q_{k, \bom}$.  The equation is the same as  $\De \la_0 = R_{k, \bom} \de \cA $
which has a unique solution since $\De$ is invertible on $ \ker Q_{k, \bom}$.  At the minimum $R_{k, \bom}  (\de \cA^{\la_0} ) =0$.
Thus our Landau gauge condition is   (with $R_{k, \bom} =0$ on $\Om_1^c$)
\be
R_{k, \bom}  (\de \cA ) =0
\ee

 \begin{lem} \label{sumpter2} { \ }
 
 \begin{enumerate}
 \item For any $\cA$ there exists a restricted $\la$ such that $\cA - d \la$ is Landau.
 \item 
  Every restricted orbit has a unique  Landau gauge representative.
\end{enumerate}
\end{lem}
\bigskip

\pr \cite{Bal84b} (1.)
We  seek a restricted gauge transformation
taking any $\cA$  to Landau gauge.
So we  look for a  gauge function $\la$ such that  $Q_{k, \bom} \la = 0$  ($\la =0$ on $\Om_1^c) $ and such  that $\cA' = \cA -d \la$
satisfies $R_{k,\bom} \de \cA' =0$.
 Assuming 
  $Q_{k, \bom} \la= 0$
  the condition is 
  \be
  0 = R_{k,\bom}  \de \cA'  = R_{k,\bom} (\de \cA - \De \la )  =  R_{k,\bom} \de \cA - \De \la 
  \ee  
  Thus we look for solutions of $ \De \la =    R_{k, \bom}( \de \cA )$ which is the same as 
  \be
 (\De + Q^T_{k, \bom} \ba Q_{k, \bom} )  \la =    R_{k, \bom} \de \cA 
 \ee  
for any sequence $\ba = (a^{(k)}_1, \dots, a^{(k)}_k)$ with $a^{(k)}_j$  constant on  $\de \Om_j$.
If we take  $a^{(k)}_j = L^{2(k-j)}a_j$ and $a_{j+1}  = aa_j/(a_j + aL^{-2})$ with $a_1 =a > 0 $ fixed,  then this is a well-studied operator.
It is invertible on $\Om_1$ with Dirichlet boundary conditions  and we define  $G_{k,\bom} = [\De + Q^T_{k, \bom} \ba Q_{k, \bom} ]_{\Om_1} ^{-1}$.
  Then  a tentative solution to our problem is   
\be
\la  = G_{k, \bom}   R_{k, \bom} \de \cA 
\ee
This is an actual solution since it does satisfy the condition $Q_{k, \bom} \la = 0$ as we see by inserting the explicit representation for $R_{k, \bom}$
from equation (217) in  \cite{Bal84b}: 
\be    \label{ppp}
R_{k, \bom}= I - P_{k, \bom} =   I - G_{k,\bom}Q_{k, \bom}^T (  Q_{k, \bom}G^2_{k,\bom}Q_{k, \bom}^T )^{-1}Q_{k, \bom}    G_{k,\bom} 
\ee

(2.) There is at least one Landau representative by (1.).
Suppose $\cA, \cA'$ are both Landau and  $\cA =\cA' - d\la$ with $Q_{k, \bom}  \la = 0$.   Then $R_{k, \bom} (\de d \la ) =  R_{k, \bom} (\De \la )=0$.  
Since  $Q_{k, \bom}  \la = 0$ we have $\De \la $ in the range of $R_{k, \bom}$ and so $\De \la =0$.  This together with $Q_{k, \bom}  \la = 0$ implies $\la =0$ as
above.  Hence $\cA = \cA'$.  This completes the proof.
\bigskip

\subsubsection{minimizers}

  We are interested in finding the minimizer of $\|dA \|^2$ subject to the conditions  $\cQ_{k, \bom} \cA = A_{k, \bom}$.
Since both $\|dA \|^2$  and  $\cQ_{k, \bom} \cA $ are invariant under restricted gauge transformations this is a minimization problem on restricted orbits.
From the axial gauge results we know that there is a unique minimizer which in that gauge is given by  $\cA^{\sx}_{k, \bom} = \cH^{\sx}_{k, \bom}A_{k, \bom} $.
The equivalent Landau gauge minimizer is denoted $\cA_{k, \bom}$  which we now consider.

The definition is  
\be  \label{sloppy}
  \cA_{k, \bom} =      \textrm{    minimizer  of  }     \| d \cA \|^2   
  \textrm{  subject to    } \cQ_{k, \bom} \cA = A_{k, \bom}   \textrm{   and   }  R_{k, \bom}  (  \de  \cA  ) =0    
   \ee 
   Let $\ba$ be as before but now with $a_j =a>0$
The problem can be reformulated as finding the minimizer of
\be  \label{int8}
    \blan \cA, \B( \de d + d R_{k, \bom} \de + \cQ^T_{k, \bom} \ba \cQ_{k, \bom} \B) \cA \bran
   -  \| \ba^{\frac12}A_{k, \bom} \|^2
\ee   
subject to the same constraints. This is the same  since the extra terms are zero.   ( In $\Om_1^c$ this is taken to be  $<\cA, ( \de d  + L^{2k}a) \cA > - L^{2k} \|A_0\|^2$.)
This has the advantage that the quadratic form in $\cA$ is  positive definite  and we define a  Green's function by 
\be  \label{gk}
\cG_{k, \bom} =    \B( \de d + d R_{k, \bom} \de + \cQ^T_{k, \bom} \ba \cQ_{k, \bom} \B)^{-1}
\ee
Balaban  \cite{Bal84b} computes the constrained minimum using Lagrange multipliers.    Using also the identities 
\be
R_{k, \bom} \de \cG_{k, \bom} d R_{k, \bom} =R_{k, \bom} \hs    \cQ_{k, \bom}\cG_{k, \bom} d R_{k, \bom} =0
\ee
the  minimum $\cA_{k, \bom}  =   \cA_{k, \bom} (A_{k, \bom}) $     is found to be  
\be \label{fortune1}
  \cA_{k, \bom}   = \cH_{k, \bom} A_{k, \bom} \equiv \cG_{k, \bom} \cQ^T_{k,\bom}  \B( \cQ_{k,\bom}     \cG_{k, \bom} \cQ^T_{k,\bom} \B)^{-1}    A_{k, \bom} 
  \ee
  In $\Om^c_1$ nothing is happening:  $\cA_{k, \bom} = A_{k, \bom} = A_0$.    

The minimizer  $\cA_{k, \bom}$  has   certain locality property .    On the set  $\Om^c_1$ it   is fixed at $\cA_{k,\bom} = A_0$.
At bonds in $\Om_1$ the claim is that  the field $\cA_{k, \bom}$ only depends on the $A_{k, \bom}$ on a neighborhood of  $\Om_1$ and
not on $A_0$ off this neighborhood. 
 Indeed referring to the definition (\ref{sloppy})  note that  with
$\cA$ fixed on $\Om_1^c$ all the players in this drama, namely   $\|d \cA \|^2$   and the conditions $\cQ_{k, \bom} \cA = A_{k, \bom}$ and   $\cR_{k, \bom}( \de \cA ) =0$, 
have this property.   

\begin{lem} 
Let $\cA_1, \cA_2$ be minimizers of $\|d \cA\|^2$ with $\cQ_{k, \bom} \cA_1  = A_{k, \bom } = \cQ_{k, \bom} \cA_2$.  Then $\cA_1, \cA_2$ are on the same 
restricted orbit.
\end{lem}
\bigskip

\pr Choose a restricted $\la_1$ so that $\cA' _1 = \cA_1 - d \la_1 $ satisfies $\cR_{k,\bom} (\de \cA'_1 ) =0$.  Then $\cA'_1$ is still a minimizer and  still
satisfies  $\cQ_{k,\bom} \cA'_1 =0$.   Hence $\cA'_1 =\cA_{k, \bom}$.    Similarly  choose restricted $\la_2$ so that $\cA' _2 = \cA_2 - d \la_2 $ satisfies $\cR_{k,\bom} (\de \cA'_2 ) =0$.
Again $\cA'_2 =\cA_{k, \bom}$.  Hence  $\cA'_1 = \cA'_2$  and $\cA_1 = \cA_2 - d \la$ with $\la = \la_2- \la_1$.

\subsection{random walk expansions}

 \subsubsection{Green's functions}
 We  quote  some results  about estimates  and random walk expansions   from  Balaban \cite{Bal84a},\cite{Bal84b},\cite{Bal85b}.
First    consider  the     Landau  gauge   Green's functions on $\tk$  which can be written
\be   
  \cG_{k, \bom}  =   \B(  \De  -   d P_{k, \bom} \de  +  \cQ^T_{k,\bom}  \ba \cQ_{k,\bom}  \B)^{-1}  
\ee
One  begins by constructing a local  inverse  $ \cG_{k, \bom}( \sq)$.
Recall that in the Dirac case we took   $ L^{-(k-j)}M$  cubes $\sq \in \de \Om_j$ and  denoted  the set of all such $\pi_j(\de \Om_j)$.    In a cosmetic modification we now take 
$ L^{-(k-j)}2M$  cubes $\sq $ centered on the same points in  $\de \Om_j$ and the set of all such is denoted $\pi'_j(\de \Om_j)$.  This is a cover
rather than a partition.    The operator $ \cG_{k, \bom}(\sq)  $  is the inverse of the operator restricted to $ \tilde  \sq^3$ but now with periodic boundary conditions on $\tilde \sq^3$. We write it as 
\be   
  \cG_{k, \bom}(\sq)  =  \B(  \B[  \De   -   d P_{k, \bom}(\tilde \sq^3) \de  +  \cQ^T_{k,\bom} \ba \cQ_{k,\bom}  \B]^P_{\tilde \sq^3} \B)^{-1} 
\ee
  The choice of periodic boundary conditions means 
we are treating $\tilde \sq^3$ as a small torus and can use the global results of \cite{Bal84a} at least in the case where $\tilde \sq^3$ is entirely contained
in some $\de \Om_j$.   One also has to allow the possiblility that   $\tilde \sq^3$ to crosses the  boundary  between some  $\de \Om_j$ and $\de \Om_{j+1}$  and 
this requires some rather extensive modifications.   See  \cite{Bal84b} for the full treatment.  The result  however is that the same bounds hold in this case as well.

To generate the expansion one starts with a parametrix $\sum_{\sq} h_{\sq} \cG_{k, \bom}(\sq) h_{\sq}$ where the sum is
over $\sq $ in the cover $\pi'(\bom) = \cup_i \pi'(\de \Om_i)$ and $h_{\sq}^2$ is a partition of unity with $\supp h_{\sq} \subset \sq$.
The expansion  involves the commutator   $K_{\sq}  = [ \De -   d P_{k, \bom} \de  +a \cQ^T_{k,\bom}   \cQ_{k,\bom}, h_{\sq} ]$ and unlike the Dirac case it  is not strictly localized  in $\sq$ due to the projection  operator $P_{k, \bom}$ on scalars.   However a preliminary random walk expansion for $P_{k, \bom}$ using (\ref{ppp})  shows that it is exponentially decaying in the distance from $\sq$.    Then one can define a localized commutator $K_{\sq', \sq}$ 
which is also  exponentially decaying.  
The expansion  then has the form  
 \be 
  \cG_{k, \bom}   =    \sum_{\om }   \cG_{k, \bom, \om} 
  \ee
 where
  $   \om   =  (\sq_0, \sq_1, \sq_2\dots,   \sq_{2n-1} \sq_{2n} )$ is a random walk 
 consisting of cubes in  $\pi'(\bom) $  and 
 \be
 \begin{split}
  \cG_{k, \bom, \om}   =   & \B( h_{\sq_0} \cG_{k, \bom}( \sq_{0}) h_{\sq_0}\B) \B(K_{\sq_1,\sq_2}  \cG_{k, \bom}( \sq_2) h_{\sq_2} \B)
 \cdots
\B( K_{\sq_{2n-1}, \sq_{2n}}  \cG_{k,\bom}( \sq_{2n}) h_{\sq_{2n}}\B)  \\
 \end{split}
 \ee
 Here  $\sq_i, \sq_{i+1}$ must overlap for $i$ even, but  $\sq_i, \sq_{i+1}$ can be a long jump for $i$ odd.

  Let     $  \De_y$  be  a  $L^{-(k-j) }$ cube in   $\de \Om_j$  and     $\De_{y'}$  be  a  $L^{-(k-j') }$ cube in   $\de \Om_{j'}$, and let  $\zeta_y$   be a smooth partition on unity with $\supp\  \zeta_y \subset \tilde  \De_{y} $. 
    One shows that \cite{Bal84b}
 \be  \label{day0}
 |1_{\De_y} K_{\sq', \sq} \cG_{k, \bom}(\sq)   1_{\De_{y'}} f| 
   \leq     CM^{-1} e^{  -  \ga  d_{\bom} (  y,y') }  \|f \|_{\infty}
\ee
 Combined with the basic bound on  $ \cG_{k, \bom}( \sq)$
 this yields
      \be   
 \begin{split}
    \label{day}
  |1_{\De_y}  \cG_{k, \bom, \om}   1_{\De_{y'}} f| 
   \leq   
   &  C  L^{-2(k-j')} (CM^{-1} )^n e^{  -  \ga  d_{\bom} (  y,y') }  \|f \|_{\infty}\\
   |1_{\De_y} \pa  \cG_{k, \bom, \om}   1_{\De_{y'}} f|
     \leq 
     &    C  L^{-(k-j')} (CM^{-1} )^n e^{  -  \ga  d_{\bom} (  y,y') }  \|f \|_{\infty} \\
    \| \zeta_y \pa   \cG_{k,\bom, \om }   1_{\De_{y'} }   f \|_{(\al)}  
  \leq 
   & C L^{-(1+\al)(k-j') } (CM^{-1} )^n e^{-\ga  d_{\bom}  (y,y')   }  \| f \|_{\infty} 
    \\
 \end{split}
 \ee
 Here $\| \cdot \|_{(\al) }$ is a norm on the H\"{o}lder derivative of order $\al <1$.  For a function $f$ defined on bonds, and with $f_{\mu}(x) = f(x, x+ L^{-k} e_{\mu})$, etc.,
 the norm is 
 \be
 \| f \|_{(\al) }  = \sup_{\mu, 0 < |x-y| \leq 1} \frac{ |f_{\mu}(x) - f_{\mu}(y)|}{ |x-y|^{\al} }
 \ee
 The bounds enable  convergence for $M$ sufficiently large  and since   $L^{|j-j'|} \leq e^{\cO(M^{-1}) d_{\bom}(y,y')}$ we can write this as 
    \be      \label{standard2}  
  \begin{split}  
&  |1_{\De_y} \cG_{k,\bom} 1_{\De_{y'}}    f |, \   L^{-(k-j) }  |1_{\De_y} \pa  \cG_{k,\bom} 1_{\De_y}    f |, \ 
 L^{-(1 + \al) (k-j) }  \|  \zeta_y \pa   \cG_{k,\bom}   1_{\De_{y'} }   f \|_{(\al)}  \\
& \hs  \leq   CL^{-2(k-j') } e^{-\ga  d_{\bom}(y,{y'})   }  \| f \|_{\infty}    \\
\end{split}
  \ee
 This can be further modified  using   $L^{\beta|j-j'|} \leq  Ce^{\cO(M^{-1}) d_{\bom}(y,y')}$  for any   $\beta =\one $.
 We also allow weight factors satisfying 
 \be \label{weight} 
 \frac{p_{j'}}{p_j} \leq L^{|j-j'|}
 \ee
  to obtain 
   \be      \label{standard2.5}  
  \begin{split}  
& L^{-\beta(k-j)} |1_{\De_y} \cG_{k,\bom} 1_{\De_{y'}}    f |, \   L^{-(1+ \beta)(k-j) }  |1_{\De_y} \pa  \cG_{k,\bom} 1_{\De_y}    f |, \\
& 
 L^{-(1 + \al + \beta) (k-j) }   \|  \zeta_y \pa   \cG_{k,\bom}   1_{\De_{y'} }   f \|_{(\al)}  
 \ \   \leq   CL^{-(2+ \beta)(k-j') }p_j p_{j'}^{-1}e^{-\ga  d_{\bom}(y,{y'})   }  \| f \|_{\infty}    \\
\end{split}
  \ee
Summing over $y'$ and using  (\ref{multisum})  yields the bounds on $\de \Om_j$   (with $\|  \cdot \|_{(\al)} $ on $\de \Om_j$)
   \be      \label{standard2.6}  
  \begin{split}  
& L^{-\beta(k-j)}   | \cG_{k,\bom}    f |, \   L^{-(1+ \beta)(k-j) }  | \pa  \cG_{k,\bom}    f |, \ 
 L^{-(1 + \al + \beta) (k-j) }  \|  \pa   \cG_{k,\bom}     f \|_{(\al) }  \\
& \hs  \leq   C p_j  \sup_{j'}  L^{-(2+ \beta)(k-j') }  p_{j'}^{-1} \sup_{\de \Om_{j'}} |f|   \\
\end{split}
  \ee

The bounds on $\pa \cG_{k, \bom}$ hold as well for $\cG_{k, \bom} \pa^T$. Specializing to $\beta = -\frac12   $ (the only case we need) 
 we have on $\de \Om_j$
  \be      \label{standard2.7}  
     L^{- \frac12 (k-j) }    |  \cG_{k,\bom} \pa^T   f |, \ 
 L^{-(\frac12  + \al ) (k-j) }   \|     \cG_{k,\bom}  \pa^T   f \|_{(\al)}  
   \leq   C p_j   \sup_{j'}  L^{-\frac32 (k-j') } p_{j'}^{-1}  \sup_{\de \Om_{j'}} |f|  
  \ee
We also have the bounds on $\de \Om_j$ for $\al + \ep <1$ \cite{Bal84b}
 \be  \label{standard2.8}
 \begin{split}
L^{-\frac32(k-j)}   | \pa \cG_{k, \bom} \pa^T  f | & 
   \leq C p_j \sup_{j'} L^{-\frac32(k-j')}p_{j'}^{-1}  \B(L^{-\ep(k-j')}  \|  f\|_{(\ep), \de \Om_{j'}}  + \| f\|_{\infty, \de \Om _{j'}} \B) \\
L^{-(\frac32 + \al )(k-j)}    \|   \pa \cG_{k, \bom} \pa^T  f \|_{(\al)} &
    \leq C p_j\sup_{j'}L^{-\frac32(k-j')} p_{j'}^{-1} \B( L^{-(\al +\ep)(k-j')}  \| f\|_{(\al + \ep), \de \Om_{j'}}  + \| f\|_{\infty, \de \Om_{j'}} \B) \\
 \end{split}
 \ee    
     
The choice of weight factors needs further comment.  The choice $p_j=1$ is possible but  in the sequel it will be convenient to 
take for some positive integer $p$
\be
p_j =  (- \log e_j )^p = \B (\frac12  (N-j) \log L - \log e \B )^p
\ee
To see that this satisfies (\ref{weight}) note that $|\log e_j - \log e_{j'} | \leq \frac12 |j-j'| \log L$, hence for $e$ sufficiently small
   $| \log e_{j'}/\log e_j |  \leq  1 + \frac12 |j-j'| \log L / |\log e_j | \leq e^{|j-j'|/p }$. Raising this to the $p^{th}$ power gives
   $p_{j'}/p_j \leq  e^{|j-j'| } $ which suffices. 
     
   \subsubsection{minimizers}

Next consider random walk expansions for the Landau gauge  minimizer which can be written  
\be
\cH_{k, \bom} =   \cG_{k, \bom}  \cQ_{\bom}^T  \cN_{k, \bom}    \hs
    \cN_{k, \bom}  \equiv ( \cQ_{k,\bom} \cG_{k, \bom}\cQ^T_{k, \bom} )^{-1}
\ee
Here  $ \cQ_{k,\bom} \cG_{k, \bom}\cQ^T_{k, \bom} $ and   $ \cN_{k, \bom} $ acts on   multiscale objects like $A_{k, \bom}$.    
The operator  $ \cQ_{k,\bom} \cG_{k, \bom}\cQ^T_{k, \bom} $ has a kernel 
defined   for $b,b' \in \cup_j \de \Om_j^{(j)}$ by 
\be   
( \cQ_{k,\bom} \cG_{k, \bom}\cQ^T_{k, \bom} A_{k, \bom} ) (b) 
= \sum_{j'=1}^k  \sum_{b' \in \de \Om_{j'}^{(j')}} \B( \cQ_{k,\bom} \cG_{k, \bom}\cQ^T_{k, \bom}\B)(b,b') L^{-3(k-j')} A_{j', \de \Om_j}(b')
\ee
Suppose that
$b \in \de \Om_j^{(j)},  b \in \de \Om_{j'}^{(j')}$.    Let $ \de_b'$ be the delta functions on bonds  in $    \bbT^{-(k-j')}_{N-k}$.  Then 
\be
\begin{split}
  \B( \cQ_{k,\bom} \cG_{k, \bom}\cQ^T_{k, \bom}\B)(b,b') 
 &=  \B(\cQ_{k,\bom} \cG_{k, \bom}\cQ^T_{k, \bom} \de_{ b'}\B)(b) \\
 & =  \sum_{y,y'} \B( \cQ_{k,\bom}1_{\De_y} \cG_{k, \bom}1_{\De_{y'}}\cQ^T_{k, \bom} \de_{ b'} \B)(b) \\
  & =  \sum_{y,y'}  \B( \cQ_{j}1_{\De_y} \cG_{k, \bom}1_{\De_{y'}}\cQ^T_{j'} \de_{ b'} \B)(b) \\
\end{split} 
\ee
For fixed $b,b'$ the number of terms in the sum over $y,y'$ is $\one$  and they are
all near $b,b'$.  So  using the bound (\ref{standard2}) we have
\be
 | \B( \cQ_{k,\bom} \cG_{k, \bom}\cQ^T_{k, \bom}\B)(b,b') |
\leq  C   \|\cQ^T_{j'} \de_{b'} \|_{\infty}  e^{ - \ga  d_{\bom}  (b,b') } 
\ee
But from the explicit formula  (\ref{lunge})  for $\cQ^T$ we have  $ \|\cQ^Tf \|_{\infty} \leq \| f\|_{\infty}$, hence
$ \|\cQ^T_jf \|_{\infty} \leq \| f\|_{\infty}$, and hence $ \|\cQ^T_{j'} \de_{b'} \|_{\infty} \leq  \|\de_{b'} \|_{\infty}  =\cO(L^{3(k-j')})$.
  Thus 
for $b \in \de \Om_j^{(j)},  b \in \de \Om_{j'}^{(j')}$
\be \label{coke}
|  \B( \cQ_{k,\bom} \cG_{k, \bom}\cQ^T_{k, \bom}\B)(b,b') | \leq C  L^{-2(k-j)} L^{3(k-j') } e^{ - \ga  d_{\bom}  (b,b') } 
\ee

Next one considers local inverses  of the form 
\be 
\cN_{k, \bom} (\sq)=    \B[ \cQ_{k,\bom} \cG_{k, \bom}(\sq)\cQ^T_{k, \bom} \B]_{\sq}^{-1}  \hs \sq \in \pi'(\bom)
\ee
This satisfies a bound of the same form as (\ref{coke}) but with $ L^{2(k-j)}$ rather than $L^{-2(k-j)}$; see Balaban \cite{Bal84b} for details.
The local inverses can be used to generate an expansion
 for $\cN_{k, \bom}$ of the form 
 \be
   \cN_{k, \bom} = \sum_{\om}     \cN_{k, \bom, \om}
 \ee   
 where for $\om= (\om_0, \dots, \om_{2n} )$
  \be
   \cN_{k, \bom, \om} =
 \B( h_{\sq_0} \cN_{k, \bom}( \sq_{0}) h_{\sq_0}\B) \B(K_{\sq_1,\sq_2}  \cN_{k, \bom}( \sq_2) h_{\sq_2} \B)
 \cdots
\B( K_{\sq_{2n-1}, \sq_{2n}}  \cN_{k, \bom}( \sq_{2n}) h_{\sq_{2n}}\B)  
  \ee
Here $K_{\sq, \sq'}$ is a local version of the commutator $K_{\sq}=[ \cQ_{k,\bom} \cG_{k, \bom}\cQ^T_{k, \bom}, h_{\sq} ]$.
 One establishes that   for $b \in \de \Om_j^{(j)},  b' \in \de \Om_{j'}^{(j')}$
 \be  
 | \cN_{k, \bom, \om}(b,b')|  \leq     C (CM^{-1} )^n   L^{2(k-j)} L^{3(k-j') }      e^{ - \ga  d_{\bom}  (b,b') } 
 \ee
It follows  that the expansion converges and satisfies 
\be \label{listless}
|\cN_{k, \bom}(b,b')| \leq  C   L^{2(k-j)} L^{3(k-j') }      e^{ - \ga  d_{\bom}  (b,b') } 
\ee

Combining  the expansion for $\cG_{k, \bom}$ and the expansion for $\cN_{k, \bom}$ and using  the fact that    $\cQ^T _{\bom}$ is local   we generate a random walk  expansion for
$\cH_{k, \bom} $.  This leads to estimates 
  for  $b, \tilde b  \in \de \Om_j ,  b' \in \de \Om_{j'}^{(j')}$   
 \be      \label{standard3}  
  \begin{split}  
  | \cH_{k,\bom} (b,b') |, \   L^{- (k-j) }     |\pa  \cH_{k, \bom }(b,b') |, \  L^{-(1 + \al) (k-j) }    | \de_{\al}  \pa    \cH _{k, \bom}( (b,\tilde b); b')  |
  &  \leq    CL^{3(k-j')} e^{-\ga  d_{\bom}  (b,b')   }  
  \\
\end{split}
  \ee 
   Note the cancellation of the factors  $L^{-2(k-j')}$ in (\ref{standard2})     and  $L^{2(k-j)}$ (here $L^{2(k-j')}$)  in (\ref{listless}). 
   Again using $L^{ \frac12 |j-j'|} \leq  Ce^{\cO(M^{-1}) d_{\bom}(y,y)} $ and weight factors $p_j$ satisfying (\ref{weight}) we have
   \be      \label{standard3.5}  
  \begin{split}  
& L^{- \frac12 (k-j)}   | \cH_{k,\bom} (b,b') |, \   L^{- \frac32  (k-j) }      |\pa  \cH_{k, \bom }(b,b') |, \  L^{-(\frac32 + \al ) (k-j) }    | \de_{\al}  \pa    \cH _{k, \bom}( (b,\tilde b); b')  |\\
  &  \hs \leq    C  L^{-\frac12 (k-j')}  L^{3(k-j')} p_j p_{j'}^{-1} e^{-\ga  d_{\bom}  (b,b')   }  
  \\
\end{split}
  \ee 
   Summing over $b'$ and  using  the estimate  (\ref{multisum})  we obtain the bounds on $\de \Om_j$   
 \be      \label{standard4}  
 \begin{split}
  & L^{-\frac12(k-j)}   | \cH_{k,\bom}    A_{k, \bom} |, \ 
  L^{ - \frac32 (k-j) }  | \pa  \cH _{k, \bom}   A_{k, \bom} |, \ 
 L^{-(\frac32+ \al )(k-j) }    \|  \pa    \cH _{k, \bom}    A_{k, \bom} \|_{(\al)}  \\
& \hs  \leq     Cp_j  \sup_{j'} L^{- \frac12  (k-j') }p_{j'}^{-1} \sup_{\de \Om_{j'}} | A_{j'} | \\
  \end{split}
  \ee

\subsubsection{fluctuation covariance}

The fluctuation covariance  $C_{k, \bom^+}  $ acts on a special space as explained in section \ref{nextstep}.   But closely related is the 
operator on     $\Om^{(k)}_{k+1} \subset   \tz$   defined as   
\be
C'_{k, \bom^+} =   C C_{k, \bom^+}  C^T
   = C  \B(   C^T [  \De_{k, \bom} ]_{\bom_{k+1} } C \B)^{-1}  C^T
\ee
This has the representation   \cite{Bal85b}, \cite{Dim15}
\be   
C'_{k, \bom^+}   = ( 1 + \pa \cM )\B[ \cQ_{k}  \tilde \cG_{k+1, \bom^+}  \cQ_{k}^T  \B]_{\Om_{k+1}} ( 1 + \pa \cM ) ^T
\ee
Here $\cM$ is a local operator defined by specifying that   $\la  = \cM A$  is the unique solution of  $\tau ( A + d \la ) =0$ and  $Q \la  = 0$.
The operator $\tilde \cG_{k+1, \bom^+}  $ is another Green's function defined on  $ \Om_0 \subset \tk$  by 
\be    \label{lorenzo}
\tilde \cG_{k+1, \bom^+ }  =   \cG^0_{k+1} 
-  \cG^0_{k+1, \bom^+} \cQ_{k+1, \bom^+}^T   \B(\cQ_{k+1,\bom^+} \cG^0_{k+1, \bom^+} \cQ_{k+1,\bom^+}^T\B)^{-1}   \cQ_{k+1,\bom^+} \cG^0_{k+1, \bom^+} 
\ee
where    for any  $a >0$
\be   
\cG^0_{k+1, \bom^+}  = \B(   \de  d   +    d R^0_{k+1, \bom^+} \de  \  +  \cQ^T_{k+1, \bom^+} \ba \cQ_{k+1, \bom^+}   \B)^{-1}
\ee
Here $R^0_{k+1, \bom^+}$ is the projection onto $\De \ker (Q_{k+1, \bom^+})$.
Both     $ \cG^0_{k+1, \bom^+}$  and    $ (\cQ_{k+1,\bom^+} \cG^0_{k+1, \bom^+} \cQ_{k+1,\bom^+}^T)^{-1}  $ have convergent   random walk expansions
as was the case for     $ \cG_{k, \bom}$  and    $ (\cQ_{k,\bom} \cG_{k, \bom} \cQ_{k,\bom}^T)^{-1}  $. 
Hence  the same is true  for   $\tilde \cG_{k+1, \bom^+ } $  and    $C'_{k, \bom^+}  $.  With some care as in \cite{Dim15b} one can arrange that
the leading term  in  the expansion for   $C'_{k, \bom^+} $  (with no jumps)  is positive definite.   These expansions  
  give the bound  on the kernel
\be \label{standard6}
  |  C'_{k, \bom^+} (b,b')|  \leq  C e^{- \ga  d(b,b') } 
\ee
There is a similar representation of    $ (C'_{k, \bom})  ^{\frac12}$  in terms of Green's functions \cite{Bal88a}, \cite{Dim15}, and it leads
to a random walk expansion and   bounds  
\be 
  | ( C'_{k, \bom^+} )^{\frac12}(b, b')|  \leq  C e^{- \ga  d(b,b') } 
\ee

 \subsubsection{weakened operators}
\label{newly}

All these random walk expansions allow long jumps.  In particular in the expansion for $\cG_{k,\bom}$ the term  $  [  d P_{k, \bom} \de  , h_{\sq} ]$ in
$ K_{\sq', \sq}  $ is not short range. For weakening the expansion  it is useful to have all walks  consisting of short steps.
This is accomplished by inserting the  random walk expansion for $P_{k, \bom}$ based on the representation  (\ref{ppp}).   For details  see \cite{Bal85b}.
The result is a expansion of the form  
\be
  \cG_{k, \bom}  = \sum \cG_{k,\bom,  \om} 
\ee 
The sum is over indexed walks 
\be \label{crackerjack}
\om = \B( \sq_0, (\al_1, X_1), \dots, (\al_n, X_n) \B)  
\ee
where $X$ is a localization domain consisting of  a small  connected union  of cubes  from $\pi'(\bom)$ and the index $\al$ labels a finite number of possibilities for
each $X$.  The combination $(\al,X)$ is called a \textit{unit}. The contribution of a walk $\om$ is
\be
\cG_{k, \om}   =  R_0(\sq_0) R_{\al_1}(X_1) \cdots  R_{\al_n}(X_n)
\ee
 Here $R_0(\sq_0) = h_{\sq_0} \cG_{k, \bom}(\sq_0) h_{\sq_0}$ as  before
while  $R_{\al} (X)$ is localized in $X$. 
Not all choices actually occur, but those that do have $X_i \cap X_{i+1} \neq \emptyset$ so we only have short jumps.  
The $R_{\al}(X)$   satisfy the bound (\ref{day0}) and so the expansion generates the same bound (\ref{day}) (\ref{standard2}) as before.

As in the Dirac case in  section \ref{study it}  we introduce weakening parameters $s =\{s_{\sq} \}$ for $\sq \in \pi'(\bom)$ and define 
for $|\om| \geq 1$
\be
s_{\om} = \prod_{\sq \subset X_{\om}} s_{\sq}  \hs X_{\om} = \bigcup_{i=0}^n   X_i 
\ee
If  $|\om| =0$ so $\om$ is a single cube then $s_{\om} = s_{\sq} =1$.
The  weakened operator is 
 \be
 \cG_{k, \bom}(s)  = \sum s_{\om}\cG_{k,\bom,  \om} 
\ee
Again for $\al_0 <1$ and   complex $|s| \leq M^{\al_0}$  the operator   $ \cG_{k, \bom}(s) $   satisfies the same bounds as the original.

Similarly $\cN_{k, \bom}, \cH_{k, \bom}, C'_{k,\bom^+}$
have random walk expansions  with short jumps,  and weakened versions  $\cN_{k, \bom}(s), \cH_{k, \bom}(s), C'_{k,\bom^+}(s)$.
See \cite{Bal85b} for more details.

\subsection{more estimates}
 
 One of the main issues in our problem is using bounds on the field strength $d \cA$ to generate bounds on some  gauge potentials $\cA$.
 Locally one can accomplish this  with a gauge transformation.    If we add some hypotheses about averages of $\cA$ then we can 
 get a more global result.  The bound in \cite{Bal84b}  is an $L^2$ version of this but we need pointwise bounds.  In this section we prove  some bounds of
 this form  for  the axial gauge, specifically the unsymmetrized axial gauge.   These are abelian versions of non-abelian results proved  by Balaban 
 \cite{Bal85aa}.

 \begin{lem}  \label{orange}  
  Let $A$ be an axial gauge function on a unit lattice. Then  for some constant $c_0 = \cO(1)$ 
   \be
|A| \leq   c_0 L^2\| dA \|_{\infty} + L   \|  \cQ A \|_{\infty} 
\ee
\end{lem} 
\bigskip

\pr   
If $(x, x+ e_{\mu})$ is contained  in some $B(y)$, then by the axial gauge condition $\cA( \Ga(y,x)) =0$
and Stoke's theorem
\be \label{gunison1}
\begin{split}
A(x, x+ e_{\mu}) 
=&  A\B( \Ga(y, x, x+ e_{\mu}, y )\B)
=  dA \B( \Si ( y, x, x+ e_{\mu}, y)\B ) 
\end{split}
\ee
 Here $\Ga(y, x, x+ e_{\mu}, y ) \equiv  \Ga(y, x) )+[x, x+ e_{\mu}]  +  \Ga( x+e_{\mu}, y )$ is a closed 
 path and   $ \Si ( y, x, x+ e_{\mu}, y)$ is the surface bounding the path.
This has  at most  $ \one L$  plaquettes  and so 
\be \label{gunison2}
|A(x, x+ e_{\mu}) | \leq  \one L \|dA\|_{\infty} 
\ee
Now suppose $(x, x+ e_{\mu})$ is a surface bond in $B^s(y, y + L e_{\mu})$.
Then by the axial gauge condition
 \be \label{28palms}
\begin{split}
A(x, x+ e_{\mu}) 
=&  A\B( \Ga(y, x, x+e_{\mu}, y+ L e_{\mu} ) \B)\\
= & dA \B( \Si ( y, x, x+ e_{\mu}, y+Le_{\mu},y )\B )  + A\B(\Ga(y, y+ Le_{\mu} ) \B)
\end{split}
\ee
The surface in the first term has at most  $\one L^2$ plaquettes so it is bounded by $\one L^2 \|dA\|_{\infty}$.   For the second term it suffices
to bound  the quantity $A(\Ga(y, y+ Le_{\mu} ) ) -L ( \cQ A )(y, y+ Le_{\mu} ) $ since $L ( \cQ A )(y, y+ Le_{\mu} )| \leq L  \|\cQ A\|_{\infty}$
We have
\be  \label{29palms}
\begin{split}
A(\Ga(y, y+ Le_{\mu} ) ) -L (\cQ A )(y, y+ Le_{\mu} ) 
 = &  \sum_{x \in B(y)} L^{-3}\B( A(\Ga(y, y+ Le_{\mu} )) -A(\Ga(x, x+ Le_{\mu} ) ) \B)\\
= &  \sum_{x \in B(y)} L^{-3} A\B(\Ga(y,y+ Le_{\mu}, x+ Le_{\mu}, x,y )  \B)  \\
= &  \sum_{x \in B(y)} L^{-3} dA\B(\Si(y,y+ Le_{\mu}, x+ Le_{\mu}, x,y ) \B)   \\
\end{split}
\ee
This is bounded by $ \one L^2  \| dA \|_{\infty}$   and  hence the result. 
\bigskip
 
\rem
The  bound for this result is  local.   For the  bound on $|A(x,x+e_{\mu})|$ with $x \in B(y)$ and $x+ e_{\mu} \in B(y')$ it suffices to 
take the supremum on the right over $B(y) \cup B(y')$.    Similar remarks apply to the next result. 
\bigskip

 \begin{lem}  \label{orange2}  
   Let $\cA$ be an axial gauge  function on a  $L^{-k}$ lattice, $k \geq 1$.
  Then for $j=0,1, \dots k$
\be
|\cQ_j \cA| \leq  L^{k-j}\B(  2c_0   \| d \cA \|_{\infty} + \| \cQ_k \cA \|_{\infty} \B) 
\ee
\end{lem} 
\bigskip

\pr The unsymmetrized axial gauge condition is that for  $x_0, x_1, x_2, \dots, x_k$ with $x_i$  in the $L^{-(k-i)}$ lattice and   
and $x_{i-1} \subset B(x_i)$  we have
\be
 (\cQ_{i-1}\cA)(\Ga(x_i,x_{i-1} ) )=0  \hs i=1, \dots, k 
 \ee

 First consider  $\cQ_{k-1}\cA$ on the $L^{-1}$ lattice.  Then $(\cQ_{k-1}\cA)_L$ (scaling without scaling factors)  is on a unit lattice,
 satisfies the axial gauge condition  of the previous lemma, and satisfies $\cQ (\cQ_{k-1}\cA_L) =  (\cQ_{k}\cA)_L$.
Hence by the previous lemma  
\be
|(\cQ_{k-1}\cA)_L | \leq c_0 L^2 \| d (\cQ_{k-1}\cA)_L \|_{\infty}  + L  \|(\cQ_k \cA)_L \|_{\infty}
\ee
The derivative gives a factor $L^{-1}$ and $ d \cQ_{k-1}\cA = \cQ^{(2)}_{k-1} d\cA$ is bounded by $ \|d \cA \|_{\infty}$.  Thus 
\be \label{ten}
|\cQ_{k-1}\cA | \leq  L \B(c_0 \| d \cA \|_{\infty}  +    \| \cQ_k \cA \|_{\infty}\B)
\ee

Next consider  $\cQ_{k-2}\cA$ on an  $L^{-2} $ lattice.  Then $(\cQ_{k-2}\cA)_{L^2}$ is on a unit lattice, is axial gauge,  and $\cQ(\cQ_{k-2}\cA)_{L^2}=(\cQ_{k-1}\cA)_{L^2}$. Hence by the previous lemma:
\be
|(\cQ_{k-2}\cA)_{L^2} | \leq c_0 L^2 \| d (\cQ_{k-2}\cA)_{L^2} \|_{\infty}  + L \|(\cQ_{k-1} \cA)_{L^2} \|_{\infty}
\ee
The derivative gives a factor $L^{-2}$ and  $ d \cQ_{k-2}\cA = \cQ^{(2)}_{k-2} d\cA$.  Using also (\ref{ten})
\be
\begin{split}
|\cQ_{k-2}\cA | \leq  &  c_0  \| d \cA \|_{\infty}  + L \| \cQ_{k-1} \cA  \|_{\infty} \\
 \leq &  L^2 \B( c_0  ( 1+ L^{-2})  \| d \cA \|_{\infty}  +  \| \cQ_k \cA  \|_{\infty} \B)\\
\end{split}
\ee

Continuing in this fashion we  get after $n$ steps
\be
| \cQ_{k-n} \cA| \leq   L^n \B( c_0 (1+ L^{-2} + \dots + L^{-2(n-1)}) \|d \cA \|_{\infty} +      \| \cQ_k \cA  \|_{\infty} \B)
  \leq L^n \B(  2c_0   \|d \cA \|_{\infty} + \|  \cQ_k \cA \|_{\infty} \B)
\ee
With $j=k-n$  this is the stated result for $j<k$.  The case  $j=k$ is trivial.
\bigskip

\subsection{regularity}

The bounds  (\ref{standard4}) on $\cH_{k, \bom}$   give estimates on the Landau gauge minimizer  $\cA_{k, \bom} =  \cH_{k, \bom}  A_{k, \bom} $
and its derivatives   depending on   $  \|A_{k, \bom} \|_{\infty}$.   Locally we can do better and get bounds depending
only  on    $  \| dA_{k, \bom} \|_{\infty}$.   This is important since we  will have better control over    $  \| dA_{k, \bom} \|_{\infty}$.  Bounds with this improvement are what we are calling 
\textit{regularity} 

In \cite{Dim15b} the regularity bounds for the global version were proved  by a technique which involved the use of axial gauge minimizers with improved smoothness.
This does not work so well in the multiscale setting due to the fact that identities like $\cQ_k \cQ^{s,T}_k = I$ and $d \cQ^{s,T}_k = \cQ^{e,T}_k d$
no longer hold for the multiscale averaging operators $\cQ_{k, \bom}, \cQ^{s,T}_{k, \bom}$. (Something which was overlooked in an earlier version of this paper.)  
So here we give a different proof which also works for the global case.

 We  more or less follow an intricate  strategy developed by Balaban in 
\cite{Bal85aa},  \cite{Bal85d}.   The following are  abelian versions  of non-abelian results in these references.

\begin{thm} \label{REG}  (regularity) Consider a  restricted  orbit  $\cQ_{k, \bom} \cA = A_{k, \bom}$ minimizing $\| d \cA\|^2$
with  $A_{j, \de \Om_j}$ axial for $j \leq k-1$.    Let $\sq$ be an $L^{-(k-j)} M$ cube in $\de \Om_j$, $j=1, \dots, k$ and
consider  weight factors $p_j$ satisfying (\ref{weight}). 
Then there is a constant $C_0$ depending only on $L$ such that  for $M$ sufficiently large  any 
 $\cA$ in  the   orbit   is gauge equivalent on the enlargement  $\tilde \sq^{\da}  $  to a field  $\cA'$ satisfying
 (with  $\| \cdot \|_{(\al)}$ on $\tilde \sq^{ \da}$)
\be   \label{tingting}
\begin{split}
& L^{-\frac12(k-j)}|\cA'| ,\  L^{-\frac32(k-j)}|\pa \cA'|, \   L^{-(\frac32 + \al)(k-j)} \|  \pa \cA'\|_{(\al)}  \\
&  \leq   C_0 M p_j \sup_{1 \leq  j'\leq k} L^{-\frac32(k-j')}p_{j'}^{-1}  \sup_{p \in \de \Om_{j'}^{(j')}} |dA_{j'}(p)|   \\
  \end{split}
  \ee 
In particular this is true for the axial  gauge  $\cA^{\sx}_{k, \bom}$ and the Landau gauge  $\cA_{k ,\bom}$.
\end{thm}

\rems
\begin{enumerate}
\item As in \cite{Dim15b} if $\sq$ is  an  $L^{-(k-j)} M$  cube then $\sq^{\da} \equiv \tilde  \sq^{c_0L}$ is the enlargment by
$c_0L$ layers of  $L^{-(k-j)} M$ cubes where $c_0 = \one$. 
\item
  In the sequel   we will have enough control over $dA_j$ to use this result to conclude that 
 $\cA_{k, \bom}$ is in the space $\tilde \fG_{k, \bom}$  needed as background for Dirac propagators.  
\item  In the course of the proof the following definition will be useful.  Let
  $\fA_{k, \bom}(C)$  be the space of  all gauge fields $ \cA$  on $\tk$ satisfying on $\de \Om_j$ 
\be 
\label{oscar1}
|d\cA| \leq  CL^{\frac32(k-j)} p_j  \hs \|  d \cA\|_{(\al)} \leq  C L^{(\frac32+ \al )(k-j)} p_j 
\ee
This is a condition   on orbits.   The bounds get tighter as $j$ increases, so  the bounds for $\{ \de \Om_j \}$ 
imply the same bounds for $\{  \Om_j \}$. 
\end{enumerate}
\bigskip

\pr  It suffices to prove the result for the Landau minimizer $\cA_{k, \bom} $.
Since  $\cA_{k, \bom} $ is a linear function of  $A_{k, \bom} $ it is equivalent to assume that  
 \be   \label{oops}
\sup_{1 \leq j \leq k}  L^{-\frac32(k-j)}  p_j^{-1} \sup_{p \in \de \Om_{j}^{(j)}} |dA_{j}(p)|    \leq 1 
\ee
 and show 
that there is an  $\cA' \sim \cA_{k+1, \bom^+}$ on $\sq^{\da} $  satisfying
\be   \label{tingting2}
L^{-\frac12(k-j)}|\cA'| ,\  L^{-\frac32(k-j)}|\pa \cA'|, \   L^{-(\frac32 + \al)(k-j)} \|  \pa \cA'\|_{(\al)}   \leq   C_0 M p_j
  \ee
  
The proof is by induction on $k$.  We assume it is true for $k$ and prove it for $k+1$.  The first step from $k=0$ to $k=1$
is  a special case.
We are assuming therefore that $A_{j, \de \Om_j}$ is axial   for $j=1, \dots k$ and that
\be \label{except1}
|dA_j| \leq L^{ \frac32 (k+1-j)}p_j  \textrm{  on } \de \Om_j \hs j=0, 1, \dots, k+1
\ee
  We want to show that for an  $L^{-(k+1-j)} M$ cube $\sq$ in $\de \Om_j$ the 
field $\cA_{k+1, \bom^+}$ on $\bbT^{-k-1}_{N-k-1}$ is gauge equivalent  on $\sq^{\da}$ to $\cA'$  satisfying
(\ref{tingting2}) with $k+1$ instead of $k$
 \bigskip

 \noindent
 \textbf{Part I}:  
Start with the axial minimizer $\cA^{\sx}_{k+1, \bom^+}$  on $\bbT^{-k-1}_{N-k-1}$  which has   $  A_{k+1, \bom^+}=\cQ_{k+1, \bom^+} \cA^{\sx} _{k+1, \bom^+}$.  To make contact
with our earlier discussion we scale 
up to  $\cA^{0,\sx}_{k+1, \bom^+}$ on $\bbT^{-k}_{N-k}$.
Recall  that 
$  \cA^{0, \sx}_{k+1, \bom} = \cH^{0,\sx}_{k,\bom} A^{\min}_{k, \bom^+} $ where $A^{\min}_{k, \bom^+} $   
\be
A^{\min}_{k, \bom^+}  = (A_{1, \de\Om_1}, \dots, A_{k \de \Om_k}, A^{\min}_{k, \Om_{k+1}} )
\ee
Each $A_{j, \de\Om_j }$ is a scaling of the original and now every entry is axial.  The bound (\ref{except1}) becomes
\be \label{except2}
|dA_j| \leq  C L^{\frac32(k-j)} p_j \textrm{  on } \de \Om_j \hs j=0, 1, \dots, k+1
\ee
As an approximation to $A^{\min}_{k, \bom^+} $   we take  we take something  over which we have more control and which is still axial gauge namely  
\be
A^{\st}  =(A_{1, \de\Om_1}, \dots, A_{k, \de \Om_k}, \cQ^{s,T} A_{k+1, \Om_{k+1}}  ) 
\ee
Note that for either  $A^{\min}_{k, \bom^+} $ or  $A^{\st}$ the second to last entry on $\de \Om^{(k)}_k$ and the last entry on $\Om^{(k)}_{k+1}$
combine to give a function $A^{\st}_k$   on $\Om^{(k)}_k$.  In this sense  $A^{\st} $ is a sequence associated with $\bom = (\Om_1, \dots, \Om_k)$
 not $\bom^+ = (\bom, \Om_{k+1})$.
  Let $\cA^{\st}$ on $\tk$  be the Landau gauge minimizer of $\|\cA \|^2$
subject to  $\cQ_{k,\bom} \cA = A^{\st}$.   Thus
\be
\cA^{\st}= \cH_{k, \bom} A^{\st}    
\ee
\bigskip

\noindent
\textbf{Part II: }
We digress to  prove some bounds.  First we claim that on $\de \Om_j$ for $j=1, \dots, k$
\be \label{starbound}
|dA^{\st}_j | \leq  C L^{\frac32(k-j)}p_j  \textrm{  on } \de \Om_j \hs j=0, 1, \dots, k
\ee
 This holds for $j=1, \dots, k-1$  by  (\ref{except2}).  The region $\de \Om_k$ which is here $  \Om_k$  has two parts $\Om_k - \Om_{k+1}$ and
 $\Om_{k+1}$.    In the former the bound again holds by (\ref{except2}).    It also holds on plaquettes with all bonds  in $\Om_{k+1}$
 since $|dA_{k+1, \Om_{k+1}}| \leq  C$ and so on such plaquettes
 \be
|dA^{\st}_k| =  |d \cQ^{s,T} A_{k+1, \Om_{k+1}}| =| \cQ^{e,T} dA_{k+1, \Om_{k+1}}| \leq L^2  \|dA_{k+1, \Om_{k+1}} \|_{\infty} \leq  Cp_{k+1}
\leq Cp_k 
 \ee
But there are plaquettes which have some bonds in $\Om_{k+1}$ and some in $\Om^c_{k+1}$ which require special treatment.
For example suppose $p = < x, x + e_{\mu}, x+ e_{\mu} + e_{\nu}, x + e_{\nu},x>$ is an edge block
with the points contained in the $L$-blocks $B(y), B(y +L e_{\mu}), B(y+ L e_{\mu} +L e_{\nu}), B(y +L e_{\nu})$ respectively.
Further suppose that $x+ e_{\mu} + e_{\nu}$ and $B(y+ Le_{\mu} +L e_{\nu})$ are contained in $\Om_{k+1}$ but the others
are in  $\de \Om_k \subset \Om^c_{k+1}$. 
Then  since
$(\cQ^{s,T}A_{k+1} )(x+ e_{\mu} , x+ e_{\mu} + e_{\nu}) = L  A_{k+1} (y+L e_{\mu}, y+L e_{\mu} + L e_{\nu}) $ 
\be
\begin{split}
dA^{\st}(p) = &
A_k (x, x + e_{\mu}) + L A_{k+1}( y +L e_{\mu}, y+ Le_{\mu} +L e_{\nu} )\\
& +  L A_{k+1}( y+ Le_{\mu} +L e_{\nu},y +L e_{\nu})   
+ A_k ( x + e_{\nu},x) \\
\end{split}
\ee
Now let $p' =  < y, y + Le_{\mu}, y+ Le_{\mu} + Le_{\nu}, y + Le_{\nu},y>$ and compare $dA^{\st}(p)$ with 
\be
\begin{split}
L^2dA_{k+1, \bom^+}(p') =  &
\B(\cQ A_k (y, y +L e_{\mu}) +  A_{k+1}( y +L e_{\mu}, y+L e_{\mu} +L e_{\nu} )\\
 & +   A_{k+1}( y+ Le_{\mu} +L e_{\nu},y + Le_{\nu})   
+\cQ  A_k ( y + Le_{\nu},y ) \B)L \\
\end{split}
\ee
which we are assuming is bounded by a constant.    The difference has terms like  $A_k (x, x + e_{\mu}) - L \cQ A_k (y, y +L e_{\mu}) $. 
Since $A_k$ is axial  we can use  (\ref{28palms}),  (\ref{29palms})  to estimate this by   $ \one L^2  \|d A_{k, \de \Om_k} \|_{\infty} \leq  Cp_k$. Arguing in this fashion
$|dA^{\st}_k| \leq Cp_k$ on any plaquette in  $\Om_{k}$ and hence
(\ref{starbound}) is established.  
\bigskip

Next we define
\be
B^{\st}   = \cQ_{k+1, \bom^+} (\cA^{0, \sx}_{k+1, \bom^+} -\cA^{\star} )=  A_{k+1, \bom^+} -\cQ_{k+1, \bom^+} \cA^{\star} 
\ee
which will come into play shortly.
We claim that 
\be \label{Bbound}
|B^{\st}| \leq C p_k
\ee
Since $ \cQ_{k, \bom} \cA^{\star} = A^{\st} $ and since  $ \cQ_{k+1, \bom^+} $ has
 an extra averaging operator $\cQ$ in $\Om_{k+1}$ we have
 \be
 \cQ_{k+1, \bom^+} \cA^{\star}  = \B(A_{1, \de\Om_1}, \dots, A_{k, \de \Om_k},[ \cQ  (A_{k, \de \Om_k}, \cQ^{s,T} A_{k+1, \Om_{k+1}}  )]_{\Om_{k+1}}\B) 
 \ee
 and therefore
 \be
B^{\st}   =  \B(0, \dots, 0, A_{k+1, \Om_{k+1}}- [ \cQ  (A_{k, \de \Om_k}, \cQ^{s,T} A_{k+1, \Om_{k+1}} )]_{\Om_{k+1}}\B) 
 \ee
 In the interior of $\Om_{k+1}$ the $A_{k, \de \Om_k}$ does not contribute and we have  $ \cQ  \cQ^{s,T} A_{k+1, \Om_{k+1}}   = A_{k+1, \Om_{k+1}}$
 for a net contribution  to $B^{\st}  $ of zero.   Thus $B^{\st}  $ is supported on the boundary of $\Om_{k+1}$.
 On  boundary bonds we write 
\be
 \cQ \B (A_{k, \de \Om_k}, \cQ^{s,T} A_{k+1, \Om_{k+1}} \B) 
= \cQ \B (A_{k, \de \Om_k}, 0 \B)  +  \cQ \B (0, \cQ^{s,T} A_{k+1, \Om_{k+1}} \B)
 \ee
The second term is again $A_{k+1, \Om_{k+1}}$ for a net contribution of zero.
Thus the only contribution  to $B^{\st} $ is the first term.   We evaluate it on a boundary bond $y, y+ Le_{\mu}$ with
say $y \in  \de\Om^{(k+1)}_k$ and $y +L e_{\mu} \in \Om^{(k+1)}_{k+1}$ by 
\be
\label{taco}
 \cQ \B (A_{k, \de \Om_k}, 0 \B)(y, y + L e_{\mu} )  =  \sum_{x \in B(y)} L^{-4} A_k( \Ga( x, x+ L e_{\mu} ) \cap \de \Om_k )
 \ee
 Since $A_k$ is axial  we have by (\ref{gunison2})  on bonds entirely in  $B(y) \subset \de \Om_k$ that  $|A_k| \leq  C \| d A_{k, \de \Om_k}\|_{\infty} \leq Cp_k$. 
 Using this in (\ref{taco}) gives  $ |\cQ  (A_{k, \de \Om_k}, 0 )(y, y + L e_{\mu} )  | \leq Cp_k$.
So  (\ref{Bbound}) holds. 
\bigskip

\noindent
\textbf{Part III: } We now return to the main argument. By (\ref{starbound}) and   the theorem for $k$ applied to $\cA^{\star} $ we conclude that for $\sq \subset \de \Om_j,  j=1,\dots, k$ we have  on $\sq^{\da}$
that  $\cA^{\st} \sim \cA'$   where
 \be  \label{peter1}
 L^{- \frac12(k-j)}|\cA'| ,\  L^{-\frac32(k-j)}|\pa \cA'|, \   L^{-(\frac32 + \al)(k-j)} \| \pa \cA'\|_{(\al) } 
 \leq CC_0M  p_j
\ee
Now scale $\cA^{\st}$  back  down from $\tk$  to $\bbT^{-(k+1)}_{N-k-1}$ but keep the same notation.  
The bound on the old $\de \Om_k = \Om_k$ is split into a bound on the new $\de \Om_k = \Om_k - \Om_{k+1}$ and $\Om_{k+1}$. 
Then with  an adjustment in the constant 
$C$ we still have the bound (\ref{peter1}) with $k+1$ instead of $k$.   Thus  for $\sq \subset \de \Om_j,  j=1,\dots, k+1$ we have  on $\sq^{\da}$
that  $\cA^{\star} \sim \cA'$   where
\be  \label{peter2}
\begin{split}
& L^{-(1 + \beta)(k+1-j)}|\cA'| ,\  L^{-\frac32(k+1-j)}|\pa \cA'|, \   L^{-(\frac32 + \al)(k+1-j)} \| \pa \cA'\|_{(\al) } 
   \leq CC_0M p_j  \\
\end{split}
\ee
This implies the global bounds on $d\cA^{\st}$ namely  $ \cA^{\st} \in \fA_{k+1, \bom^+}(CC_0M)$ as defined in (\ref{oscar1}).
 We also scale $A_{k+1, \bom^+},A^{\st}, B^{\st}$ and  keep the same notation.
We still have that  $B^{\st}    =  A_{k+1, \bom^+} -\cQ_{k+1, \bom^+} \cA^{\star} $
and  $|B^{\st}   | \leq C p_k$.

\bigskip 
We are studying 
 \be 
   \cA_{k+1, \bom^+}  = \textrm{ minimizer of  } \frac12 \| d \cA\|^2   \textrm{ subject to  }  
 \cQ_{k+1, \bom^+}\cA   =     A_{k+1, \bom^+} \textrm{ and }   R_{k+1, \bom^+}(\de \cA) =0  
 \ee  
We    reformulate  the problem relative to $\cA^{\star}$. We seek a  
critical point  of $ \cB \to \| d(   \cA^{\star} + \cB)\|^2$ subject to the conditions   
   $\cQ_{k+1, \bom^+}(\cB +  \cA^{\star}) =A_{k+1, \bom^+}$  and $\cR_{k+1, \bom^+} (\de (\cB +  \cA^{\star}) ) =0$
which is  the same as    $\cQ_{k+1, \bom^+}\cB   =     B^{\st}   $ and   $\cR_{k+1, \bom^+} (\de \cB ) =0$.   We have
$
 \cA_{k+1, \bom^+}=    \cA^{\st} + \cB^{\st} 
$
  where 
  \be 
   \cB^{\st}  = \textrm{ minimizer of  } \frac12  \| d(   \cA^{\star} + \cB)\|^2   \textrm{ subject to  }  
   \cQ_{k+1, \bom^+}\cB   =  B^{\st}    \textrm{ and }  R_{k+1, \bom^+}(\de \cB) =0  
 \ee  
 This is the same as finding the minimizer of    $ \frac12  \| d  \cB\|^2 + < \cB, \de d \cA^{\st} >$ with the same constraints.
 Again we compute the minimum with Lagrange multipliers,  but now with the extra term  $< \cB, \de d \cA^{\st} >$.  One finds
 \be
 \label{lemony}
 \cB^{\st} =   \cH_{k+1, \bom^+} B^{\st}   -  \tilde \cA +  \cH_{k+1, \bom^+}\cQ_{k+1, \bom^+} \tilde  \cA  +\cG_{k+1, \bom^+} (dR_{k+1, \bom^+} \de)  \tilde \cA 
 \ee
 where
 \be
 \tilde \cA \equiv   \cG_{k+1, \bom^+} \de d \cA^{\st} 
 \ee

 Now we claim that on $\de \Om_j$ for $j=1,2, \dots, k+1$
  \be    \label{yoyo}
  L^{-\frac12(k+1-j)}| \cB^{\st} | ,  L^{-\frac32(k+1-j)}|\pa  \cB^{\st} |,    L^{-(\frac32 + \al)(k+1-j)} \|\pa  \cB^{\st}\|_{\al}  
 \leq C C_0M p_j
\ee

The first term $ \cH_{k+1, \bom^+} B^{\st}$ satisfies these bounds  by  (\ref{standard4}) and the bound on $ B^{\st}$ which is supported on $\Om_{k+1}$.   

For the second term $\tilde \cA$
we use the fact that   $\cA^{\st} \in \fA_{k+1, \bom^+}(CC_0M)$  which says 
   on $\de \Om_j$  
 \be
 L^{-\frac32(k+1-j)}|d \cA^{\st}|,L^{-(\frac32+ \al)(k+1-j)}\|d \cA^{\st}\|_{(\al)}  \leq     CC_0M p_j
\ee
Now the bounds 
  (\ref{standard2.6}), (\ref{standard2.7}) and (\ref{standard2.8})  on $\cG_{k+1,\bom^+} \de $ yield the desired  bounds on $\tilde \cA$ on $\de \Om_j$:
  \be      \label{legion}  
  \begin{split}  
  \   L^{-\frac12(k+1-j) }  | \tilde \cA |  
   \leq &  C p_j  \sup_{j'}  L^{-\frac32(k+1-j') } p_{j'}^{-1}  \sup_{\de \Om_{j'}} |d \cA^{\st}| \leq CC_0M p_j \\
L^{-\frac32(k+1-j)}  | \pa \tilde \cA  |  
   \leq &  C  p_j   \sup_{j'} L^{-\frac32(k+1-j')} p_{j'}^{-1} \B(L^{- \ep(k+1-j')} \|d \cA^{\st} \|_{(\ep), \de \Om_{j'}}  + \| d \cA^{\st} \|_{\infty, \de \Om _{j'}} \B) \\
   \leq  &  CC_0M p_j \\
L^{-(\frac32 + \al )(k+1-j)}   \|   \pa \tilde \cA  \|_{(\al)} 
    \leq &  C  p_j  \sup_{j'}L^{-\frac32(k+1-j')} p_{j'}^{-1} \B( L^{-(\al+ \ep)(k+1-j')}\| d \cA^{\st}\|_{(\al + \ep), \de \Om_{j'}}  + \| d \cA^{\st}\|_{\infty, \de \Om_{j'}} \B) \\
    \leq & CC_0M  p_j\\
 \end{split}
 \ee

 For the third term in  (\ref{lemony}) has the desired bound by $|\cQ_{k+1, \bom^+} \tilde \cA| \leq  L^{\frac12(k+1-j) }CC_0M p_j$
 and (\ref{standard4}).  
 
 The last term in (\ref{lemony}) splits into   $\cG_{k+1, \bom^+} d\de   \tilde \cA$   
 and $- \cG_{k+1, \bom^+} dP_{k+1, \bom^+} \de  \tilde \cA $.  The first term here is estimated in the same way as $\tilde \cA$. 
 For the second term  we use the bound in $\de \Om_j$  (this is (2.88) in \cite{Bal84b} modified by weight factors )
 \be
L^{-\frac52(k+1-j)} |dP_{k+1, \bom^+} \de  \tilde \cA |
 \leq  p_j\sup_{j'} L^{- \frac12(k+1-j') } p_{j'}^{-1} \sup_{\de \Om_{j'}}|\tilde \cA| \leq CC_0 M p_j
\ee 
 Then by  (\ref{standard2.6}) with $\beta = \frac12 $  
   \be     
  \begin{split}  
& L^{- \frac12 (k+1-j)}  | \cG_{k+1,\bom^+}     dP_{k+1, \bom^+} \de  \tilde \cA |, \   L^{-\frac32(k+1-j) }  | \pa  \cG_{k+1,\bom^+}    dP_{k+1, \bom^+} \de  \tilde \cA|, \\ 
&  L^{-(\frac32  + \al  ) (k+1-j) }  \|  \pa   \cG_{k+1,\bom^+}      dP_{k+1, \bom^+} \de  \tilde \cA \|_{(\al) }  \\
& \leq   C  p_j \sup_{j'}  L^{- \frac52 (k+1-j') } p_{j'}^{-1} \sup_{\de \Om_{j'}} | dP_{k+1, \bom^+} \de  \tilde \cA|   \leq CC_0 M p_j  \\
\end{split}
  \ee
Thus (\ref{yoyo}) is established.

Combining  (\ref{peter2}) and (\ref{yoyo}) we have  that $\cA_{k+1, \bom^+}=   \cA^{\star} +\cB^{\st} $
is equivalent    in $\sq^{\da} \subset \bbT^{-k-1}_{N-k-1}$  to  a field  $\cA'$ satisfying 
 \be   \label{peter4}
  L^{-\frac12(k+1-j)}| \cA' | ,  L^{-\frac32(k+1-j)}|\pa  \cA' |,    L^{-(\frac32 + \al)(k+1-j)} \|\pa  \cA' \|_{(\al)}  
 \leq C C_0M p_j 
\ee
This is just what we want except that the constant on the right is too big. We need $C_0Mp_j$. 
\bigskip

 \noindent
\textbf{part IV }:  
We have to improve the last bound.  Of the developments so far                  
we  will use only  a consequence of  the crude bound (\ref{peter4}) which is 
$\cA_{k+1, \bom^+}  \in \fA _{k+1, \bom^+}(C C_0 M ) $.  
 To put it another way we are  reduced to showing that there is 
 a constant $C_0$ such that  for any $k$ if $\cA_{k, \bom}  \in \fA_{k, \bom}(C C_0 M) $ (and (\ref{oops}) holds)
then the conclusions of the theorem hold.

In general $\sq$ is an $L^{-(k-j)}M$ cube in $\de \Om_j$.  We take the special case $j = k$ so that $\sq$ is an $M$ cube in $\Om_{k}$.
The other cases can be obtained by scaling.

 Consider the buffer  of cubes  centered on $\sq$
 \be
 \bom(\sq) = ( \sq_1, \dots, \sq_k  ) \hs \sq_{k} \supset \sq^{\da}
 \ee
 defined with  $\sq_{k} \supset  \sq^{\da}$  and   $d(\sq^c_{k}, \sq^{\da}) = M$ and with 
  $\sq_j \supset \sq_{j+1}$ 
 and  $d( \sq^c_j, \sq_{j+1}) =L^{-(k-j)} M$.   Then $\sq_1$ has width less than $ CM$.  We  allow the possibility that $\sq \cap \de \Om_{k-1} \neq \emptyset$.  

\bigskip
Next define $\uA \sim \cA_{k, \bom}$  so that $\cQ_k \uA$ satisfies a generalized  axial gauge condition on all of $\sq_1$. That is we require
$(\cQ_{k} \uA)(\Ga(y,x) ) =0$ for $y$ the center of $\sq, \sq_1$ and $x \in \sq_1$.   We accomplish this  by $\uA = \cA_{k, \bom} - d\la$ with any $\la$
satisfying 
\be
(Q_k\la)(x) - (Q_k\la)(y)  = ( \cQ_{k} \cA_{k, \bom})(\Ga(y,x) ) 
\ee
Here $\la$ is supported on $\sq_1$ and outside of $\sq_1$ we have $\uA  = \cA_{k, \bom}$. 
Then $\uA$  leaves  the restricted  orbit $\cQ_{k,\bom} \cA = A_{k,\bom}$ and is now on the  restricted orbit   $\cQ_{k,\bom} \cA = A^{\sx}$
where
$ A^{\sx}=\cQ_{k,\bom} \uA$.

Next define $\cA^{\sx} \sim \uA    \sim \cA_{k, \bom}$ to be the axial representative of the orbit $\cQ_{k,\bom} \cA = A^{\sx}$ as in lemma \ref{sumpter} so  
\be  
A^{\sx} =  \cQ_{k,\bom} \cA^{\sx}
\ee
Then $\cQ_{k,\bom} \cA^{\sx} = \cQ_{k, \bom} \uA$.  In particular  $\cQ_{k-1} \cA^{\sx} = \cQ_{k-1}\uA$ on $\de \Om_{k-1}$
and $\cQ_{k} \cA^{\sx} = \cQ_{k} \uA$ on $ \Om_k$.  Hence   $ \cQ_{k} \cA^{\sx} =   \cQ_{k} \uA  $ is  generalized  axial on $\sq_1$.

  Since  $d\cA^{\sx} = d \uA =  d \cA_{k, \bom} $ we have $\cA^{\sx}  \in \fA_{k, \bom}(C C_0 M) $.
 Because $\cQ_{k}\cA^{\sx}$ is generalized  axial on a set of width $ \one M$ and from the bound on $d \cA^{\sx}$ we have on $\sq_1 $
 as in (\ref{gunison1}),(\ref{gunison2}):
 \be
  |\cQ_{k} \cA^{\sx} | \leq    \one M \| d \cA^{\sx} \|_{\infty}  \leq  CC_0  M^2
 \ee
Then by lemma  \ref{orange2}  we have on $ \de \Om^{(j)}_j $ 
\be 
\begin{split}
|A^{\sx}| = |\cQ_j \cA^{\sx} | \leq L^{k-j}\B(  2c_0L \| d\cA^{\sx} \|_{\infty}   + \|\cQ_{k} \cA^{\sx} \|_{\infty} \B)
\leq  L^{k-j}C C_0M^2
\end{split}
\ee

We want to get  a similar estimate with $\bom(\sq)$ instead of $\bom$.
But first we make a  small modification.   Split 
   $\sq_k$ 
by  
 \be
 \sq_{k}  =  \sq'_{k} \cup \sq_{k}'' \hs \hs  \sq'_{k} = \sq_{k} \cap \de \Om_{k-1} \hs \sq_{k}''  = \sq_{k} \cap  \Om_{k}
 \ee 
and then include the $\sq'_{k}$ piece with $\sq_{k-1}$.   So we define
\be
 \bom^{\bullet} (\sq) = (\sq_1, \dots, \sq_{k-1} \cup  \sq'_{k},  \sq_{k}'')
 \ee
and 
\be
 A'=   \cQ_{k, \bom^{\bullet}(\sq)} \cA^{\sx}
\ee
Again by lemma   \ref{orange2}  have a bound on $[\de \Om_j^{\bullet}(\sq)]^{(j)}$ 
\be  \label{purdy1}
  |A'| \leq    L^{k-j}C C_0M^2
\ee
 We also have that      $A^{\sx} = A'$ on $\sq_{k}$ since  both $ \cQ_{k, \bom} \cA^{\sx}$ and $  \cQ_{k, \bom^{\bullet}(\sq)}\cA^{\sx}$
have averaging $\cQ_{k-1}$ in $  \sq'_{k+1}$ and $\cQ_{k}$ in  $   \sq''_{k}$. 
Also  $ \Om^{\bullet}_j (\sq)  \subset  \Om_j$.
These are  the reasons for introducing $ \bom^{\bullet} (\sq) $.

Note  that since     $\cA^{\sx} \sim \cA_{k,\bom}$ it is still on a minimizing orbit.   Hence it is a minimizer on its restricted orbit and so 
 \be 
   \cA^{\sx}  \textrm{ is a minimizer of  }  \| d\cA\|^2    \textrm{ subject to  } \cQ_{k, \bom}\cA   =      A^{\sx}    
 \ee  
 We also   note  that     $  \Om^{\bullet}_j (\sq)  \subset  \Om_j$ implies that
\be 
 \label{certain}
   \cA^{\sx}   \textrm{ is a  minimizer of  }  \| d\cA\|^2    \textrm{ subject to  } 
   \cQ_{k, \bom^{\bullet}(\sq)}\cA   =      A'    
 \ee  
This follows since   $\{ \cA: \cQ_{k, \bom^{\bullet}(\sq)} \cA = A'\}$ is contained
in   $\{ \cA: \cQ_{k, \bom }\cA = A^{\sx}\}$ as can be seen by adding  the same  extra averaging operators
to each side of $\cQ_{k, \bom^{\bullet}(\sq)} \cA = A'$. Hence the minimizer in the latter set gives a minimizer in the former set. 

We need improved estimates on $A^{\sx}= A'$ on $\sq_{k}$.  From (\ref{oops})
 we  are assuming  $|dA_{k-1}| \leq L^{2 + \beta}$ on $\de  \Om_{k-1}$ and $|dA_{k}| \leq 1$ on $ \Om_{k}$.
However  
\be
dA^{\sx} = d\cQ_{k, \bom}\cA^{\sx}=   \cQ^{(2)}_{k, \bom}d\cA^{\sx} =
\cQ^{(2)}_{k, \bom}d\cA_{k, \bom} = d\cQ_{k, \bom}\cA_{k, \bom} = dA_{k, \bom}
\ee
Hence $|dA^{\sx}| \leq L^{2+ \beta}$ on $\de  \Om_{k-1}$ and $|dA^{\sx} | \leq 1$ on $ \Om_{k}$.

Let   $A^{\#}$  on $\sq^{(k)}_{k}$ be the generalized axial function  $A^{\#} =  \cQ_{k} \cA^{\sx}$.  This  is the same as
 \be
 A^\#(b) =   \begin{cases}  \cQ A^{\sx}(b)  &  b \in    \sq'^{(k)}_{k} \\
 A^{\sx}  (b)  &  b \in    \sq''^{(k)}_{k} \\
\end{cases}
\ee
Then we have  (even for boundary plaquettes) 
 \be
 dA^\#(p) =   \begin{cases}  (\cQ^{(2)} dA^{\sx})(p)  &  p \in    \sq'^{(k)}_{k} \\
 dA^{\sx}  (p)  &  p \in    \sq''^{(k)}_{k} \\
\end{cases}
\ee
It follows that 
\be
|d A^{\#}| \leq    \one 
\ee

Because $A^{\#}$ is generalized axial gauge  on a set of size $C M$ we have as in as in (\ref{gunison1}),(\ref{gunison2}) that 
$ | A^{\#}|  \leq C M \|dA^{\#}\|_{\infty}$ and so 
\be 
| A^{\#}| \leq C  M  
\ee
The last bound implies on $ \sq_{k}'' $ that  $|A^{\sx} | =|A^\#|  \leq C M $. Furthermore on $\sq'_{k}$  the bounds $|dA^{\sx} |  \leq L^{2 + \beta}$ 
 and $|\cQ A^{\sx}|  = |A^\#| \leq  C M$  and the axial gauge condition imply by  lemma \ref{orange}  that 
$|A^{\sx}  |   \leq  CM $.    
Altogether then we have the  bound  on $\sq_{k}$
\be  \label{purdy2}
|A^{\sx} | \leq  CM \hs  \textrm{ and hence } \hs    |A'| \leq  CM 
 \ee
(This bound on $A'$ would not work on all of $\sq_1$ since $A^{\sx}= A'  $ is not true there.)

Now let $\cA' \sim \cA^{\sx}$ be the Landau gauge representative  of the restricted  minimizing orbit $\cQ_{k, \bom^{\bullet}(\sq)} \cA = A'$. 
We  are ready for a sharp bound on $\cA'$  which  is given by 
\be 
\cA' = \cH_{k, \bom^{\bullet}(\sq)} A'
\ee
We  make   the following  estimate on a unit cube  $ \De_x \subset  \sq^{\da}$.      By  (\ref{standard3}),  then (\ref{purdy1}) and  (\ref{purdy2}), and then (\ref{multisum}) with $\de \Om'_j =  [\de \Om^{\bullet}_j(\sq)]^{(j)}$
\be
 \begin{split}
    \label{peter5}
&  | \cA' | ,  |\pa   \cA'  |,    \|  \pa \cA'    \|_{\al} \\
 & \leq  C     \sum_j \sum_{b \in  \de \Om'_j } e^{- \ga d_{\bom^{\bullet}(\sq)}(x, b)} |A'(b)| \\
&  \leq C M \sum_{b \in (\de \Om'_{k-1} \cup \Om'_k ) \cap \sq_k } e^{- \ga d_{\bom^{\bullet}(\sq)}(x, b)}    
  + C C_0 M^2 \sum_j  \sum_{b \in  \de \Om'_j - \sq_k} e^{-\ga  d_{\bom^{\bullet}(\sq)}(x, b) }L^{k-j} 
  \\
  & 
  \leq CM  +    CC_0M^2    e^{-\ga M}  \leq  C_0M 
\end{split}
\ee
Here we used for $b \in \sq_k^c$  that   $d_{\bom^{\bullet} (\sq)}(x, b) \geq M$ to extract a factor $e^{-\ga M} $.  We also  used for $b \in \de \Om'_j $
$L^{k-j'}  \leq  e^{ \one M^{-1} d_{\bom^{\bullet}(\sq)}(x,b) } $. 
 In  the last step we have chosen  $C_0 = \frac12 C$  and  $M$ sufficiently large so  $CM    e^{-\ga M} < \frac12$.  This is the desired result for $j=k$.

 To summarize since   $
  \cA_{k, \bom} \sim  \cA^{\sx}  \sim   \cA' $
 on $  \sq^{\da}$  we have the  desired result (\ref{tingting2})  for any $k$ with the extra $\fA_{k, \bom}(CC_0M)$ assumption.
 Hence the result  (\ref{tingting2})  holds for $k+1$ in our inductive argument.  This completes the proof.

 \appendix

 \section{Identities for Dirac operators} \label{A}

We develop two identities for the operator 
\be  
 \Ga_{k, \bom^+, y }(\cA)   = \B[    D_{k, \bom} (\cA)  +  bL^{-1} P(\cA)  +  i\ga_3 y  \B]^{-1} _{\Om_{k+1}}   
\ee
 on  $\Om^{(k)}_{k+1} \subset \tz$ and the operator
\be   \label{stinger}  
S_{k, \bom^+,y}(\cA)  = \B[ \fD_{\cA}     + \bar m _k  +    P_{k,\bom}(\cA)    -  b_k^2  [   Q^T_k(-\cA)   \sB_{k,y} (\cA)  Q_k(\cA)]_{\Om_{k+1}}    \B]^{-1} _{\Om_1} 
 \ee
 on $\Om_1 \subset \tk$,
 where  
\be
\sB_{k,y}(\cA)   =         \Big(   b_k  +  bL^{-1} P(\cA) + i\ga_3y  \Big) ^{-1}=      \frac{1}{b_k+ i\ga_3y}  (I -  P(\cA) )   +   \frac{1}{ b_k + bL^{-1}+ i\ga_3y }  P(\cA)  \\
 \ee
There is an alternate representation of  $S_{k, \bom^+,y}(\cA) $.   To derive it 
write  $  P_{k,\bom}(\cA)  =      [P_{k, \bom}(\cA)]_{\Om^c_{k+1} }+ b_k [P_k(\cA)]_{\Om_{k+1}}$  
and calculate  in   $\Om_{k+1}$ as in \cite{Dim15b}, Appendix B      
\be
\begin{split}  
&  b_kP_k(\cA)   -   b_k^2  Q^T_k(-  \cA)  \sB_{k,y}(\cA )      Q_k(  \cA)  =    \al_k   P_k(\cA)   +\beta_k  P_{k+1} (\cA)
  \\
 \end{split}
\ee
where
\be
 \al_k =    \frac{b_k  i\ga_3y  }{b_k + i\ga_3y}  
    \hs    \beta_k  = \frac{b_k^2 bL^{-1} }{( b_k + bL^{-1}+ i\ga_3y) (b_k + i\ga_3y )}
 \ee
Thus we have  
\be  
\begin{split}
&     S_{k, \bom^+,y}(\cA) = \B[ \fD_{\cA}     + \bar m _k  +  [P_{\bom}(\cA)]_{\Om^c_{k+1} } + \al_k[  P_k(\cA)  ]_{\Om_{k+1}}
   + \beta_k  [P_{k+1} (\cA)]_{\Om_{k+1} } \B]^{-1}_{\Om_1}  \\
   \end{split}
  \ee   

The following identities generalize results in \cite{BOS91}.

\begin{lem}  
\be
  \Ga_{k, \bom^+,y }(\cA)    =   
\B[   \sB_{k,y}(\cA)   +   b_k^2  \sB_{k,y}(\cA) Q_k(\cA)   S_{k, \bom^+,  y}(\cA)Q_k^T(-\cA) \sB_{k,y}(\cA)  \B]_{\Om_{k+1} }
\ee
\end{lem}
\bigskip

\rem  In case $y=0$   we have $\al_k = 0$ and $\beta_k = b_{k+1}L^{-1}$ and so $S_{k, \bom^+,  0}(\cA) = 
 S^0_{k+1, \bom^+}(\cA) =  [ \fD_{\cA}     + \bar m _k  +  P^0_{k+1, \bom^+}(\cA)]^{-1}_{\Om_1} $.
The identity relates this to   $ \Ga_{k, \bom^+ }(\cA)   = [    D_{k, \bom} (\cA)  +  bL^{-1} P(\cA)  ]^{-1} _{\Om_{k+1}}   $
  by
\be
  \Ga_{k, \bom^+ }(\cA)    =   
\B[   \sB_{k}(\cA)   +   b_k^2  \sB_{k}(\cA) Q_k(\cA)   S^0_{k+1, \bom^+}(\cA)Q_k^T(-\cA) \sB_{k}(\cA)  \B]_{\Om_{k+1} }
\ee
where   now  
\be  \sB_{k}(\cA)   =b_k^{-1}  (I -  P(\cA) )   +  ( b_k + bL^{-1} )^{-1}  P(\cA) \ee
\bigskip  

\pr  Let $\bar J, J$ and $\bPsi_k, \Psi_k$ be Grassmann variables indexed  by $\Om^{(k)}_{k+1} \subset \tz$.
 Then 
\be  \label{urgent}
\begin{split}
& \exp  \B(   < \bar J,  \Ga_{k, \bom^+,y }(\cA)  J  >  \B)  \\
=  &  \const \int     \exp \B(  \blan \bPsi_k,  J \bran +  \blan \bar J, \Psi_k \bran     -  \blan  \bPsi_k, \B(  D_{k, \bom} (\cA)+bL^{-1}P(\cA)+  i\ga_3y  \B) \Psi_k \bran 
\ D\Psi_{k}    \\
\end{split}
\ee
By (\ref{lurch0}) with $F=1$ and  (\ref{fable}) and (\ref{tootootoo}) we have
\be \begin{split}
&        \exp \B(   -  \blan  \bPsi_{k, \bom},   D_{k, \bom} (\cA)  \Psi_{k, \bom} \bran  \B)  =  \const    \\
 &  \int   \exp \left(   - \blan  (\bPsi_{ k,  \bom}- Q_{ k,  \bom} (- \cA ) \bpsi),   \bb^{(k)}  (  \Psi_{ k,  \bom}- Q_{ k, \bom} (\cA ) \psi )  \bran_{\Om_1}
  -  \blan   \bpsi,(\fD_{ \cA } + \bar   m_k) \psi  \bran  \right)   \  D \psi_{\Om_1}  
    \\
\end{split}  
 \ee  
Specializing to $\Psi_{k, \bom} = \Psi_k$  (i.e. just looking at parts only depending on $\Psi_k$)  this can be written 
 \be
\begin{split}
&  \exp \B(    -   \blan \bPsi_k,  D_{k, \bom} (\cA) \Psi _k  \bran  \B)  \\  
 =&    \const    
 \int  \exp \left(   - \blan  (\bPsi_k- Q_{k, \bom} (- \cA ) \bpsi),   \bb  (  \Psi_{k}- Q_{k, \bom} (\cA ) \psi )  \bran
  -  \blan   \bpsi,(\fD_{\cA }+ \bar   m_k) \psi  \bran  \right)    \   D \psi_{\Om_1}   \\
  =&       \const 
 \int   \exp \B(   - b_k\blan  \bPsi_k,   \Psi_{k} \bran
  +  b_k  \blan  \bPsi_k,   Q_{k, \bom} (\cA ) \psi   \bran \\
 &  \hs \hs+  b_k \blan   Q_{k, \bom} (- \cA ) \bpsi,    \Psi_{k} \bran
  -  \blan   \bpsi,(\fD_{ \cA }+ \bar   m_k + P_{k, \bom}(\cA) ) \psi  \bran  \B)    \  D \psi_{\Om_1}   \\
\end{split}  
 \ee  
Insert this into (\ref{urgent})   and  do the integral  over  $\Psi_k$  which is 
\begin{equation}
\begin{split}
& 
    \int   \exp \B(  \blan\bPsi, J +b_k Q_k(\cA) \psi\bran   +    \blan\bar  J  +b_k Q_k(-\cA) \bpsi  , \Psi_k   \bran 
     - \blan \bPsi_k,  \Big(   b_k  +   bL^{-1}P(\cA) +  i\ga_3y \Big)  \Psi _k \bran     \B)  \   D\Psi_{k}
      \\
 &=  \const    \exp   \Big(   \blan (\bar  J  +b_k Q_k(-\cA) \bpsi) ,  \sB_{k,y} (\cA)  (J +b_k Q_k(\cA) \psi) \bran_{\Om_{k+1}} \B)  
 \\
\end{split} 
\end{equation}
This gives  
\begin{equation}
\begin{split}
&\exp  \B(  < \bar J,  \Ga_{k}(\cA) J  >  \B)
=  \const \int    \ D \psi_{\Om_1} \\
& \exp   \Big(       \blan (\bar  J  +b_k Q_k(-\cA) \bpsi) ,  \sB_{k,y} (\cA)  (J +b_k Q_k(\cA) \psi)    \bran_{\Om_{k+1} } 
 -       \blan \bpsi,  \B( \fD_{\cA}     + \bar m _k     +        P_{k, \bom}(\cA) \B) \psi  \bran  \Big)  
 \\
= &     
 \const
 \int    \exp   \Big(   <  \bar  J ,  \sB_{k,y}(\cA)   J  > + \blan b_k Q_k^T(\cA)  \sB^T_{k,y}(\cA)\bar  J  ,   \psi \bran_{\Om_{k+1} }   
+   \blan \bpsi ,b_k  Q_k^T(-\cA) \sB_{k,y}(\cA)   J \bran_{\Om_{k+1} }   \Big)
  \ 
 \\
& \hs   \hs    -     \blan \bpsi,  \B( \fD_{\cA}     + \bar m _k +     P_{k, \bom }(\cA)    -  b_k^2  [   Q^T_k(-\cA)   \sB_{k,y} (\cA)  Q_k(\cA)]_{\Om_{k+1}}    \psi  \bran  \B)   \ D \psi_{\Om_1} \\
=& \const \exp \Big(    <  \bar  J ,  \sB_{k,y} (\cA)  J  > 
+   b_k^2  \blan   \bar J  , \sB_{k,y}(\cA)Q_k(\cA)  S_{k,\bom^+,y}(\cA) Q_k^T(-\cA) \sB_{k,y}(\cA)  J    \bran   \Big)  \\
\end{split}
  \end{equation}
  and the result follows.
 \bigskip

    \begin{lem} \label{deer}
\be  \label{oshkosh1}
  S_{k, \bom^+, y}(\cA)    =  S_{k, \bom}(\cA)   +   \cH_{k, \bom}(\cA)  \Ga_{k,\bom^+,y} (\cA)  \cH^T_{k, \bom}( \cA)   
 \ee
 \end{lem}  
 \bigskip
 
 \rem In case $y=0$ this says
 \be \label{oshkosh2}
 S^0_{k+1, \bom^+}(\cA)    =  S_{k, \bom}(\cA)   +   \cH_{k, \bom}(\cA)  \Ga_{k, \bom^+} (\cA)  \cH^T_{k, \bom}( \cA)   
\ee

 \pr    
Start with     
 \be
\begin{split}
& \exp \B(   \blan \bar J,    S_{k,\bom^+, y}(\cA)   J \bran  \B) \B) =  \const \int \  D\psi_{\Om_1}    \\
 &   \exp \left(  \blan \bpsi,  J \bran +  \blan \bar J, \psi \bran    -  \blan \bpsi, \B(   P_{k, \bom}(\cA)  -    b_k^2 [Q_k^T(-\cA) \sB_{k,y}(\cA)  Q_k(\cA)]_{\Om_{k+1}}  \B) \psi \bran 
  -   \blan \bpsi, \B(  \fD_{\cA} +    \bar m_k  \B) \psi \bran  \right)  \\
\end{split}
\ee
Make the split $ P_{k, \bom}(\cA) = [  P_{k, \bom}(\cA) ]_{\Om^c_{k+1} }+  [b_kP_k(\cA)]_{\Om_{k+1}}$
and use the identity  
\begin{equation}
\begin{split}
&       \exp   \Big(   -    \blan    \bpsi,  \B [ b_k P_k(\cA)  -    b_k^2 Q_k^T(-\cA) \sB_{k,y}(\cA)  Q_k(\cA) \B]_{\Om_{k+1}} \bpsi  \bran 
   \Big) \\
=&  \const    \int   \exp \left(   -  \blan  \bPsi_k, \B(b L^{-1}P(\cA)+  i\ga_3y  \B) \Psi_k \bran    - b_k \blan   \bPsi_k    -   Q_k(-\cA)  \bpsi,     \Psi _k   -   Q_k(\cA)  \psi \bran_{\Om_{k+1}}    \right)
 D\Psi_{k, \Om_{k+1}}   \\
\end{split} 
\end{equation}
  Substitute this   above and use 
\be
\begin{split}
& \blan \bpsi,  [P_{k, \bom}(\cA)]_{\Om^c_{k+1} } \psi \bran
+ b_k \blan  \bPsi_k    -   Q_k(-\cA)  \bpsi,   \Psi _k   -   Q_k(\cA)  \psi  \bran_{\Om_{k+1}}\\
 = &\blan  (\bPsi_k    -   Q_{k, \bom}(-\cA)  \bpsi ),    \bb^{(k)} (  \Psi _k   -   Q_{k, \bom} (\cA)  \psi ) \bran \\
 \end{split}
\ee
  Then the  $\psi$ integral is      
\be   
\begin{split}
&   \int   \exp \left(  \blan \bpsi,  J \bran +  \blan \bar J, \psi \bran    -\blan  (\bPsi_k    -   Q_{k, \bom}(-\cA)  \bpsi ),    \bb^{(k)} (  \Psi _k   -   Q_{k, \bom} (\cA)  \psi ) \bran  -   \blan \bpsi, \B(  \fD_{\cA} +    \bar m_k \B) \psi \bran  \right)      D \psi   \\  
  &   =    \const      \exp \left( \blan    \cH_{k, \bom}(\cA)  \bPsi_{k},  J \bran +  \blan \bar J, \cH_{k, \bom}(\cA)  \Psi_{k} \bran  
      -     \blan   \bPsi_{k} , D_{k, \bom}(\cA)     \Psi_{k}    \bran   +      \blan \bar J,    S_{k, \bom}(\cA)   J \bran  \right)  \\
\end{split}
 \ee
 which follows  by expanding around the critical point     $\psi   =   \cH_{k, \bom}(\cA)  \Psi_{k} + \cW$
 and     $\bpsi  =  \cH_{k, \bom}(\cA)  \bPsi_{k} + \bar \cW$ as in section \ref{2.1}  (here with $\Psi_{k, \bom} = \Psi_{k, \Om_{k+1}}$).
 Then we have             
  \be
\begin{split}
&    \exp \B(   \blan \bar J,    S_{k,\bom^+,y}(\cA)   J \bran  \B)     \B)    = \const    \exp \B(      \blan \bar J,    S_{k, \bom}(\cA)   J \bran \B)  \\
&\int    
   \exp \left( \blan    \cH_{k, \bom}(\cA)  \bPsi_{k},  J \bran +  \blan \bar J, \cH_{k, \bom}(\cA)  \Psi_{k} \bran  
      -     \blan   \bPsi_{k} ,  \B[ D_{k, \bom}(\cA)  +    bL^{-1} P(\cA)  + i \ga_3 y \B]    \Psi_{k}    \bran    \right)  
D  \Psi_{k, \Om_{k+1}}     \\    
=   &       \const    \exp \B(        \blan \bar J,    S_{k, \bom}(\cA)   J \bran  
+         \blan  \cH_{k, \bom}^T(\cA)    \bar J ,  \Ga_{k,\bom^+, y} (\cA)       \cH^T_{k, \bom}(\cA)  J  \bran        \B)
\end{split}
\ee
 and the result follows.

\end{document}